\documentclass[aps,reprint,superscriptaddress,nofootinbib,twocolumn]{revtex4-2}
\usepackage{amsmath, latexsym, amssymb, hyperref, graphicx, slashed, color, amsfonts, bbold}
\usepackage[T1]{fontenc}
\usepackage{multirow}

\hypersetup{colorlinks, citecolor=nicegreen,linkcolor=nicered, urlcolor=darkblue}
\definecolor{nicered}{rgb}{.7,.1,.1}
\definecolor{nicegreen}{rgb}{.2,.7,.1}
\definecolor{lightgreen}{rgb}{.6,.9,.5}
\definecolor{darkblue}{rgb}{0,0,.5}
\definecolor{el}{rgb}{.9, .8, .7}
\definecolor{mu}{rgb}{.8, .7, .8}

\newcommand\eV{\text{eV}}

\newcommand\GeV{\text{GeV}}
\newcommand\TeV{\text{TeV}}

\begin{document}

\title{Enabling Precise Predictions for Left-Right Symmetry at Colliders}

\author{Jonathan Kriewald}
\email{jonathan.kriewald@ijs.si}
\affiliation{\normalsize \it  Jo\v zef Stefan Institute, Jamova 39, 1000 Ljubljana, Slovenia}

\author{Miha Nemev\v{s}ek}
\email{miha.nemevsek@ijs.si}
\affiliation{\normalsize \it  Jo\v zef Stefan Institute, Jamova 39, 1000 Ljubljana, Slovenia}
\affiliation{\normalsize \it  Faculty of Mathematics and Physics, University of Ljubljana, 
Jadranska 19, 1000 Ljubljana, Slovenia}

\author{Fabrizio Nesti}
\email{fabrizio.nesti@aquila.infn.it}
\affiliation{\normalsize \it 
  Dipartimento di Scienze Fisiche e Chimiche, Universit\`a dell'Aquila, via Vetoio, I-67100, L'Aquila, Italy}
\affiliation{\normalsize \it INFN, Laboratori Nazionali del Gran Sasso, I-67100 Assergi (AQ), Italy}

\date{\today}

\vspace{1cm}

\begin{abstract} \noindent 
  We investigate the structure of the minimal Left-Right symmetric model  that enables precise predictions in the gauge, scalar and neutrino sector.
  We revisit the complete set of mass spectra and mixings for the charged and neutral gauge bosons, 
  would-be-Goldstones and gauge fixing, together with the ghost Lagrangian.
  In the scalar sector, we analytically re-derive all the massive states with mixings and devise a non-trivial physical 
  input scheme, expressing the model couplings in terms of masses and mixing angles.
  Fermion couplings are also determined in closed form, including the Dirac mixing in the neutrino sector, 
  evaluated explicitly using the Cayley-Hamilton theorem.
  These analytic developments are implemented in a comprehensive FeynRules model file.
  We calculate the one loop QCD corrections and provide a complete UFO file for NLO studies, 
  demonstrated on relevant hadron-collider benchmarks.
  We provide various restricted variants  of the model file with different gauges, massless states,
  neutrino hierarchies and parity violating $g_L \neq g_R$ gauge couplings.
\end{abstract}


\pacs{12.60.Cn, 14.70.Pw, 11.30.Er, 11.30.Fs}

\maketitle


%
%
\section{Introduction} \label{sec:Intro}

\noindent Understanding the microscopic nature of forces and the spontaneous origin of mass are the two core 
attributes of the Standard Model (SM)~\cite{Weinberg:1967tq}.
Neutrino oscillations have now firmly established that neutrinos are massive, in contrast to the SM, and 
recent significant experimental progress determined their mixing properties to unprecedented precision.
Still, the nature of neutrino mass (Dirac vs.\ Majorana) and even more importantly its origin, remain
an unsolved mystery of particle physics.
Contrary to charged fermions, whose origin is tightly connected to the Higgs mechanism, as the LHC data 
keeps confirming~\cite{ATLAS:2022vkf, CMS:2022dwd}, we do not know which theoretical framework 
is responsible for neutrino mass generation.
Another unsolved puzzle in the SM is the glaring parity asymmetry of weak interactions.
With the enduring SM experimental success, we got used to the chiral nature of weak interactions and are 
taking it for given, but the reason behind it remains as unclear as it was at the time of its discovery.

Left-Right (LR) symmetry addresses both of these shortcomings within a single framework by postulating
parity invariance of weak interactions via interchangeable $SU(2)_L \times SU(2)_R$ local symmetries, 
featuring new right-handed gauge bosons $W_R$ and $Z_R$.
The need for anomaly cancellation automatically brings in three right-handed neutrinos $N_{i = 1,2,3}$, 
whose mass is tied to the spontaneous breaking scale of $SU(2)_R$.
Moreover, the $U(1)$ charge assignment becomes very simple and parity symmetric (or vector-like) and is 
given by $U(1)_{B-L}$.

Gauging lepton number indicates that the Higgs sector may (and in fact does) lead to lepton number 
violation (LNV), if neutrinos couple to the Higgs with a Majorana-type Yukawa coupling.
The original works addressed the issue of parity restoration~\cite{Pati:1974yy, Mohapatra:1974gc} with 
soft breaking, but it was soon realized that spontaneous breaking works with both 
doublets~\cite{Senjanovic:1975rk} and $\Delta_{L,R}$ triplets~\cite{Senjanovic:1978ev, Mohapatra:1979ia, Senjanovic:1979cta}. 
The latter option allows for Majorana mass terms for heavy and light neutrinos, which led to 
the celebrated seesaw mechanism~\cite{Minkowski:1977sc, Mohapatra:1980yp} and features LNV.
We shall refer to it as the minimal Left-Right symmetric model (LRSM).
Unlike GUT-inspired seesaw scenarios~\cite{Gell-Mann:1979vob, Glashow:1979nm}, the LRSM scale 
need not be much above the electroweak scale, since the interactions do not mediate proton decay.

Indeed, prior to the advent of LHC, the most stringent constraints on the LRSM scale came from
flavor physics.
The point here is that, as usual in gauge theories, flavor asymmetry comes from the Yukawa sector
when the bi-doublet $\phi$ couples to quark doublets $Q_{L,R}$.
When LR symmetry, in either $\mathcal P$ or $\mathcal C$ form, is imposed on the Lagrangian, Yukawa 
couplings are constrained to be nearly hermitian or symmetric.
This implies similar flavor mixings in the quark sector $V^q_R \sim V^q_L$ and has led to sig\-ni\-fi\-cant
constraints since the Tevatron era~\cite{Senjanovic:1979cta, Beall:1981ze, Mohapatra:1983ae, Ecker:1985vv}
that pushed the scale beyond direct detection capabilities of colliders for some time.
The increase of center-of-mass energy at the LHC suffices to probe the scales beyond
the flavor constraints~\cite{Zhang:2007da, Zhang:2007fn, Maiezza:2010ic, Bertolini:2013noa, Maiezza:2014ala,
Bertolini:2019out, Bertolini:2020hjc, Bernard:2020cyi, Dekens:2021bro}, which spurred a 
number of collider studies, see reviews in~\cite{Atre:2009rg, BhupalDev:2012zg, Cai:2017mow, Abdullahi:2022jlv}.

Arguably the most fundamental signal one can look for is the Keung-Senjanovi\'c(KS)~\cite{Keung:1983uu} 
production of heavy Majorana neutrinos through $W_R$-mediated gauge interactions.
With even a tiny amount of data, the recast of leptoquark searches led to a significant 
bound~\cite{Nemevsek:2011hz} that was extensively characterized across the $M_{W_R}-m_N$
plane~\cite{Nemevsek:2018bbt} from well separated $\ell^\pm \ell^\pm j j$ to boosted
neutrino jets~\cite{Mitra:2016kov, Dey:2022tbp} with a displaced vertex~\cite{Nemevsek:2018bbt, Alimena:2019zri}
and finally invisible $N$~\cite{Nemevsek:2018bbt}.
(Some of these bounds are relaxed if CKM is different in left- and right-handed 
interactions~\cite{Langacker:1989xa, Frank:2018ifw, Solera:2023kwt}.)
With enough data it may be possible to characterize the chirality~\cite{Rizzo:2007xs, Han:2012vk}, 
disambiguate production channels~\cite{Dev:2015kca}, look for CP phases in case of 
degeneracy~\cite{Gluza:2016qqv} and include top final states~\cite{Frank:2023epx}.
At even smaller masses, decays of hadrons and mesons become important when a significant amount
of decaying states is available~\cite{Mandal:2017tab, Godbole:2020jqw, Alves:2023znq}.
At very high $W_R$ masses, off-shell production takes place~\cite{Ruiz:2017nip}.
Looking into the future, the reach of an FCC-hh was estimated to be around 
30 TeV~\cite{Nemevsek:2023hwx}, in agreement with other studies~\cite{Rizzo:2014xma, Ng:2015hba, 
Golling:2016gvc, Ruiz:2017nip}.
Similarly, the potential sensitivity of displaced vertices at lepton collider can probe the $M_{W_R}$ up 
to $\sim 26$ TeV at FCC-ee and $\sim 70$ TeV for a muon collider~\cite{Urquia-Calderon:2023dkf}.

At present, many experimental searches by CMS and ATLAS are targeting $W'$ and $Z'$ resonances.
Here we focus on $W_R$, which is lighter than $Z_R$ and thus sets the most stringent constraint 
on the LR breaking scale and we only review the most recent experimental updates.
The KS channel~\cite{CMS:2021dzb, ATLAS:2023cjo} searches are looking for resolved as well as boosted 
(but prompt) topologies of $N$ decays.
In the low $m_N$ regime, the $W_R \to \ell N$ channel is equivalent to $W' \to \ell \nu$, because 
$N$ becomes long-lived enough to decay outside of the detector.
One can then recast both leptonic ($e, \mu$)~\cite{ATLAS:2019lsy, CMS:2022krd} as well as $\tau$ final 
states~\cite{ATLAS:2018ihk, ATLAS:2021bjk, CMS:2022ncp}.
In the LR symmetric case, the CKM in the left and right-handed sectors are similar and one has direct 
limits from dijet~\cite{ATLAS:2019fgd, CMS:2019gwf} and $t b$ resonance~\cite{CMS:2023gte, ATLAS:2023ibb} 
searches.
The final state topologies here are clearly independent of $m_N$ and so is the resulting limit on 
$M_{W_R}$, apart from the slight suppression of the Br when $N$ channels open up.
Finally, $W_R$ can decay to SM gauge bosons $W,Z$ and Higgs $h$, leading to further dedicated 
searches~\cite{ATLAS:2022enb, CMS:2022pjv}.

The imposition of LR-symmetry has important ramifications in the leptonic sector, such as $\mu \to e \gamma$, 
which was the original motivation for seesaw~\cite{Minkowski:1977sc}.
When the triplet Yukawa couplings are restricted to a near-hermitian or symmetric form, Dirac and Majorana 
couplings become directly related.
In the case of $\mathcal C$~\cite{Nemevsek:2012iq} this connection allows one to calculate the Dirac mass 
in terms of Majorana masses, via a square root of a matrix.
The new analytic solution found in this work gives a direct insight into the Dirac-Majorana connection from 
colliders, where Majorana masses may be measured, and connect it to any Dirac-mediated process.
These processes appear either at colliders~\cite{Nemevsek:2012iq, Chen:2013foz, Arbelaez:2017zqq, Li:2022cuq}, 
nuclear $0\nu2\beta$~\cite{Mohapatra:1980yp, Tello:2010am, Nemevsek:2011aa, Barry:2013xxa, deVries:2022nyh}, 
electron EDMs or radiative decays, and may be relevant for dark matter 
searches~\cite{Nemevsek:2012cd, Nemevsek:2022anh, Nemevsek:2023yjl}.
The connection becomes rather involved in the case of $\mathcal{P}$~\cite{Senjanovic:2018xtu, Senjanovic:2020int} 
and one can resort to numerical algorithms~\cite{Kiers:2022cyc} to invert it.

The primary focus of LHC studies has been on the gauge sector, looking for a massive TeV scale $W_R$ resonance 
in the $s$-channel and the associated Majorana neutrino~\cite{Maiezza:2010ic, Nemevsek:2011hz, BhupalDev:2012zg, 
Dev:2015kca, Nemevsek:2018bbt}.
On the other hand, the Higgs sector is of fundamental importance for experimentally establishing the spontaneous 
mass origin of heavy neutrinos from $\Delta_R$ Yukawa coupling to $N$.
The scalar sector was introduced in~\cite{Senjanovic:1975rk, Senjanovic:1978ev} and analyzed in follow-up 
studies~\cite{Olness:1985bg, Gunion:1989in, Deshpande:1990ip, Kiers:2005gh, Khasanov:2001tu} to the more recent 
works~\cite{Dekens:2014ina, Bambhaniya:2015wna, Maiezza:2016ybz, Dev:2016dja} with emphasis on
perturbativity~\cite{Maiezza:2016bzp, Chauhan:2018uuy, Mohapatra:2019qid},
vacuum stability~\cite{Nemevsek:2016enw, BhupalDev:2018xya, Chauhan:2019fji} and gravitational 
waves~\cite{Brdar:2019fur, Li:2020eun}.
Our approach is to start from physical quantities, masses and mixing angles and use them to compute 
the potential couplings.
We therefore revisit and re-derive the minimization conditions and compute the spectrum to give a closed 
form solution for the potential parameters in terms of physical inputs.

There are two simplifying assumptions that we make in the entire analysis.
The first regards the vev of $\Delta_L$, which has to be small for phenomenological reasons, limited to 
$v_L \lesssim \text{ GeV}$ by the $\rho$ parameter.
For the most part of our discussion we set $v_L = 0$, which simplifies the diagonalization of the scalar mass spectra
 without a significant impact on the eigenvalues and phenomenology.
The other assumption regards the two quartic couplings of the bi-doublet, which we take to be related by 
$\lambda_2 = 2 \lambda_3$.
This removes a part of the mixing between the scalar and pseudo-scalar components and drastically simplifies the
neutral mass matrix, thus allowing for analytic diagonalization without significant phenomenological impact of 
additional CP violation in the heavy scalar sector.

The most fundamental conceptual goal in the Higgs sector, at least in relation to neutrino mass origin, 
is to prove the spontaneous origin of the heavy $N$ masses.
In this mechanism the decay rate of $\Delta_R \to N N$ is proportional to $m_N$, which can be measured 
kinematically, just like $\Gamma_{h \to f \overline{f}} \propto m_f^2$ in the SM, where $m_f$ are already known.  
The two scalars $h$ and $\Delta_R$ can mix and we get both $h \to N N$ from the 125 GeV resonance~\cite{Maiezza:2015lza}
and/or a sizeable production of $\Delta_R \to N N$ at the unknown $m_\Delta$~\cite{Nemevsek:2016enw}.
Such signals of Majorana Higgses would finalize the program of determining the neutrino mass origin and 
complete the microscopic picture of massive neutrinos.

The rest of the Higgs sector is also conceptually interesting but phenomenologically 
challenging~\cite{Maiezza:2016bzp, Dev:2016dja}.
The scalar $H$, pseudoscalar $A$ and singly charged $H^+$ from the bi-doublet $\phi$ can all mediate FCNCs
at tree level and therefore have to be quite above the \TeV\ scale.
They are unlikely to be accessible at the LHC but might be visible at future 
colliders~\cite{Mohapatra:2019qid}.
If $W_R$ were light enough to be seen at the LHC, the entire $\Delta_L$ would be somewhat split~\cite{Maiezza:2016bzp} 
and would appear mainly in cascade decays~\cite{Melfo:2011nx}.
The doubly charged $\Delta_R^{++}$ may be produced in association with $W_R$~\cite{Maalampi:2002vx} and would 
lead to additional LNV final states~\cite{Roitgrund:2020cge,Alloul:2013raa} or long-lived $\text{d}E/\text{d}x$ 
signatures~\cite{Akhmedov:2024rvp}.

Clearly, there is plenty of rich and conceptually relevant phenomenology to be studied in the context of 
the LRSM, at colliders, low energies (T2K~\cite{Alves:2023znq}, SHiP~\cite{Alekhin:2015byh} and 
FASER~\cite{Kling:2018wct}) and in cosmology.
To ease such studies it is desirable to have a complete analytical model solution and possibly a consistent implementation in a model file.
This is precisely the purpose of this work, where we solve analytically the model  and provide an implementation 
within~\textsf{FeynRules} (FR)~\cite{Christensen:2008py,Alloul:2013bka} and related~\textsf{UniversalFeynmanOutput} (UFO) 
versions~\cite{Degrande:2011ua,Darme:2023jdn}.

We collect and re-derive all the parts of the LRSM, starting from covariant derivatives and gauge eigen-states, 
together with would-be-Goldstones and ghosts, implementing a switch between the Feynman and unitary gauge.
A study of renormalization was performed in the gauge sector~\cite{Duka:1999uc} and in the Higgs 
sector~\cite{Maiezza:2016ybz} at one loop.
An earlier tree-level implementation of the LRSM can be found in~\cite{Roitgrund:2014zka} with arbitrary Yukawa 
couplings and no mixing in the scalar sector.
In the context of seesaw, there are available model files at NLO for type I seesaw/singlets~\cite{Degrande:2016aje},
type II~\cite{Fuks:2019clu} and type III~\cite{AH:2023hft}, as well as effective $W'/Z'$
model~\cite{Mattelaer:2016ynf, Fuks:2017vtl} and the Weinberg operator~\cite{Fuks:2020zbm}.

In this work we pay special attention to the scalar sector and derive anew analytic expressions 
for masses and mixings, providing a new physical input scheme and perturbativity constraints.
We take into account the symmetry-imposed flavor structure in the quark sector, leading to symmetric
CKM mixings and calculable Yukawa matrices of the extended Higgs sector to fermions.
In the neutrino sector we focus in detail on the Majorana-Dirac connection and implement the heavy-light 
neutrino mixing by devising an explicit closed form solution for the root(s) using the Cayley-Hamilton theorem.
This is the first time that a complete and calculable heavy-light neutrino mixing is implemented in a
model file, including the current neutrino oscillation data.

Finally, we obtain the QCD one loop counter-terms and provide, for the first time, the complete
LRSM at NLO, which  enhances and simplify future collider studies.
We demonstrate the use of the model file on a number of benchmarks that exemplify these developments 
and quantify, in some cases for the first time, the reduction of uncertainties in LNV signals at the 
LHC and beyond.

The paper is organized as follows: in section~\ref{sec:LRSM} we introduce the minimal LRSM with
the relevant fields.
We discuss each of its sectors in a separate sub-section, going from gauge bosons and would-be-Goldstones 
in~\ref{subsec:Gauge} to scalars~\ref{subsec:Scalar} (including the inversion of 
couplings~\ref{subsec:Inversion}), fermions~\ref{subsec:Fermions} and finishing with ghosts in~\ref{subsec:Ghost}.
The section~\ref{sec:ParityBreak} is dedicated to the variant of the model with different gauge couplings.
The core of the implementation of the model file is laid out in~\ref{sec:FRImplement} and the QCD NLO 
corrections are computed in~\ref{subsec:NLO}.
Some phenomenologically relevant benchmarks are given in~\ref{sec:Benchmarks}, while the summary and outlook 
is done in~\ref{sec:Summary}.
The appendix~\ref{app:SqrtMatrix} explains the Cayley-Hamilton approach used to obtain an analytic closed 
form of the square root of a matrix.

\section{The minimal Left-Right symmetric model} \label{sec:LRSM}

\noindent
The minimal Left-Right symmetric model is based on a parity symmetric gauge group
\begin{equation} \label{eq:LRGroup}
  G_{LR} = SU(3)_c \otimes SU(2)_L \otimes SU(2)_R \otimes U(1)_{B-L} \, ,
\end{equation}
where some discrete symmetry interconnects the two $SU(2)_{L,R}$ gauge groups.
These two groups are assumed to have the same gauge couplings at some energy scale.
We shall be somewhat agnostic about this and first take $g_L = g_R = g$ for simplicity.
Later on we generalize to $g_L \neq g_R$, which turns out not to be a very drastic change.
The gauge coupling of the remaining $U(1)_{B-L}$ is $g'$.

In the minimal LRSM, parity and the gauge symmetry are broken spontaneously in the scalar sector, featuring a bi-doublet $\phi = (\mathbf{1}, \mathbf{2}, \mathbf{2}, \mathbf{0})$ and two triplets 
$\Delta_L = (\mathbf{1}, \mathbf{3}, \mathbf{1}, \mathbf{2})$, 
$\Delta_R = (\mathbf{1}, \mathbf{1}, \mathbf{3}, \mathbf{2})$ under $G_{LR}$, 
%
%
\begin{align} \label{eq:DefScalar}
  \phi &= \begin{pmatrix} \phi_1^{0 *} & \phi_2^+ 
  \\ 
  \phi_1^- & \phi_2^0 \end{pmatrix} , 
  &
  \Delta_{L,R} &= \begin{pmatrix} \frac{\Delta^+}{\sqrt 2} & \Delta^{++} 
  \\ 
  \Delta^0 & -\frac{\Delta^+}{\sqrt 2} \end{pmatrix}_{L,R} .
\end{align}
The spontaneous symmetry breaking, which leaves unbroken only the electric charge generator
\begin{equation}
  Q = T_{3 L} + T_{3 R} + \frac{B-L}{2} \, ,
\end{equation}
follows from the vevs of the neutral components,
\begin{align} \label{eq:ScalarVevs}
  \langle \phi \rangle &= \begin{pmatrix} v_1 & 0 \\ 0 & -e^{i \alpha} v_2 \end{pmatrix}  ,
  &
  \langle \Delta_{L,R} \rangle &= \begin{pmatrix} 0 & 0 \\ v_{L,R} & 0 \end{pmatrix} ,
\end{align}
where 
\begin{align}
  v^2 &= v_1^2 + v_2^2 & &\text{and} & 0 \leq \tan \beta &= \frac{v_2}{v_1} < 1 \, ,
\end{align}
and we assumed $v_1 > v_2$ without loss of generality.

These vevs are quite hierarchical
\begin{align}
  v_L &\lesssim \text{GeV} \, ,
  &
  v &= 174 \text{ GeV} \, ,
  &
  v_R \gtrsim \text{TeV} \, ,
\end{align}
and generate the masses of all the states: the gauge and Higgs bosons as well as fermions.
The SM-like gauge bosons $W_L, Z$ are on the order of $v$ and so is the Higgs boson $h$.
The RH gauge bosons $W_R, Z_{LR}$ are heavy, on the order of $v_R$ and so are most
of the remaining scalars in $\Delta_{L,R}$ and the mass eigenstates $H, A, H^\pm$ from
the bi-doublet.
Of course, the photon remains massless, while the Goldstones are initially massless and then 
get their unphysical masses from gauge fixing.

The fermions come in LR symmetric doublet representations
\vspace*{-1ex}
\begin{align}
  Q_{L, R}  &= \begin{pmatrix}
    u \\ d
  \end{pmatrix}_{L, R}  ,
  &
  L_{L, R}  &= \begin{pmatrix}
    \nu \\ \ell
  \end{pmatrix}_{L, R} ,
\end{align}
with the usual three family copies.
The appealing feature of this setup is that the three RH neutrinos $\nu_R$ are required by anomaly
cancellation.
In the LRSM, all the fermion masses come from spontaneous breaking of the two $SU(2)$ groups, meaning
their masses are a product of Yukawa couplings and vevs.
For $O(1)$ Yukawa couplings, the Dirac masses are on the order of $v$, the Majorana mass for 
$\nu_R$ is $\mathcal O(v_R)$, while the light neutrinos $\nu_L$ get their Majorana mass as a combination of
$v^2/v_R$ and $v_L$ terms (combination of type I and type II seesaw).

In the following section~\ref{subsec:Gauge} we compute the gauge boson masses and mixings and derive
the required gauge fixing terms.
Then we focus on the scalar sector, which includes the Goldstones and the physical states.
We shall calculate the mass spectra, define an input scheme and compute the parameters of the potential 
in terms of masses and mixings.

\subsection{Gauge sector} \label{subsec:Gauge}

\noindent
{\em Gauge boson masses and mixings.}
Let us begin with the calculation of the gauge boson mass spectrum in the LRSM.
The canonical kinetic terms for complex scalar fields are
\vspace*{-1ex}
\begin{align}
  \mathcal L_{\text{kin}} &= \left| D \phi \right|^2 + \left| D \Delta_L \right|^2 + 
  \left| D \Delta_R \right|^2 \, ,
\end{align}
with
\vspace*{-2ex}
\begin{align}
\label{eq:LagKinCovDerPhi}
  D \phi &= \partial \phi - i g \left(A_L \phi - \phi A_R \right) \, , \\[1ex]
  \label{eq:LagKinCovDerDelta}
  D \Delta_i & = \partial \Delta_i - i g \left[A_i, \Delta_i \right] - i g' B \Delta_i \, ,
\end{align}
where $i = L, R$ and the gauge  and Lorentz indices are suppressed.
The gauge bosons fields in the flavor basis are
%
$ A_i = A_i^a \,\sigma^a/2$, 
$ A_i^\pm = (A^1_i \mp i A^2_i)/\sqrt{2}$,
%
with implied sum over the repeated indices $a = 1, 2, 3$.

The signs of the gauge coupling terms in the covariant derivatives~\eqref{eq:LagKinCovDerPhi} 
and~\eqref{eq:LagKinCovDerDelta} are not uniquely defined, as is the case for the SM~\cite{Romao:2012pq}.
The convention above is chosen to match the SM-like decoupling limit, such that when $\tan \beta \to 0$, 
the $\phi_1$ takes on the role of the SM Higgs and has the usual covariant derivative.
One also has to keep a consistent choice of signs between the bi-doublet $\phi$ and $\Delta_R$.
None of these considerations affect the gauge boson mass calculation, they only come into play when
we consider the gauge fixing that has to remove the mixed generic $V \partial \phi$ terms.

\medskip
\noindent{\em Gauge boson masses and mixings.}
Covariant derivatives in~\eqref{eq:LagKinCovDerPhi} and~\eqref{eq:LagKinCovDerDelta} generate the 
gauge boson mass matrices for charged and neutral gauge bosons
\begin{equation}
\begin{split}
  &\begin{pmatrix}
    A_L^- & A_R^-
  \end{pmatrix}
  M_{W_{LR}}^2
  \begin{pmatrix}
    A_L^+ \\ A_R^+
  \end{pmatrix} 
  =\\
  &\quad = 
  \begin{pmatrix}
    W_L^- & W_R^-
  \end{pmatrix}
  \begin{pmatrix}
    M_{W_L}^2 & 0
    \\
    0 & M_{W_R}^2
  \end{pmatrix} 
  \begin{pmatrix}
    W_L^+ \\ W_R^+
  \end{pmatrix}  .
\end{split}
\end{equation}
The hermitian mass matrix $M_{W_{LR}}^2$ in the gauge basis is
\begin{align} \label{eq:MW2gauge}
  M_{W_{LR}}^2 &= \frac{g^2}{2} v_R^2
  \begin{pmatrix} 
    \epsilon^2 & e^{-i \alpha} \epsilon^2 s_{2\beta}  
    \\[1ex] 
    e^{i \alpha} \epsilon^2 s_{2\beta} & 2 + \epsilon^2
  \end{pmatrix} , 
\end{align}
with the short-hands $s_x = \sin(x), c_x = \cos(x)$ and $t_x = \tan(x)$.
We introduced the dimensionless ratio of vevs
\begin{align}
  \epsilon &= \frac{v}{v_R} \, ,
\end{align}
which needs to be small on phenomenological grounds.

We compute the eigenvalues and expand them in $\epsilon$
\begin{align} \label{eq:MWs}
  M_{W_L} &\simeq \frac{g v}{\sqrt 2} \, ,
  & 
  M_{W_R} &\simeq g v_R \left( 1 + \frac{\epsilon^2}{4} \right) \, .
\end{align}
We drop higher order corrections of the order of $\epsilon^4$ in both the $M_{W_L}^2$ and $M_{W_R}^2$.
Specifically, the $\epsilon^2$ term in $M_{W_R}^2$ is of the same order as $M_{W_L}^2$.

To get from the gauge/flavor to the mass basis
\begin{align} 
    \begin{pmatrix}
    A_L^+ \\ A_R^+
  \end{pmatrix} 
  &= U_W
  \begin{pmatrix}
    W_L^+ \\ W_R^+
  \end{pmatrix}  ,
\end{align}  
we have the following $2D$ unitary rotation matrix
\begin{align} \label{eq:UW}
  U_W &=
  \begin{pmatrix}
    c_\xi & s_\xi e^{-i \alpha}
    \\[1ex] 
    - s_\xi e^{i \alpha}  & c_\xi
  \end{pmatrix}  ,
\end{align}
with the mixing angle given by
\begin{align}  
  s_\xi &\simeq \frac{\epsilon^2}{2} s_{2 \beta} \simeq \frac{M_{W_L}^2}{M_{W_R}^2} s_{2 \beta} \, .
\end{align}
The $W^-$ states get rotated with $U_W^*$, such that
\begin{align} 
  U_W^\dagger M_{W_{LR}}^2 U_W
  &= 
  \begin{pmatrix} 
    M_{W_L}^2  & 0
    \\ 
    0 & M_{W_R}^2
  \end{pmatrix} . 
\end{align}

We move on to neutral gauge bosons between the following bases
\begin{equation}
    \begin{split}
  &\begin{pmatrix}
    A_{3L} & A_{3R} & B
  \end{pmatrix}
  M_{Z_{LR}}^2
  \begin{pmatrix}
    A_{3L} \\ A_{3R} \\ B
  \end{pmatrix} 
  =
  \\
  &\quad = 
  \begin{pmatrix}
    A & Z_L & Z_R
  \end{pmatrix}
  \begin{pmatrix}
    0 & 0 & 0
    \\
    0 & M_Z^2 & 0
    \\
    0 & 0 & M_{Z_{LR}}^2
  \end{pmatrix} 
  \begin{pmatrix}
    A \\ Z_L \\ Z_R
  \end{pmatrix}  ,
\end{split} 
\end{equation}
where the symmetric mass matrix in the gauge basis is given by
\begin{align} \label{eq:MZ2Def}
  M_{Z_{LR}}^2 &= \frac{g^2}{2} v_R^2 
  \begin{pmatrix} 
    \epsilon^2      & - \epsilon^2	 & 0
    \\ 
    - \epsilon^2    & 4 + \epsilon^2 & -4 r
    \\ 
    0 		        & -4 r 			 & 4 r^2
  \end{pmatrix} ,
  &
  r &= \frac{g'}{g} \, .
\end{align}
This is a real matrix (no CP phases) that does not depend on $t_\beta$, and its eigenvalues can 
be expanded in $\epsilon$
\begin{align} \label{eq:MZs}
  M_A &= 0 \, ,\qquad 
  \,
  M_Z \simeq \frac{g v}{\sqrt{1 + \frac{1}{1 + 2 r^2}}} \, , 
  \\ 
  M_{Z_{LR}} &\simeq g v_R \sqrt{2 (1 + r^2)} \left( 
  1 + \frac{\epsilon^2}{8 \left(1 + r^2 \right)^2 } \right) .
  \label{eq:mzlr1}
\end{align}
One way to define the weak mixing angle is through the ratio of gauge boson masses
at leading order in $\epsilon$
\begin{align}
  c_w &= \frac{M_{W_L}}{M_{Z_L}} \biggr|_{\mathcal O(\epsilon^0)} \, ,
\end{align}
which in turn determines the ratio
\begin{align}
  r &= \frac{s_w}{\sqrt{c_{2w}}} \simeq
  \sqrt{\frac{M_{W_L}^2}{2 M_{W_L}^2 - M_{Z_L}^2} - 1}\simeq 0.63 \, .
\end{align}
From (\ref{eq:mzlr1}) one then recovers the prediction
\begin{align}
  \frac{M_{Z_{LR}}}{M_{W_R}} &\simeq  \sqrt{\frac{2c_w^2}{c_{2w}}}\left(1+
  \frac{\epsilon^2}{8}\frac{c_{2w}^2}{c_w^4}\right)\simeq 1.67 \, .
  \label{eqn:MZRMWR}
\end{align}

With these definitions, the real orthogonal matrix $O_Z$ that takes us from the flavor to 
the mass basis,
\begin{align}
  \begin{pmatrix} 
    A_{3L} \\ A_{3R} \\ B
  \end{pmatrix} & = O_Z
  \begin{pmatrix} 
    A \\ Z_L \\ Z_R
  \end{pmatrix} ,
\end{align}
is found at second order in $\epsilon$ to be
\begin{align} \label{eq:OZ}
  O_Z = &
  \begin{pmatrix}
    s_w & -c_w & 0 
    \\
    s_w & s_w t_w & \!\!-\frac{\sqrt{c_{2 w}}}{c_w} 
    \\
    \sqrt{c_{2 w}} & \sqrt{c_{2 w}} t_w & t_w 
  \end{pmatrix} 
  \nonumber
  \\[1ex]
  &+ \frac{\epsilon^2}{4}
  \begin{pmatrix}
    0 & 0 & \frac{c_{2 w}^{3/2}}{c_w^3} 
    \\
    0 & -\frac{c_{2 w}^2}{c_w^5} & -\frac{c_{2 w}^{3/2} t_w^2}{c_w^3} 
    \\
    0 & \frac{c_{2 w}^{3/2} t_w}{c_w^4} & -\frac{c_{2 w}^2 t_w}{c_w^4} 
  \end{pmatrix} .
\end{align}
This concludes the analysis of the gauge boson spectrum.

\medskip

\noindent
{\em Goldstones and gauge fixing.}
To complete the description of massive gauge bosons with spontaneous symmetry breaking, we need to 
address the would-be-Goldstone (wbG) boson sector.
Prior to gauge fixing, these appear as massless states of the scalar mass matrices.
In the SM there is only one neutral and one charged wbG, however in theories with an extended scalar 
sector, like the LRSM, there are more such states and they typically mix.
Thus we get degenerate zeroes and additional freedom of rotation between massless states.

In the case of the LRSM, we are dealing with two charged wbG modes for the $W_{L,R}$ and two neutral
ones for $Z, Z_{LR}$ gauge bosons, thus this freedom comes as one mixing angle for each sector.  
These angles are not arbitrary, because the proper wbG mode has to couple derivatively and
in proportion to the corresponding gauge boson mass, i.e.\ $M_V (V \partial \phi)$.
In other words, one needs to isolate the proper wbG mode for each gauge boson.
Eventually, we will add the gauge fixing terms in the mass basis, which cancel the derivative mixed 
terms and generate the usual gauge dependent propagators for vectors and wbGs.

We discuss first the singly charged states and derive their mass matrix.
It comes from the mixed second derivatives of the potential $\text{d}^2V/(\text{d} \phi_i^+ \text{d}\phi_j^-)$, 
evaluated at its minimum. 
The minimization is described in the scalar sector section below.
We use the following basis
\vspace*{-1ex}
  \begin{equation}
    \begin{split} \label{eq:Mp2Def}
  &\begin{pmatrix}
    \phi_1^- & \phi_2^- & \Delta_R^-
  \end{pmatrix}
  M_+^2
  \begin{pmatrix}
    \phi_1^+ \\ \phi_2^+ \\ \Delta_R^+
  \end{pmatrix} 
  = \\
  &\quad= 
  \begin{pmatrix}
    \varphi_L^- & H^- & \varphi_R^-
  \end{pmatrix}
  \begin{pmatrix}
    0 & 0 & 0
    \\
    0 & m_{H^+}^2 & 0
    \\
    0 & 0 & 0
  \end{pmatrix} 
  \begin{pmatrix}
    \varphi_L^+ \\ H^+ \\ \varphi_R^+
  \end{pmatrix}  ,
\end{split}
\end{equation}  
where we dropped the $\Delta_L^\pm$, because of the negligible $v_L$.

The singly charged mass matrix and its only nonzero eigenvalue are given by
\begin{align} \label{eq:Mp2}
  M_+^2 &= \alpha_3 v_R^2
  \begin{pmatrix}
     \frac{s_\beta^2}{c_{2\beta}} & -e^{-i \alpha} \frac{t_{2 \beta}}{2 }& -\epsilon  
     e^{-i \alpha} \frac{s_\beta}{\sqrt{2}}
    \\
    -e^{i \alpha} \frac{t_{2 \beta}}{2} & \frac{c_\beta^2}{c_{2\beta}} & \epsilon \frac{c_\beta}{\sqrt{2}}
    \\[2ex]
    -\epsilon e^{i \alpha} \frac{s_\beta}{\sqrt{2}} & \epsilon  \frac{c_\beta}{\sqrt{2}} & 
    \epsilon^2 \frac{c_{2 \beta}}{2}
  \end{pmatrix}  ,
  \\[1.5ex]
  m_{H^+} &\simeq \sqrt{\frac{\alpha_3}{c_{2 \beta}}} v_R 
  \left( 1 + \epsilon^2 \frac{c_{2 \beta}^{3/2}}{4} \right) \, .
\end{align}
As expected, there are two zero modes in~\eqref{eq:Mp2Def} that are eaten up by the 
SM-like $W_L$ and the new $W_R$.

One can diagonalize $M_+^2$ in~\eqref{eq:Mp2} by two subsequent rotations.
However, this is not enough: after this diagonalization and after performing the $U_W$ rotation among $W_{L,R}$, 
the mixed derivative terms, coming from~\eqref{eq:LagKinCovDerPhi} and~\eqref{eq:LagKinCovDerDelta}, are still off-diagonal.
In particular, while both $W_{L,R}^- \partial H^+ $ terms vanish as they should, the other four combinations 
$W_{L,R}^- \partial \varphi_{L,R}^+$ between incompletely rotated $\varphi^+_{L,R}$ are all there.
We need to perform one final rotation between wbGs to get rid off the two off-diagonal terms, which finally extracts the canonical would-be-Goldstone modes. 

Combining all the rotations, we get the complete transformation of the charged scalars 
\begin{align} \label{eq:UpFields}
  \begin{pmatrix}
    \phi_1^+ \\ \phi_2^+ \\ \Delta_R^+
  \end{pmatrix}
  = U_+ \,
  \begin{pmatrix}
    \varphi_L^+ \\ H^+ \\ \varphi_R^+
  \end{pmatrix}  ,
\end{align}
where the unitary matrix $U_+$ is given by three subsequent rotations and can also be expanded in $\epsilon$, 
\begin{widetext}
  \vspace*{-1.5ex}
  \begin{equation} \label{eq:Up}
\begin{split}
  U_+ &= 
  \begin{pmatrix}
    c_\beta & -e^{-i \alpha} s_\beta & 0 
    \\
    e^{i \alpha} s_\beta & c_\beta & 0 
    \\
    0 & 0 & 1
  \end{pmatrix}
  \begin{pmatrix}
    1 & 0 & 0 
    \\
    0 & -1 + \frac{1}{4} \epsilon ^2 c_{2 \beta}^2 & 
    \frac{\epsilon c_{2 \beta}}{\sqrt{2}} 
   \\
    0 & -\frac{\epsilon  c_{2 \beta}}{\sqrt{2}} & 
    -1 + \frac{1}{4} \epsilon ^2 c_{2 \beta}^2   
  \end{pmatrix}
  \begin{pmatrix}
    1 - \frac{1}{4} \epsilon ^2 s_{2 \beta}^2 & 0 & \frac{\epsilon s_{2 \beta}}{\sqrt{2}} 
    \\
    0 & 1 & 0 
    \\
    -\frac{\epsilon  s_{2 \beta}}{\sqrt{2}} & 0 & 1 - \frac{1}{4} \epsilon^2 s_{2 \beta}^2    
  \end{pmatrix}
  \\[.5ex] 
  & \simeq
  \begin{pmatrix}
    c_\beta & e^{-i \alpha} s_\beta & 0 
    \\
    e^{i \alpha} s_\beta & -c_\beta & 0 
    \\
    0 & 0 & -1
  \end{pmatrix} 
  + \frac{\epsilon}{\sqrt{2}}
  \begin{pmatrix}
    0 & 0 & e^{-i \alpha} s_\beta
    \\
    0 & 0 & c_\beta
    \\
    e^{i \alpha} s_{2 \beta} & - c_{2 \beta} & 0 
  \end{pmatrix} 
  + \frac{\epsilon^2}{2}
  \begin{pmatrix}
    -4 c_\beta s_\beta^4 & -e^{-i \alpha} s_\beta c_{2 \beta}^2 & 0 
    \\
    -4 e^{i \alpha} s_\beta c_\beta^4 & c_\beta c_{2 \beta}^2 & 0 
    \\
    0 & 0 & 1
  \end{pmatrix} ,
\end{split} 
\end{equation}
which is unitary up to $\mathcal{O}(\epsilon^2)$.
\end{widetext}

After performing all the rotations with $U_W$ and $U_+$, we end up with canonical wbGs that have 
derivative couplings proportional to the corresponding gauge boson masses
\begin{align} \label{eq:MixedWdPhi}
  \mathcal L_{\text{kin}} \ni -i \sum_{i = L,R} M_{W_i} W_i^- \partial \varphi_i^+ \, .
\end{align}
The final step is to add the gauge fixing terms, which come in the usual form, written in the mass 
basis
\begin{align} \label{eq:GaugeFixP}
  \mathcal L_{\text{gf}} &\ni - \sum_{i = L,R} \frac{1}{\xi_{W_i}} F_i^+ F_i^- \, ,
  \\[.7ex]
  F_i^+ &= \partial W_i^+ + i \xi_{W_i} M_{W_i} \varphi^+_i \, .
\end{align}
After expanding all the terms in $\mathcal L_{\text{gf}}$ and integrating by parts, the mixed 
$V \partial \varphi$ terms in~\eqref{eq:MixedWdPhi} cancel away.
The remaining terms in~\eqref{eq:GaugeFixP} give the standard $R_\xi$ propagators for the $W$'s 
and wbGs.

\smallskip

In the neutral sector, the procedure of setting up the wbG rotations and fixing the gauge
goes along the same lines.
The major additional complication is that there are more neutral states that mix and 
computing their eigenvalues and mixings becomes a bit more tedious.
However, it is fairly simple to isolate the wbG modes away from the massive scalars 
and complete all the rotations in the gauge-wbG sector.
The following field substitution in~\eqref{eq:DefScalar} gets the job done
\begin{align} \label{eq:demixneutrals}
  \phi_{10} &=  v_1 + \frac{1}{\sqrt 2} \left( c_\beta \varphi_{10} + 
  s_\beta e^{-i \alpha}  \varphi_{20} \right) \, ,
  \\
  \phi_{20} &= -v_2 e^{i \alpha} + \frac{1}{\sqrt{2}} \left( c_\beta \varphi_{20} - 
  s_\beta e^{i \alpha} \varphi_{10} \right) \, .
\end{align}
With this rotation, the $\Im(\varphi_{10})$ and $\Im(\Delta_R^0)$ decouple from the
other neutral scalars and they both have a zero eigenvalue.
However, their derivative couplings are still off-diagonal, so we need to perform one 
final $2D$ rotation to get to the diagonal neutral wbG states
\begin{align} \label{eq:Ophi}
  \begin{pmatrix} 
    \Im(\varphi_{10}) \\ \Im(\Delta_R^0)
  \end{pmatrix} & = O_\varphi
  \begin{pmatrix} 
    \varphi_L^0 \\ \varphi_R^0
  \end{pmatrix} ,
  \\
  \label{eq:OphiN}
  O_\varphi &= I_2 - \frac{i \sigma_2}{2 \left(1 + r^2 \right)} \epsilon
  - \frac{I_2}{8 \left(1+r^2\right)^2} \epsilon^2 \, .
\end{align}
Here, $I_2$ is the 2D identity matrix and $\sigma_2$ is the usual Pauli matrix.
The resulting mixed derivative terms are now  diagonal  and proportional to
the neutral gauge boson masses
\begin{align} \label{eq:MixedZdPhi0}
  \mathcal L_{\text{kin}} \ni - \sum_{i = L,R} M_{Z_i} Z_i \partial \varphi_i^0 \, .
\end{align}
Again, they get cancelled by the terms that complete the gauge fixing
\begin{align} \label{eq:GaugeFixN}
  \mathcal L_{\text{gf}} &\ni - \sum_{i = L,R} \frac{1}{2 \xi_{Z_i}} F_{Z_i}^2 \, ,
  &
  F_{Z_i} &= \partial Z_i + \xi_{Z_i} M_{Z_i} \varphi^0_i \, ,
\end{align}
which also give the canonical $R_\xi$ propagators for the neutral gauge bosons and wbGs.

\bigskip
\subsection{Scalar sector} \label{subsec:Scalar}
\noindent
The scalar potential of the LRSM comprises all possible bi-linear and quartic terms of the scalar 
fields, obeying the gauge symmetry, and one discrete LR symmetry, either $\mathcal{P}$ or $\mathcal{C}$:
\begin{widetext}
\begin{align} 
  \mathcal{V} &= -
  \mu_1^2 \left[ \phi ^\dagger \phi \right] -
  \mu_2^2 \left( \left[\tilde{\phi} \phi ^\dagger \right] + \left[ \tilde{\phi }^\dagger \phi \right]\right) -
  \mu_3^2 \left( \left[\Delta_L^{{\color{white}\dagger}} \Delta _L^\dagger\right] + \left[\Delta_R^{{\color{white}\dagger}} \Delta_R^\dagger\right]\right)
  \nonumber 
  \\[0ex]
  &+ \lambda_1 \left[\phi ^\dagger \phi \right]^2 +
  \lambda_2 \left(\left[\tilde{\phi } \phi ^\dagger\right]^2 + \left[\tilde{\phi }^\dagger \phi \right] ^2\right) +
  \lambda_3 \left[\tilde{\phi } \phi ^\dagger\right] \left[\tilde{\phi }^\dagger \phi \right] + 
  \lambda_4 \left[\phi ^\dagger \phi \right] \left(\left[\tilde{\phi } \phi ^\dagger\right] + \left[\tilde{\phi }^\dagger \phi \right]\right)
  \nonumber
  \\[0ex]
  &+
  \rho_1 \left(\left[\Delta_L^{{\color{white}\dagger}} \Delta _L^\dagger\right]^2 + \left[\Delta_R^{{\color{white}\dagger}} \Delta_R^\dagger\right]^2\right) + 
  \rho_2 \left(\left[\Delta_L^{{\color{white}\dagger}} \Delta _L^{{\color{white}\dagger}} \right]
  \left[\Delta _L^\dagger \Delta _L^\dagger\right]+\left[\Delta_R^{{\color{white}\dagger}} \Delta_R^{{\color{white}\dagger}} \right] \left[\Delta _R^\dagger \Delta _R^\dagger\right]\right) + 
  \rho_3 \left[\Delta_L^{{\color{white}\dagger}} \Delta _L^\dagger\right] \left[\Delta_R^{{\color{white}\dagger}} \Delta _R^\dagger\right]
  \nonumber
  \\[1ex]
  &+ \rho _4 \left(\left[\Delta_L^{{\color{white}\dagger}} \Delta _L^{{\color{white}\dagger}} \right] \left[\Delta_R^\dagger \Delta _R^\dagger\right] + \left[\Delta _L^\dagger \Delta _L^\dagger\right] \left[\Delta_R^{{\color{white}\dagger}} \Delta_R^{{\color{white}\dagger}} \right]\right)
 + \alpha _1 \left[\phi ^\dagger \phi \right] \left(\left[\Delta_L^{{\color{white}\dagger}} \Delta _L^\dagger\right]+\left[\Delta_R^{{\color{white}\dagger}} \Delta _R^\dagger\right]\right)
 \label{eq:VP}
 \\[1ex]
 &+ \left( \alpha_2 \left(\left[\tilde{\phi } \phi ^\dagger\right] \left[\Delta_L^{{\color{white}\dagger}} \Delta _L^\dagger\right]+\left[\tilde{\phi }^\dagger \phi \right] \left[\Delta_R^{{\color{white}\dagger}} \Delta _R^\dagger\right]\right) + \text{h.c.} \right)
   +\alpha _3 \left(\left[\phi  \phi ^\dagger \Delta_L^{{\color{white}\dagger}} \Delta _L^\dagger\right]+\left[\phi ^\dagger \phi  \Delta_R^{{\color{white}\dagger}} \Delta _R^\dagger\right]\right)
\nonumber
  \\[1.3ex]
  &+ \beta _1 \left(\left[\phi  \Delta_R^{{\color{white}\dagger}} \phi ^\dagger \Delta _L^\dagger\right]+\left[\phi ^\dagger \Delta_L^{{\color{white}\dagger}} \phi  \Delta _R^\dagger\right]\right)+\beta _2 \left(\left[\tilde{\phi } \Delta_R^{{\color{white}\dagger}} \phi ^\dagger \Delta_L^\dagger\right]+\left[\tilde{\phi }^\dagger \Delta_L^{{\color{white}\dagger}} \phi  \Delta _R^\dagger\right]\right)
+ \beta _3 \left(\left[\phi  \Delta_R^{{\color{white}\dagger}} \tilde{\phi}^\dagger \Delta _L^\dagger\right]+\left[\phi ^\dagger \Delta_L^{{\color{white}\dagger}} \tilde{\phi } \Delta _R^\dagger\right]\right)  ,
\nonumber
\end{align}
\end{widetext}
where $\tilde{\phi} = \sigma_2 \phi^{*} \sigma_2$ and the square brackets imply the trace 
over field components. 
In the following, we set for simplicity $\beta_i \simeq 0$, which sets $v_L = 0$ and remains
small; this is a technically natural assumption.

The $\mathcal{P}$ or $\mathcal{C}$ discrete symmetries further constrain the coupling constants.

\paragraph{The case of $\mathcal{P}$.} 
With this choice, the scalars transform as
\begin{align}
  \mathcal P &: \phi \to \phi^\dagger \, , & \Delta_L &\leftrightarrow \Delta_R \, ,
\end{align}
and all the couplings in the potential must be real except for $\alpha_2$, which carries a phase 
${\rm e}^{i\delta_2}$.

The minimization conditions can be interpreted as determining the $\mu_i$ in terms of the vevs and quartic 
couplings:
\begin{align}
  \mu_1^2 =& 2 \left( \lambda_1 + s_{2 \beta} c_\alpha \lambda_4 \right) v^2 + 
  \left(\alpha_1 - \alpha_3 \frac{s_\beta^2}{c_{2 \beta}}\right) v_R^2 \, ,
  \\[1ex]
  \mu_2^2 =& \left( s_{2 \beta} \left(2 c_{2 \alpha} \lambda_2 + \lambda_3 \right) + 
  \lambda_4 \right) v^2
  \nonumber\\ 
  &{} + \frac{1}{2 c_\alpha} \left(2 c_{\alpha +\delta_2} \alpha_2 + 
  \alpha_3 \frac{t_{2 \beta}}{2 c_\alpha} \right) v_R^2 \, ,
  \\[1.5ex]
  \mu_3^2 =& \left(\alpha_1 + 
  \left(2 c_{\alpha + \delta_2} \alpha_2 s_{2\beta} +
  \alpha_3 s_\beta^2 \right) \right) v^2 + 2 \rho_1 v_R^2 \, .
\end{align}
where $v_L$ is neglected, since $v_L \ll v \ll v_R$.
In fact, one has
\begin{equation} \label{eqVevssaw}
\begin{split}
  v_L =& \frac{\epsilon^2 v_R}{\left(1 + t_\beta ^2\right) \left(2 \rho_1 - \rho_3\right)} \biggl(
    - \beta_1 t_\beta \cos (\alpha - \theta_L) 
    \\
    &+ \beta_2 \cos (\theta_L) + \beta_3 t_\beta^2 
    \cos (2 \alpha - \theta_L) \biggr) \,.    
\end{split}
\end{equation}

The derivative with respect to $\alpha$ connects it to the only CP phase $\delta_2$: 
\begin{align} \label{eqRelD2} 
  \alpha_2 s_{\delta_2} &= \frac{s_\alpha}{4} \left(\alpha_3 t_{2 \beta} + 
  4 \left(\lambda_3 - 2 \lambda_2\right) s_{2 \beta} \epsilon^2 \right)  . 
\end{align}

\bigskip

\paragraph{The case of $\mathcal{C}$.} This acts as
\begin{equation}
  \mathcal C : \phi \to \phi^T\,, \qquad \Delta_L \leftrightarrow \Delta_R^*\,.
\end{equation}
This choice allows for additional complex phases in the potential.  
In particular, the couplings $\mu_2$, $\lambda_2$, $\lambda_4$, $\rho_4$ and $\beta_i$ are now complex, 
introducing seven new CP violating phases. 
The minimization conditions described above also change accordingly. 

However, for collider studies, these additional phases do not lead to relevant new features, while spoiling 
the exact solvability of the scalar spectrum that we achieve below.  
Therefore, in the present work we retain the single CP phase $\delta_2$ as in the case of $\mathcal{P}$. 
Similarly, we can consider the limit of vanishing $v_L$, and accordingly we set $\beta_i \to 0$.

\subsubsection{Doubly charged \texorpdfstring{$\Delta_R^{++}$}{DRpp} and the entire 
\texorpdfstring{$\Delta_L$}{DL} triplet}

\noindent
In the $v_L \to 0$ limit, the doubly charged $\Delta_R^{++}$ and $\Delta_L$ triplet of states do not 
mix with other scalars, and their mass matrix is diagonal to begin with.
The simplest is
\begin{align}
  m_{\Delta_R^{++}}^2 &= v_R^2 \left(4 \rho_2  + \frac{c_{2 \beta}}{c_\beta^4} \alpha_3 \epsilon^2 \right) \, ,
\end{align}  
and the $\Delta_L$ triplet masses are given by 
\begin{align}
  m_{\Delta_L^0}^2 & = v_R^2\bigg[\rho_3 - 2 \rho_1 
  +  \epsilon^2\alpha_3  \, s_{2 \beta} t_{2 \beta}\, s^2_\alpha   \bigg]= m_{\chi_L^0}^2\, ,\\[1ex]
  m_{\Delta_L^+}^2 &= v_R^2 \bigg[\rho_3 - 2 \rho_1 
  +  \epsilon^2\alpha_3  \left(s_{2 \beta} t_{2 \beta}\, s^2_\alpha+\frac{c_{2\beta}}{2}\right)
   \bigg] \,,
  \\[1ex]
  m_{\Delta_L^{++}}^2 \!&= v_R^2 \bigg[\rho_3 - 2 \rho_1  
  +  \epsilon^2\alpha_3  \left(s_{2 \beta} t_{2 \beta}\, s^2_\alpha+c_{2\beta}\right)
    \bigg] \,,
\end{align}
where clearly the four states within the $\Delta_L$ triplet are degenerate up to $\epsilon^2$ corrections. 
These relations can be readily solved for e.g.\ $\rho_2$ and $\rho_3$ in terms of two masses, $m_{\Delta_{R}^{++}}$, $m_{\Delta_{L}}$, as we do below. 
The remaining three masses within the $\Delta_L$ triplet are then predicted and follow the known sum 
rule~\cite{Melfo:2011nx}
\begin{equation}
  m_{\Delta_L^{++}}^2 -m_{\Delta_L^+}^2 = m_{\Delta_L^+}^2-m_{\Delta_L^0}^2=v^2\alpha_3 \frac{c_{2\beta}}{2}\,.
  \label{eq:sumruleDL}
\end{equation}

\subsubsection{Singly charged states (\texorpdfstring{$H^+$}{Hp} and the Goldstones of \texorpdfstring{$W_{L, R}$}{WLR})}

\noindent
From the three singly charged states $\Phi_{1}^+$, $\Phi_{2}^+$, $\Delta_R^+$, one finds two Goldstones 
relative to $W_L^+$ and $W_R^+$, and only one massive singly charged state, with mass 
\begin{align}
  m_{H^+}^2 &= v_R^2 \frac{\alpha _3}{c_{2 \beta}}  \left(1 + \frac{c^2_{2 \beta}}{2} \, \epsilon^2\right).
\label{eq:mHp}
\end{align}
The unitary rotation from the unphysical to the physical basis was described in~\eqref{eq:Mp2Def} and \eqref{eq:Up}.

\subsubsection{\texorpdfstring{
Neutral states ($h$, $\Delta$, $H$, $A$ plus the two Goldstones of $Z_L$, $Z_R$) - Restriction $\lambda_3=2\lambda_2$}{
Neutral states (h, D, H, A plus the two Goldstones of ZL, ZR) - Restriction lam3 =2 lam2}}

\noindent
After the $t_\beta$ demixing in (\ref{eq:demixneutrals}), the neutral goldstone bosons are decoupled 
from the neutral massive states $\{\Re \varphi_{10}, \Re \Delta_{R}, \Re \varphi_{20}, \Im \varphi_{20}\}$.
Their mass matrix is then diagonalized to yield the mass eigenstates $\{h, \Delta, H, A\}$.

It is convenient to restrict to $\lambda_3 = 2\lambda_2$, which removes a tiny and phenomenologically 
irrelevant CP-violating mixing between the $H$ and $A$ states, and allows for almost exact diagonalization.
The $4 \times 4$ squared mass matrix becomes, in units of $v_R^2$,
\begin{widetext}

{\small
\begin{equation}
\left(
 \renewcommand{\arraystretch}{2}\selectfont
\begin{array}{cccc}
 4 \epsilon ^2 \left(\lambda _1+\frac{4 t c_\alpha \left(\lambda _4
   \left(t^2+1\right)+4 \lambda _2 t c_\alpha\right)}{\left(t^2+1\right)^2}\right) & 2\epsilon  \left(\alpha
   _1-\frac{t X \left(t^3+t-2\eta_2 c_\alpha\right)}{\left(t^2+1\right)^2}\right) & \frac{4
   \epsilon ^2 \left(t^2 c_{2\alpha}-1\right) \left(\lambda _4
   \left(t^2+1\right)+8 \lambda _2 t c_\alpha
   \right)}{\left(t^2+1\right)^2} & \frac{4 t^2 \epsilon ^2 s_{2 \alpha}
   \left(\lambda _4 \left(t^2+1\right)+8 \lambda _2 t c_\alpha 
   \right)}{\left(t^2+1\right)^2} 
   \\
 2\epsilon  \left(\alpha
   _1-\frac{t X \left(t^3+t-2\eta_2 c_\alpha\right)}{\left(t^2+1\right)^2}\right) 
   & Y & \frac{2  X \epsilon \left(t^2 c_{2\alpha}-1\right)\eta_2
   }{\left(t^2+1\right)^2} & \frac{2 t^2
   X \epsilon  s_{2\alpha}  \eta_2  }{\left(t^2+1\right)^2} 
   \\
 \frac{4 \epsilon ^2 \left(t^2 c_{2\alpha}-1\right) \left(\lambda _4
   \left(t^2+1\right)+8 \lambda _2 t c_\alpha
   \right)}{\left(t^2+1\right)^2} & \frac{2  X \epsilon \left(t^2 c_{2\alpha}-1\right)\eta_2
   }{\left(t^2+1\right)^2} & X+ \frac{16 \lambda _2 \epsilon ^2
   \left(t^2 c_{2\alpha}-1\right)^2}{\left(t^2+1\right)^2}& \frac{16
   \lambda _2 t^2 \epsilon ^2 s_{2 \alpha} \left(t^2 c_{2 \alpha}
   -1\right)}{\left(t^2+1\right)^2} 
  \\
  \frac{4 t^2 \epsilon ^2 s_{2 \alpha} \left(\lambda _4 \left(t^2+1\right)+8 \lambda _2 t c_\alpha
  \right)}{\left(t^2+1\right)^2} & \frac{2 t^2
   X \epsilon  s_{2\alpha}  \eta_2  }{\left(t^2+1\right)^2} 
  & \frac{16 \lambda _2 t^2 \epsilon ^2 s_{2\alpha} \left(t^2 c_{2 \alpha} -1\right)}{\left(t^2+1\right)^2} 
  & X + \frac{16 \lambda_2 t^4 \epsilon^2 s^2_{2 \alpha}}{\left(t^2+1\right)^2}
\end{array}
\right)
\vspace*{-0ex}
\end{equation}
}%
where for compactness we defined $X\equiv\frac{1+t^2}{1-t^2}\alpha_3$, $Y\equiv4\rho_1$,  $t\equiv t_\beta$, and $\eta_2 \equiv t \sin(\alpha+\delta_2)/\sin{\delta_2}$. 
\end{widetext}

The diagonalization proceeds in two steps.

\begin{itemize}
    \item
We first notice that the 2-1 entry, driving the mixing $\theta$ of the SM higgs with the unphysical $\Delta$ (i.e.\ $h$--$\Re\Delta_R$) is constrained by exotic and invisible Higgs decay searches to be phenomenologically small: $|\sin \theta| < 20 \%$~\cite{Dawson:2021jcl}. 
    The two lower entries are also small, $\sim\epsilon^2$.  
    Thus, all three entries can be cleaned with  rotations at leading order. 
    This isolates the Higgs eigenvalue, which can be expressed in terms of the $h$--$\Re\Delta_R$ 
    mixing angle $\theta$ (reported below)
    \begin{equation} \label{eq:m2h}
      m^2_h = v^2 \left(4 \lambda _1+\frac{64 \lambda _2 t^2 c_\alpha^2}{\left(t^2+1\right)^2}+
      \frac{16 \lambda _4 t c_\alpha}{t^2+1}-Y  \tilde\theta^2\right) ,
    \end{equation}
    where we use the shorthand $\tilde\theta\equiv\theta/\epsilon\lesssim O(1)$. 
\item
Then, remarkably thanks to the $\lambda_3=2\lambda_2$ restriction, the remaining $3 \times 3$ 
    block is diagonalized exactly by two  rotations, in the $3-4$ and $2-3$ planes, instead of three. 
    This isolates the last three mass eigenvalues, in terms of the $3-4$ rotation.
\end{itemize}

The simplest eigenvalue belongs to the pseudoscalar, which we used to compactify the notation
\begin{align} \label{eq:m2A}
  m^2_A &=v_R^2 \, X \, , & X &\equiv \frac{1 + t^2}{1 - t^2} \alpha_3 \, .
\end{align}
It is clearly degenerate with $H^+$, see~\eqref{eq:mHp}, up to $\epsilon^2$ terms.

\begin{widetext}The  remaining states, after this and the SM Higgs in~\eqref{eq:m2h}, are the neutral triplet
and the heavy scalar
\begin{align}
  m^2_\Delta &= v_R^2\left[Y + \sec (2 \eta) \left[(Y - X) s^2_\eta + \epsilon^2 \left(Y \tilde \theta^2 c^2_\eta -
  \frac{16 \lambda _2\left(t^4 - 2 c_{2 \alpha} t^2 + 1\right)}{\left(t^2 + 1\right)^2} s^2_\eta \right)\right] \right] ,
  \\[2ex]  
  m^2_H &= v_R^2\left[X - \sec(2 \eta) \left[(Y - X) s^2_\eta + \epsilon^2 \left(Y \tilde \theta^2 s^2_\eta -
  \frac{16 \lambda_2\left(t^4 - 2 c_{2 \alpha} t^2 + 1 \right) }{\left(t^2 + 1\right)^2} c^2_\eta \right) \right] \right] .
\end{align}

Here we list the subsequent orthogonal rotations, complementing the neutral GB rotations defined in \eqref{eq:OphiN}:
%
\begin{align} \label{eq:theta}   
\theta\equiv\epsilon\,\tilde\theta\equiv{}& \theta_{21} =
       \epsilon\frac{
       2\,\alpha_1-2\, X\, t\left(\frac{t}{1+t^2}-\frac{2 \,\eta_2 \,  c_\alpha}{\left(t^2+1\right)^2}\right)
   }
   {%
   Y-4\,\epsilon^2 \left(\lambda_1+16 \lambda_2\frac{t^2\,c_\alpha^2}{(t^2+1)^2}+4\lambda_4\frac{t \,c_\alpha}{t^2+1}\right)} 
   \\[2.5ex]  
  \label{eq:phi} \phi\equiv\epsilon^2\tilde\phi\equiv{} &\theta _{31}  = \epsilon ^2 \frac{\left(t^2 c_{2 \alpha} - 1\right)}{(t^2+1)^2}\left[\frac{32\, t\, c_\alpha \lambda_2
    + 4 \,\lambda_4 (t^2+1)}{X} - 2 \, \tilde{\theta}\,\eta_2 \right],
  \\[2.5ex]
  & \theta_{41} = \epsilon ^2 \frac{t^2 s_{2\alpha}}{\left(t^2+1\right)^2}\left[\frac{{32\, t\, c_\alpha \lambda_2 } + 
  4\, \lambda_4 (t^2 + 1)}{X} - {2 \,\tilde{\theta} \, \eta_2} \right]  ,
  \\[2.5ex]
  & \theta_{34} = \tan ^{-1}\left(\frac{t^2 s_{2\alpha}
   }{t^2 c_{2\alpha}-1}\right),
  \\[2.5ex]
\label{eq:eta}
   \hphantom{\phi}\eta \equiv {}& \theta_{23} =-\frac{1}{2} \tan^{-1}\left[\frac{4\, X\, \epsilon \sqrt{t^4 - 2 c_{2 \alpha} t^2 + 1} 
  \, \eta_2}{\left(t^2 + 1\right)^2 \left(Y\, \tilde{\theta}^2 \epsilon^2 
  - \frac{16 \left(t^4 - 2 c_{2 \alpha} t^2 + 1\right) \lambda_2 \epsilon^2}{\left(t^2 + 1\right)^2} - X + Y\right)}\right] .
\end{align}

\end{widetext}

The combined rotation for neutral scalars that relates the unphysical to the mass basis is thus
\begin{equation}
  \begin{pmatrix}
    \Re \varphi_{10} \\
    \Re \Delta_{R}   \\
    \Re \varphi_{20} \\ 
    \Im \varphi_{20}      
  \end{pmatrix} = O_N 
  \begin{pmatrix}
    h\\ \Delta\\ H\\ A      
  \end{pmatrix} ,
  \label{eq:ONdef}
\end{equation}
with $O_N = O_{21} O_{31} O_{41} O_{34} O_{23}$, where each
\begin{equation}
  O_{ij} = 
  \begin{pmatrix} 
  c_{\theta_{ij}} &-s_{\theta_{ij}}
  \\
  s_{\theta_{ij}}&c_{\theta_{ij}} 
  \end{pmatrix} 
\end{equation}
acts in the $(i,j)$ subspace, with angle $\theta_{ij}$ from \eqref{eq:theta}--\eqref{eq:eta}.

The first angle has to be small by Higgs decay constraints. This implies  e.g.\ a small $\alpha_1$. In case one considers a light $m_\Delta $ to the EW scale or below, $Y\lesssim \epsilon^2$,  the $\epsilon^2$ correction in the denominator of Eq.~\eqref{eq:theta} becomes important. The numerator has clearly to be even smaller. 

The last angle $\eta$ instead can be $\mathcal{O}(1)$, provided $X-Y$ are degenerate to $\mathcal{O}(\epsilon)$, 
which corresponds to a $H - A$ degeneracy of the order of 500\,GeV or less.

In the next section all previous expressions for mixings and for masses are inverted so that couplings 
are expressed in terms of chosen physical quantities. 
In particular, with the above convention, the interesting angles can be defined as 
\begin{align}
  h \text{ part of } \Re \Delta_R : \quad & 
  \theta \equiv \theta_{21} \simeq -(O_N)_{2,1} \, ,
  \\[1ex]
  H \text{ part of } \Re \Delta_R : \quad & 
  \eta   \equiv \theta_{23} = \arcsin[(O_N)_{2,3}/c_\theta] \, ,
  \\[1ex]
  h \text{ part of } \Re \phi_{20}: \quad & 
  \phi   \equiv \theta_{31} \simeq -(O_N)_{3,1} \, ,
\end{align}
which we will use as physical inputs. 
Notice that if  $\Delta$, $H$, $A$  are left around the scale of $v_R$, one has the following 
scalings $\eta \sim 1$, $\theta \sim \epsilon $,  $\phi \sim \epsilon^2$.

It is nice to check the case of vanishing $t_\beta\to0$: in this limit, $\lambda_2$ governs the 
$A$-$H$ mass splitting, $\lambda_4$ governs the $\phi$ mixing and $\alpha_1$ sets the $\theta$ mixing. 
On the other hand, the CP-violating ($A$-$h$ and $A$-$H$) mixings $\theta_{41}$ and $\theta_{34}$ 
vanish as expected. 
The mixing $\eta$ is governed by $\eta_2$, which in this limit can remain nonzero, see below. 

\medskip

Finally, comparing the heavy scalar masses in Eqs.~\eqref{eq:m2A} and~\eqref{eq:mHp}, we find 
another sum rule like the one in~\eqref{eq:sumruleDL}: 
\begin{align} \label{eq:sumruleH}
  m^2_{H^+} - m^2_{A} &= v^2 \alpha_3 \frac{c_{2\beta}}{2}\sim O(150\,\GeV)^2 \, .
\end{align}
It implies, recalling that $A$, $H$ have to be as heavy as 20\,\TeV~\cite{Zhang:2007da, Zhang:2007fn, Maiezza:2010ic, 
Bertolini:2013noa, Maiezza:2014ala, Bertolini:2019out, Bertolini:2020hjc, Bernard:2020cyi, Dekens:2021bro}, 
and also that the splitting with $H^+$ becomes rather tiny, 1--10\,GeV. 

\subsection{Inversion: Couplings in terms of physical masses and mixings} \label{subsec:Inversion}
\label{sec:inversion}

\noindent
Equations (\ref{eq:m2h}--\ref{eq:m2A}) together with~\eqref{eq:theta}, \eqref{eq:phi} and~\eqref{eq:eta} 
can be solved for the relevant couplings of the potential, in terms of four masses ($m^2_h, m^2_\Delta, m^2_H, m^2_A$) 
and the three mixings $\theta$, $\eta$ and $\phi$.
Collecting all the previous expressions one finds:
\begin{align}
    \label{eq:X}
&  \frac{1 + t^2}{1 - t^2} \alpha_3 \equiv X  \to \frac{m^2_A}{v_R^2} \, ,  
  \\[.7ex]
&  4 \rho_1 \equiv Y  \to \frac{m^2_\Delta + \left( m^2_H - m^2_\Delta \right) s_\eta^2}{v_R^2} \, ,
  \\
&    \rho _2\to \frac{m^2_{\Delta_R^{++}}}{4v_R^2}-\frac{\epsilon
   ^2X\left
   (t^2-1\right)^2}{4\left(t^2+1\right)^2},\\[.7ex]
&   \rho _3\to \frac{m^2_{\Delta_L}}{v_R^2}+\frac{Y}
   {2}-\frac{4 \epsilon ^2 X t^2
    s_\alpha^2}{\left(t^2+1\right)^2}\,,\\[.7ex]
&  \lambda_1         \to \frac{m^2_h}{4v^2}+ \frac{Y}{4} \tilde{\theta}^2 - 4\, t c_\alpha \left(\frac{\lambda_4}{t^2+1}  + 
  \frac{4 t c_\alpha \lambda_2}{\left(t^2 + 1\right)^2}  \right)  ,
  \\[.7ex]
&  \lambda_2         \to \frac{m^2_H - m^2_A - \left(m^2_H - m^2_\Delta \right) s^2_\eta}{16 \, v^2}
  \frac{\left(t^2 + 1\right)^2}{ \left(t^4 - 2 c_{2 \alpha} t^2 + 1 \right) } \, ,  
  \\[.7ex]
&  \lambda_3         \to 2 \lambda_2 \, ,
  \\[1.8ex]
& \lambda _4\to 
\tilde\phi\frac{  X \left(t^2+1\right)} {4
   \left(t^2 c_{2 \alpha} -1\right)}
   + \tilde{\theta }\frac{ X  \eta_2} {2 \left(t^2+1\right)}
   -8 \frac{t c_\alpha\lambda _2}{\left(t^2+1\right)}
  \\[1ex]
&\eta_2\to \frac{(m^2_H-m^2_\Delta)
   \left(t^2+1\right)^2 \sin (2 \eta )}{4
   m^2_A \epsilon  \sqrt{t^4-2 c_{2
   \alpha } t^2+1}} 
   \label{eq:eta2}
   \\[1.4ex]
   &  \delta_2          \to \tan ^{-1}\left(\frac{t
   s_\alpha}{\eta_2-t c_\alpha}\right),
  \label{eq:d2}
  \\[1.4ex]
&  \alpha_1          \to \frac{Y \tilde{\theta}}{2} + t \left(\frac{t \, X}{t^2 + 1} + \frac{(Y-X) c_{\alpha} 
  t_{2 \eta}}{2 \, \epsilon \sqrt{t^4 - 2 c_{2 \alpha} t^2 + 1} }\right)  ,
  \label{eq:a1}
  \\[1.4ex]
&  \alpha_2          \to 
   \frac{X}{2}
    \frac{(\eta_2-t c_\alpha) }{
   \left(t^2+1\right)} \sqrt{\frac{t^2 s_\alpha^2}{(\eta_2-tc_\alpha)^2}+1}\,.
   \label{eq:a2}
\end{align}

These expressions  have been written by ensuring that no singularity develops for $t \to 0$ and/or $\alpha\to0$. 

On the other hand, a few comments are in order about the perturbativity of the couplings obtained, while referring
to~\cite{Maiezza:2016bzp} for a thorough study of perturbativity limits, including loop corrections.
Raising the scalar masses too much beyond $v_R$, like for instance $m_A$ in \eqref{eq:X}, 
produces couplings that enter in a non-perturbative regime.
But also, a few couplings are sensitive to scalar mass splittings in relation to mixing angles.
In particular, $\lambda_2$ becomes large in case $|m^2_H - m^2_A| > 16 v^2$, or in case 
$|m^2_H - m^2_\Delta|$ is similarly large while demanding large $\eta$ mixing.   
Similar considerations apply to $\alpha_1$,  if $(m_H^2-m_\Delta^2)/\epsilon$ becomes too large.
Next, it is obviously difficult to have a large $h-\Delta$ mixing $\theta$ near its phenomenological 
limit $0.2$, if $m_\Delta$ becomes too large.
One runs towards non-perturbativity in $\alpha_1\sim \theta Y /2\epsilon \sim m_\Delta/10\, v $.
Namely, one expectedly can not have a large  $h$-$\Delta$ mixing if $\Delta$ is heavier than circa 1--2\,\TeV. 

For all these reasons, a check on couplings shall be performed while varying input parameters, to avoid 
unphysical situations.

For
testing purposes, a function that performs the inversion along with this check is provided along with the model package. 
The inversion above together with the other coupling constants solved previously, is implemented  as a \textsf{Mathematica} function \verb!LRSMEVAL[expr,inputs]!, and provided as additional 
material~\cite{LRSM-tools,mLRSM-1.0-model}.
Here, \verb!inputs! is a set of rules for the relevant needed input parameters, where input masses are to
be expressed in TeV (or as a function of other known masses). 
For instance, 
\begin{align}
\begin{split}
\tt inputs=\big\{&\mathtt{\text{\tt MWR}\to 10 , \,  \text{\tt mH}\to 1.1 \, \text{\tt MWR}},
  \\[.4ex] &\mathtt{\text{\tt mA}\to 1.1 \, \text{\tt MWR}+0.1\,\text{\tt MW} , \, 
  \text{\tt m$\mathtt\Delta $}\to \text{\tt MWR}} , 
  \\[.4ex] &\mathtt{\text{\tt m$\mathtt\Delta $Rpp}\to 1.5\, \text{\tt MWR} , \, 
  \text{\tt m$\mathtt\Delta $L}\to 2.5\,\text{\tt MWR}} , 
  \\[.4ex] & \mathtt{\theta\to 0.05, \,\eta \to 0.01 , \, \phi \to 0.0001,}
  \\[.4ex] &\mathtt{t\to 0.01, \, \alpha \to 0.1 ,  \, \rho4\to 0.5}  \big\}.
\end{split}    
\end{align}
The function \verb!LRSMEVAL! evaluates the given \verb!expr!, function of couplings and masses, for the given physical 
inputs, while checking for perturbativity.
It prints out a warning for each scalar coupling that may turn out to be larger than 3.   
The function was checked against exact numerical diagonalization of the relevant mass matrices. 
A further implementation in~\textsf{Python} is also provided, see below. 

\subsection{Fermion sector} \label{subsec:Fermions}
\noindent
The LRSM fermions transform (in the interaction basis) under $SU(2)_L\times SU(2)_R\times U(1)_{B-L}$ as 
\begin{align}
Q^\prime_{L i} &= \begin{pmatrix}   u_L^\prime 
  \\
  d_L^\prime\end{pmatrix}_i     \sim \left(\mathbf{2}, \mathbf{1}, \mathbf{\frac{1}{3}} \right)  ,
  \\[.7ex]
  Q^\prime_{R i} &= \begin{pmatrix} u_R^\prime 
  \\ 
  d_R^\prime\end{pmatrix}_i     \sim \left(\mathbf{1}, \mathbf{2}, \mathbf{\frac{1}{3}} \right)  ,
  \\[.7ex]
  L^\prime_{L i} &= \begin{pmatrix} \nu_L^\prime 
  \\ 
  \ell_L^\prime\end{pmatrix}_i  \sim \left(\mathbf{2}, \mathbf{1}, -\mathbf{1} \right)  ,
  \\[.7ex]  
  L^\prime_{R i} &= \begin{pmatrix} \nu_R^\prime 
  \\ 
  \ell_R^\prime\end{pmatrix}_i  \sim \left(\mathbf{1}, \mathbf{2}, -\mathbf{1} \right)  ,
\end{align}
and gauge invariant Yukawa terms are written as
\begin{align}
  \mathcal L_Y^q &= \bar Q_L^\prime \left( Y_q \, \phi + \tilde Y_q \, \tilde \phi \right) Q_R^\prime + \text{H.c.} \, ,
  \\[1ex]
  \begin{split}
  \mathcal L_Y^\ell &= \bar L_L^\prime \left( Y_\ell \, \phi + \tilde Y_\ell \, \tilde \phi \right) L_R^\prime +{}  
  \\
  &+ 
  \bar L_L^{\prime c} i \sigma_2 \Delta_L Y_L^M L_L^\prime +
  \bar L_R^{\prime c} i \sigma_2 \Delta_R Y_R^M L_R^\prime + \text{H.c.} \, .
  \end{split}
\end{align}
All fermions thus get their masses exclusively from spontaneous symmetry breaking.

\subsubsection{Quarks: Yukawa terms, masses and mixings}

\noindent
After the bi-doublet acquires its vevs, see Eq.~\eqref{eq:ScalarVevs}, the mass matrices of the quarks can
be written as 
\begin{align}
  M_u &= Y_q \, v_1 - \tilde Y_q \, e^{- i \alpha} v_2 \, ,
  \\
  M_d &= -Y_q \, e^{i \alpha} v_2 + \tilde Y_q \, v_1 \, .
\end{align}
In general, these are arbitrary complex $3 \times 3$ mass matrices that can be diagonalised with 
a bi-unitary transformation
\begin{align}
  M_u &= U_{u L} \, m_u \, U_{u R}^\dagger \, ,
  &
  M_d &= U_{d L} \, m_d \, U_{d R}^\dagger \, ,
\end{align}
where $m_u$ and $m_d$ are diagonal with positive eigenvalues and $U_x U_x^\dagger = 1$.
What enters into the physical charged current interactions are the usual CKM matrix 
$V_L^\text{CKM} \equiv U_{u L}^\dagger U_{d L}$ and its right-handed analog
$V_R = U_{u R}^\dagger U_{d R}$, which in principle has different angles and five extra phases.
The additional CP phases can be extracted as
\begin{equation}
  V_R = K_u \, V_R^\text{CKM} \, K_d \, ,
\end{equation}
where $V_R^\text{CKM}$ is the right-handed analog of the CKM matrix and $K_{u,d}$ are diagonal 
matrices of complex phases
\begin{align}
  K_u &= \mathrm{diag} \left( e^{i\theta_u}, e^{i \theta_c}, e^{i\theta_t} \right) \, ,
  \\
  K_d &= \mathrm{diag} \left( e^{i\theta_d}, e^{i \theta_s}, e^{i\theta_b} \right) \, ,
\end{align}
out of which only five are physical.
The quark Yukawa couplings that enter in the Lagrangian can be re-written in terms of the physical 
masses and mixings:
\begin{align}
  Y_q &= \frac{1}{v_1^2 - v_2^2} \left(M_u \, v_1 + M_d \, e^{-i \alpha} v_2\right) \, ,
  \\
  \tilde Y_q &= -\frac{1}{v_1^2 - v_2^2} \left(M_d \, v_1 + M_u \, e^{i \alpha} v_2\right) \, .
\end{align}
Already from the third generation, one understands from here that the region of $t_\beta \lesssim 0.5$ leads
to large non-perturbative couplings and should be excluded, see Eq.~(2.14) in~\cite{Nemevsek:2023yjl} and the Appendix 
A of~\cite{Maiezza:2010ic} for details.

\subsubsection{Leptons: Yukawa terms, Dirac and Majorana masses and mixings}

\noindent
In the lepton sector, after the bi-doublet and the scalar triplets acquire their respective vevs, 
we have the usual Dirac mass for the charged leptons and a type I + II seesaw mechanism in the 
neutrino sector.
The charged lepton masses and the Dirac-Yukawa term between the left- and right-handed neutrinos 
can be cast as
\begin{align}
  M_D    &= Y_\ell v_1 - \tilde Y_\ell e^{- i \alpha} v_2 \, ,
  \\
  M_\ell &= -Y_\ell e^{i \alpha} v_2  + \tilde Y_\ell v_1 \, .
\end{align}

In the same way, we can also rewrite the lepton Yukawa couplings in terms of the charged lepton 
masses and the Dirac mass term for the neutrinos
\begin{equation}
\begin{split}
  Y_\ell &= \frac{1}{v_1^2 - v_2^2} \left(M_D \, v_1 + M_\ell\, e^{-i \alpha} v_2 \right)  ,
  \\
  \tilde Y_\ell &= -\frac{1}{v_1^2 - v_2^2}\left(M_\ell\, v_1 + M_D\, e^{i \alpha} v_2 \right)  .
\end{split}
\label{eq:Y}
\end{equation}
Similarly to the quark masses, the charged lepton mass term can be diagonalised via a bi-unitary 
transformation
\begin{equation}
  M_\ell = U_{\ell L}\, m_\ell\, U_{\ell R}^\dagger \, .
\end{equation}

In the neutrino sector on the other hand, one has to write the mass-matrix as
\begin{align}
\begin{split}
  &\mathcal L_{\nu\text{-mass}} = -\frac{1}{2} \bar n^\prime_L M_n n^{\prime c}_L 
  \\=& -\frac{1}{2} \left(\bar\nu_L^\prime\:\bar\nu_R^{\prime c}\right)
  \begin{pmatrix}
    M_L & M_D 
    \\
    M_D^T & M_R
  \end{pmatrix}
  \begin{pmatrix} \nu_L^{\prime c} \\ \nu_R^\prime \end{pmatrix} + \text{H.c.} \, ,
\end{split}  
\end{align}
in which $n^\prime_L = \left(\bar\nu_L^\prime\:\bar\nu_R^{\prime c}\right)$ and $M_L = v_L Y_L^M$ 
and $M_R = v_R Y_R^M$.
This $6 \times 6$ Majorana mass matrix is by construction (complex) symmetric and can be diagonalised via 
an Autonne-Takagi factorisation.

The full mass-matrix can be perturbatively (block-) diagonalised with a unitary rotation
\begin{equation}
  \tilde W^T M_n \tilde W = \begin{pmatrix} m_\text{light} & \mathbb{0}
  \\
  \mathbb{0} & m_\text{heavy}\end{pmatrix}  ,
\end{equation}
in which the rotation matrix can be defined as
\begin{equation}
  \tilde W = \begin{pmatrix}\sqrt{{ \mathbb 1 } - B B^\dagger} & B\\-B^\dagger & 
  \sqrt{ \mathbb{1} - B^\dagger B} \end{pmatrix}  .
\end{equation}
The light and heavy mass eigenstates are up to leading order in $M_R^{-1}$ given by
\begin{align} \label{eqn:seesaw}
  M_\nu &\simeq M_L - M_D M_R^{-1} M_D^T \, , & M_N &\simeq M_R \, .
\end{align}
The light and heavy mass matrices are then diagonalised via unitary rotations as
\begin{align}
  V_\nu^T M_\nu V_\nu &= \mathrm{diag}(m_{\nu_1}, m_{\nu_2}, m_{\nu_3}) \, , 
  \\
  V_N^\dagger M_N V_N^* &= \mathrm{diag}(m_{N_1}, m_{N_2}, m_{N_3}) \, ,
\end{align}
in which the light states are identified with mostly active neutrinos.
The full rotation matrix can be written as
\begin{align}
  W &= \tilde W \begin{pmatrix} V_\nu & \mathbb{0} \\ \mathbb{0} & V_N^* \end{pmatrix}  .
\end{align}
We can further expand the unitary matrix $W$ (and the matrix $B$) in inverse powers of $M_R$, 
resulting in
\begin{align}
  \sqrt{\mathbb{1} - B B^\dagger} &\simeq \mathbb{1} - \frac{1}{2} B_1 B_1^\dagger - 
  \frac{1}{2} \left( B_1 B_2^\dagger + B_2 B_1^\dagger \right)  ,
\end{align}
which is vaild up to $\mathcal O(M_R^{-4})$, with
\begin{align}
  B_1 &= M_D^\dagger M_R^{-1\dagger} \, ,
  &
  B_2 &= M_L^\dagger M_D^T M_R^{-1 T} M_R^{-1\dagger} \, ,
\end{align}
see~\cite{Grimus:2000vj} for further details of the perturbative expansion.
Up to leading order in $M_R^{-1}$, the full neutrino mixing matrix is then given by
\begin{align}
\begin{split}
  W &= 
  \begin{pmatrix} \sqrt{\mathbb{1} - B B^\dagger} V_\nu & B V_N^*
  \\ 
  -B^\dagger V_\nu & \sqrt{\mathbb{1} - B^\dagger B} V_N^* 
  \end{pmatrix}
  \\ 
  &\simeq 
  \begin{pmatrix} V_\nu & B_1 V_N^* 
  \\ 
  -B_1^\dagger V_\nu &  V_N^*
  \end{pmatrix}  .
\end{split}  
\end{align}
The higher order corrections to the blocks of $\tilde W$ become numerically relevant for the 
``off-diagonal'' blocks, i.e. for the ``light-heavy'' mixing terms.
With the neutrino and charged lepton mixing matrices we can now write the charged current 
interactions in the physical basis
\begin{eqnarray}
  \mathcal L_{cc}^\ell = 
  \frac{g_L}{\sqrt{2}} \bar\ell_L \gamma^\mu \mathcal U_L n_L W_L^\mu + 
  \frac{g_R}{\sqrt{2}} \bar\ell_R \gamma^\mu \mathcal U_R n_R W_R^\mu \, ,
\end{eqnarray}
where the physical neutrino mass eigenstates are $n=(\nu_1,\nu_2,\nu_3,N_1,N_2,N_3)^T$, and where we defined 
the semi-unitary $3 \times 6$ matrices 
\begin{align}
  (\mathcal U_L)_{\alpha i} &= \!\sum_{k=1}^{3} (V_{\ell L}^\dagger)_{\alpha k} W_{k i} \,,\ \,
  \\
  (\mathcal U_R)_{\alpha i} &= \!\sum_{k=1}^{3} (V_{\ell L}^\dagger)_{\alpha k} W_{(k+3) i}\,.
\end{align}
The first $3 \times 3$ block of $\mathcal U_L$ can be identified as the ``would-be'' PMNS matrix 
(which is in general no longer exactly unitary) and the second $3 \times 3$ block of $\mathcal U_R$ as its 
right-handed analog that can be directly accessed by experiments sensitive to right-handed currents.

\medskip

\paragraph{Parametrising $M_D$ in the case of $\mathcal C$.}
In the case of $\mathcal C$-invariance as the manifest LR symmetry, the scalars transform as 
$\phi \leftrightarrow \phi^T$ and $\Delta_L \leftrightarrow \Delta_R^*$, leading to convenient relations of 
the Yukawa couplings
\begin{align}
  Y_{\ell} &= Y_\ell^T \, , & \tilde Y_\ell &= \tilde Y_\ell^T \, , & Y_L^M &= Y_R^M \, ,
\end{align}
and consequently for the mass matrices
\begin{align}
  M_D &= M_D^T \, , & \quad M_L &= \frac{v_L}{v_R} M_R \, .
\end{align}
This allows us to invert Eq.~\eqref{eqn:seesaw} in order to obtain an expression of $M_D$, fully 
determined by physical parameters, in contrast to the additional freedom present in the general case.
The Dirac mass matrix is then given by
\begin{align}
\label{eq:MD}
  M_D &= M_N \sqrt{\frac{v_L}{v_R}\mathbb{1} - M_N^{-1} M_\nu } \, .
\end{align}
The square root of a matrix is in general quite a complicated expression, however it can be simplified to
a relatively simple form
\begin{align}
  \sqrt A &= c_0 \, \mathbb{1} + c_1 \, A + c_2 \, A.A \, ,
\end{align}
which holds for any (also non-symmetric) complex $3 \times 3$ matrix $A$.
The $c_i$ coefficients are calculated in terms of the three invariants of $A$ and are given in 
appendix~\ref{app:SqrtMatrix} below.%
\footnote{In the case of no manifest left-right symmetry, $M_D$ may still be parametrised via an analogue of the 
Casas-Ibarra parametrisation, uncovering the additional free parameters (see e.g.~\cite{Akhmedov:2008tb}).
In the case of $\mathcal P$ as the manifest left-right symmetry, additional relations and restrictions 
apply and the parametrisation of the lepton sector becomes significantly more complicated 
(see e.g.~\cite{Kiers:2022cyc}).}

\subsection{Ghost sector}
\label{subsec:Ghost}

\noindent
To complete the Faddeev-Popov gauge fixing procedure we have to deal with the ghosts, as 
in the case of the SM~\cite{Peskin:1995ev, Cheng:1984vwu, Romao:2012pq}.
In the LRSM we are dealing with an extended gauge group and gauge bosons that mix, therefore the
procedure is slightly more involved.
We need to introduce a set of ghost fields $c_i$ that are associated with the gauge parameters 
$\alpha_i$ in the following way
\begin{align} \label{eq:GhostAlphas}
  c_i &= \left\{ c_L^+, c_L^-, c_R^+, c_R^-, c_{Z_L}, c_{Z_R}, c_A \right \}  ,
  \\
  \alpha_i &= \left\{ \alpha_L^+, \alpha_L^-, \alpha_R^+, \alpha_R^-, \alpha_{Z_L}, \alpha_{Z_R}, \alpha_A \right \} .
\end{align}
These ghosts are written in the mass basis and enter the ghost Lagrangian in the usual way
\begin{align} \label{eq:LghostDef}
    \mathcal L_{\text{ghost}} &= \sum_{i,j} \overline c_i 
    \frac{\partial \left( \delta F_i \right) }{\partial \alpha_j} c_j \, ,
    \\
    F_i &= \left\{ F_L^+, F_L^-, F_R^+, F_R^-, F_{Z_L}, F_{Z_R}, F_A \right \}  ,
\end{align}
where the sums run over the gauge parameters in the mass basis.
The variations of the $F$s are given by
\begin{align}
  \delta F_i^+ &= \partial \delta W_i^+ + i \xi_{W_i} M_{W_i} \delta \varphi^+_i \, ,
  \\
  \delta F_{Z_i} &= \partial \delta Z_i + \xi_{Z_i} M_{Z_i} \delta  \varphi^0_i \, ,
  \\
  \delta F_A &= \partial \delta A \, ,
\end{align}
and depend on the variations of the wbGs $\delta \varphi$ and the gauge fields $\delta V$.
We get those by considering an infinitesimal expansion of the gauge transformations
\begin{align}
  \phi & \to U \phi = \phi + \delta \phi \, ,
  \\
  D \phi & \to U (D \phi) = D \phi + \delta (D \phi) \, .
\end{align}
The first term gives us the variations of the wbG and Higgs fields with respect to $\alpha_i$.
The second term gives the variation of the gauge fields, needed to restore gauge invariance of 
covariant derivatives.
The bi-doublet and the triplet transform in the following way
\begin{align}
  \phi &\to U_L \, \phi \,  U_R^\dagger \, ,
  &
  \Delta_R &\to U_R \Delta_R \, U_R^\dagger \, U_{B-L} \, ,
\end{align}
\vspace*{-4ex}
\begin{align}
    U_{L,R} &= e^{i g \alpha^a \frac{\sigma^a}{2}} \, ,
  &
  U_{B-L} & = e^{i g' \frac{B-L}{2} \alpha^7} \, .
\end{align}
The $\sigma$ matrices in $U_{L,R}$ run over their respective group indices, setting $a = 1, 2, 3$ ($a=4, 5, 6$) 
for the $SU(2)_{L(R)}$ group parameters in the gauge basis; $\alpha^7$ parametrizes the $U(1)_{B-L}$ Abelian part.
The field variations from~\eqref{eq:DefScalar} are
\begin{align}
  \delta \phi &= 
  \begin{pmatrix} 
    \delta \phi_1^{0 *} & \delta \phi_2^+ 
    \\[1ex] 
    \delta \phi_1^- & \delta \phi_2^0 
  \end{pmatrix}  , 
  &
  \delta \Delta_R &= 
  \begin{pmatrix} 
    \frac{\delta \Delta_R^+}{\sqrt 2} & 0 
    \\ 
    \delta \Delta_R^0 & -\frac{\delta \Delta_R^+}{\sqrt 2} 
  \end{pmatrix}  ,
\end{align}
where we focus only on the variations of the wbG, i.e. $\delta \varphi^\pm_{L,R}, \delta \varphi^0_{L,R}$.
We can ignore variations of all the other fields ($\delta \Delta_L$ multiplet, $\delta \Delta_R^{++}$, 
$\delta H^+$, as well as the variations of the neutral scalars) which do not enter in $F_i$.
Expanding the gauge transformations for small $\alpha^i$ and multiplying the generator and field matrices, 
we get the following variations for the bi-doublet and the triplet
\begin{align}
  \delta \phi &= i \frac{g}{2} \left(\Sigma_L
  \phi - \phi  
\Sigma_R \right)  ,
\\  
  \delta \Delta_R &= i \frac{g}{2} \left(
  \Sigma_R
  \Delta_R - \Delta_R  \Sigma_R
 \right)
  + i g' \Delta_R \alpha^7 \, .
\end{align}
with
\begin{equation}
\begin{split}
  \Sigma_L &=
  \begin{pmatrix} 
    \alpha^3 & \alpha^1 - i \alpha^2
    \\
    \alpha^1 + i \alpha^2 & -\alpha^3
  \end{pmatrix},\ 
  \\
  \Sigma_R &=
  \begin{pmatrix} 
    \alpha^6 & \alpha^4 - i \alpha^5
    \\
    \alpha^4 + i \alpha^5 & -\alpha^6
  \end{pmatrix}  .    
\end{split}  
\end{equation}

We now perform all the rotations of the fields using $U_+$ and $O_\varphi$ to go to the mass basis.
We then define the $\alpha_i$ and $c_i$ as the gauge parameters and ghost fields in the same
mass basis as the gauge bosons.
Specifically, we use the following relations
\begin{align}
  \begin{array}{l}
  \sqrt{2} \, \tilde \alpha^\pm_L = \alpha^1 \mp i \alpha^2 \, ,\quad
  \\[1.2ex]
  \sqrt{2} \, \tilde \alpha^\pm_R = \alpha^4 \mp i \alpha^5 \, ,
  \end{array}
  &
  \begin{pmatrix}
    \tilde \alpha_L^+ \\ \tilde \alpha_R^+
  \end{pmatrix} &= U_W
  \begin{pmatrix}
    \alpha_L^+ \\ \alpha_R^+
  \end{pmatrix} ,
\end{align}
to reparametrize the 'charged' gauge parameters (and the same for ghosts when $c^i \to c^\pm_{L, R}$).
Likewise, we utilize the neutral gauge rotations for the remaining 
\begin{align}
  \begin{pmatrix} 
    \alpha^3 \\ \alpha^6 \\ \alpha^7
  \end{pmatrix} & = O_Z
  \begin{pmatrix} 
    \alpha_A \\ \alpha_{Z_L} \\ \alpha_{Z_R}
  \end{pmatrix}  .
\end{align}  

To get the gauge variations $\delta V_i$ in the mass basis, we proceed along similar lines.
We start by deriving $\delta V_i$ in the flavor basis by demanding that the covariant derivatives,
defined in~\eqref{eq:LagKinCovDerPhi} and~\eqref{eq:LagKinCovDerDelta}, transform like the corresponding 
fields $\phi$ and $\Delta_R$
\begin{align}
  D \phi &\to U_L \, \left( D \phi \right)   U_R^\dagger \, ,
  \\
  D \Delta_R &\to U_R \left( D \Delta_R \right)  U_R^\dagger \, U_{B-L} \, .
\end{align}
We use the same expansions in $\alpha$ of the gauge transformations $U_{L,R,B-L}$ as above and apply 
the same rotations $U_W$ and $O_Z$ for $\delta W, \delta Z$, as we did for the $W$ and $Z$.
With both $\delta \phi$ and $\delta V$, we now have all the $\delta F_i$ written as linear functions
of the $\alpha^i$. 
It is thus trivial to take the derivatives $\partial \delta F_i/\partial \alpha^i$ over the group 
parameters and plug them into~\eqref{eq:LghostDef} to finalize the ghost Lagrangian.

\section{Left-Right asymmetric case} \label{sec:ParityBreak}

\noindent
In the previous sections we assumed that parity is manifestly respected in the gauge sector by
imposing $g_L = g_R$.
Such an equality should hold at the scale of parity restoration, which may happen at a high
scale, such that the two couplings $g_{L, R}$ run differently and $g_L \neq g_R$ at a low scale.
Although the running leads to a quite small splitting of gauge couplings, other extensions of the
LRSM may lead to different $g_{L,R}$ with no other appreciable variations. 
It is thus interesting to generalize the results to this possibility. 

Let us then set  $g_L = g$, $g_R = \zeta g$ and $g' = r g_R = r \zeta g$.
This change only affects the various covariant derivatives and does not enter in the potential,
such that the expressions for scalar masses and mixing (including the Goldstones) remain the
same.

\medskip
\noindent {\em Charged gauge bosons.}
The mass matrix for charged gauge bosons now becomes
\begin{align} \label{eq:MW2gaugePV}
  M_{W_{LR}}^2 &= \frac{g^2}{2} v_R^2
  \begin{pmatrix} 
    \epsilon^2 & e^{-i \alpha} \epsilon^2 s_{2\beta} \zeta 
    \\[1ex]
    e^{i \alpha} \epsilon^2 s_{2\beta} \zeta & \left(2 + \epsilon^2 \right) \zeta^2
  \end{pmatrix}  .
\end{align}
The light eigenvalue remains the same $M_{W_L} \simeq g v/\sqrt 2$, but the heavier one gets multiplied
with $\zeta$, such that $M_{W_R} \simeq g v_R \left( 1 + \epsilon^2/4 \right) \zeta$.
The mixing angle is also different and is given by
\begin{align} \label{eq:XiPV}
  s_\xi \simeq \frac{\epsilon^2}{2 \zeta} s_{2 \beta} \simeq \frac{M_{W_L}^2}{M_{W_R}^2} s_{2 \beta} \zeta \, .
\end{align}
Up to the relevant order $\mathcal O(\epsilon^2)$, the Goldstone rotations in Eq.~\eqref{eq:Up} are not
affected by a change in $g_R$, such that $U_+$ remains the same as in the parity conserving case.
Of course the $\zeta$ dependence enters in the actual terms, but the solution for the rotation angle
is $\zeta$ independent.
The same holds for the gauge fixing $F_i^\pm$ terms, once the $\zeta$ dependence is absorbed in $M_{W_R}$.

\medskip
\noindent {\em Neutral gauge bosons.}
With explicit parity breaking, the neutral gauge boson mass matrix is
\begin{align} \label{eq:MZ2DefPV}
  M_{Z_{LR}}^2 &= \frac{g^2}{2} v_R^2 \begin{pmatrix} 
      \epsilon^2	     & - \epsilon^2 \zeta	                 & 0
      \\[1ex]
      - \epsilon^2 \zeta & \left( 4 + \epsilon^2 \right) \zeta^2 & -4 r \zeta
      \\[1ex]
      0 		         & -4 r \zeta			                 & 4 r^2
    \end{pmatrix}  .
\end{align}
and the modified eigenvalues are
\begin{align} \label{eq:MZsPV}
  M_Z &\simeq \frac{g v}{\sqrt 2} \sqrt{1 + \frac{r^2 \zeta^2}{r^2 + \zeta^2}} \, ,
  \\
  M_{Z_{LR}} &\simeq g v_R \sqrt{2 (\zeta^2 + r^2)} \left[ 
  1 + \frac{\epsilon^2 \zeta^4}{8 \left(\zeta^2 + r^2 \right)^2 } \right] .
\end{align}
The definition of the weak mixing angle also impacts the $r$-parameter
\begin{align}
    r &= 
    \frac{t_w}{\sqrt{\zeta^2 - t_w^2}}
    \xrightarrow{\zeta \to 1} \frac{s_w}{\sqrt{c_{2w}}} \, ,
\end{align}
and the gauge boson rotation matrix gets updated to
\begin{align} \label{eq:OZPV}
  O_Z =&
  \begin{pmatrix}
    s_w & -c_w & 0 
    \\[1ex]
    \frac{s_w}{\zeta} & \frac{s_w t_w}{\zeta} & 
    -\frac{\sqrt{\zeta^2 - t_w^2}}{\zeta} 
    \\[1ex]
    \frac{c_w\sqrt{\zeta^2 - t_w^2}}{\zeta}
    & \frac{s_w\sqrt{\zeta^2 - t_w^2}}{\zeta} & \frac{t_w}{\zeta}
\end{pmatrix}\nonumber\\&+\frac{\epsilon^2}{4}
  \begin{pmatrix}
    0 & 0 & \frac{\left(\zeta^2 - t_w^2\right)^{\frac32}}{\zeta^4 } \\[1ex]
    0 & -\frac{\left(\zeta^2 - t_w^2\right)^{2}}{c_w\zeta^5 }& -\frac{t_w^2\left(\zeta^2 - 
    t_w^2\right)^{\frac32}}{\zeta^5 }
    \\[1ex]
    0 & \frac{s_w\left(\zeta^2 - t_w^2\right)^{\frac32}}{c_w^2\zeta^5 }
    & -\frac{t_w\left(\zeta^2 - t_w^2\right)^{2}}{\zeta^5 } 
  \end{pmatrix}  .
\end{align}

As with the singly charged Goldstones, also the neutral ones are rotated with the same matrix like in the 
parity conserving case at the $\epsilon^2$ order.
In particular, the $O_\varphi$ remains the same as it was in~\eqref{eq:Ophi}.
All the other parts of the Lagrangian (fermions, scalars) remain the same, except for the ghosts.
The Lagrangian terms of the ghost sector are updated with $g_L = g, g_R \to \zeta g$ and 
$g' \to r \zeta g$, and their rotations are $\zeta$-dependent, being the same as those of 
the gauge bosons.

As a final notice, the argument inside the square roots in the last expressions should always be positive, 
corresponding to the physical constraint, stemming from the symmetry breaking relations,
\begin{equation}
\frac{g_R}{g_L} \equiv \zeta > t_w \simeq 0.535 \, .
\end{equation}

\begin{table}[b]
\renewcommand{\arraystretch}{1.2}
\begin{tabular}{|l|l|}
\hline
  \verb|mLRSM.fr|               & Main file  (full)                       \\ 
  \verb|mLRSM_UG.fr|            & Main file  (unitary gauge)              \\ \hline
  \verb|LRSM_all_parameters.fr| & External inputs (w/o masses)            \\
  \verb|LRSM_potential.fr|      & Scalar Potential                        \\
  \verb|LRSM_lepton_yukawa.fr|  & Lepton Yukawa definitions               \\
  \verb|LRSM_quark_yukawa.fr|   & Quark Yukawa definitions                \\
  \verb|LRSM_matrices.fr|       & Matrix utilities                        \\
  \verb|LRSM_ONdef.fr|          & Scalar mixing matrix                    \\
  \hline
  \verb|LRSM_Massless_5f.rst|   & Restriction $m_{u,d,s,c,b,e,\mu}=0$       \\
  \verb|LRSM_Massless_5f_nu.rst|& Restriction $m_{u,d,s,c,b,e,\mu,\nu_i}=0$ \\
  \hline
\end{tabular}
\caption{\textsf{FeynRules} files included in the mLRSM  package.}
\label{tab:files}
\end{table}

\section{The FeynRules implementation} \label{sec:FRImplement}

\noindent
The model file is split for cleanliness into a main model file plus various auxiliary files, reported in Table~\ref{tab:files}. 
In particular, the file \verb|LRSM_all_parameters.fr| contains all external parameter definitions, as well all the expressions for the internal parameters calculated from the inversion described in section~\ref{sec:inversion}.
The lepton mass and mixing computations and the extraction of the square root are included in the files \verb|LRSM_lepton_yukawa.fr| and \verb|LRSM_matrices.fr|, and similarly for the quark matrices.  
The quite lengthy expression of the neutral scalars mixings (\ref{eq:ONdef}) are included in \verb|LRSM_ONdef.fr|.

In the full model \verb|mLRSM.fr| we implemented both the Feynman and the Unitary gauge.  
Either option can be adopted (e.g.\ before UFO generation) by defining the switch variable \texttt{FeynmanGauge=True}(\texttt{False}).
For usual unitary gauge studies, a simplified model file without ghosts and wbGs is provided as \texttt{mLRSM\_UG.fr}.
On the other hand, Feynman-gauge models are highly convenient for 
\textsf{CalcHEP}/\textsf{CompHEP}~\cite{Belyaev:2012qa}, where computations can become 10--100 times faster. 

In Table~\ref{tab:Particles} we list all the particles with their chosen masses for the benchmark point.  
Widths are also reported, as calculated from the leading decay diagrams, with the chosen mass spectrum.
In case one or more masses are changed, the relevant widths have clearly to be recalculated for consistency. 

Table~\ref{tab:External} lists the other external parameters, including initial benchmark values, with the LHA blocks and counters.
They include the important dimensionless parameters like $t_\beta$, $\alpha$ and scalar field mixings, as well as the mixing matrices for quarks (CKM) and leptons (PMNS).
Finally, the light-neutrino mass parameters are defined as inputs, including the unknown lightest-neutrino mass and the binary choice between normal and inverted hierarchy.

Inside the model file a number of unphysical fields are defined, in order to hold fields in gauge multiplets, flavor multiplets, collection of Majorana neutrino fields, collections of neutral or charged scalars.
The triplet fields $\Delta_{L,R}$ are arranged in two index $(\mathbf{2},\mathbf{\bar 2})$ representations, and the covariant derivative definition in \textsf{FeynRules} were generalized accordingly for them to work on reducible representations. 

For collider studies, the "light" (dynamic) quarks $u$, $d$, $s$, $c$, $b$ and leptons $e$, $\mu$, as well as the light neutrinos $\nu_i$ can usually be considered massless for all kinematical purposes. 
Setting to zero also the Dirac neutrino couplings leads to a considerable reduction of the complexity of model parameters and interactions, especially welcome at NLO order.  
Thus, along with the full model file, we provide two restriction files for such purpose, see Table~\ref{tab:files}.

\begin{table}
\renewcommand{\arraystretch}{1.2}
\begin{tabular}{|lll|lll|}
  \hline
  & \quad UFO @ LO \qquad &&& \quad  UFO @ NLO \qquad  & \\
  \hline
  & \verb|mlrsm|             &&& \verb|mlrsm-loop|    & \\
  & \verb|mlrsm-nu|          &&& \verb|mlrsm-nu-loop| & \\
  & \verb|mlrsm-full|        &&&                          & \\
  \hline
  & \verb|mlrsm-ug|          &&&  \verb|mlrsm-ug-loop| $(^*)$  & \\
  & \verb|mlrsm-ug-nu|       &&&  \verb|mlrsm-ug-nu-loop| $(^*)$    & \\
  & \verb|mlrsm-ug-full|     &&&                          & \\
  \hline
\end{tabular}
\caption{UFOs included in the mLRSM package, with various restrictions. 
  Here, the first three lines list UFO models in Feynman gauge; the last three instead list models (\texttt{-ug}) 
  where unitary gauge was enforced, stripping off ghosts and wbGs.
  While $e$, $\mu$ are always massless, we provide \texttt{-4f} and \texttt{-5f} massless flavor schemes, and \texttt{-nu} stands for further 
  simplifications with zero light neutrino masses and Dirac couplings. 
  On the other hand, \texttt{-full} stands for models where all fermions are massive.
  Finally, \texttt{-loop} stands for NLO, requiring massless light quarks.  
  $(^*)$ Notice that unitary gauge NLO UFOs are not correctly supported by \textsf{Madgraph} at 
  the moment.
  }
\label{tab:ufos}
\end{table}

\begin{table*}

\small
\renewcommand{\arraystretch}{1.1}
  \centering
  $\begin{array}{|c|c|c|c|c|c|c|c|c|}
  \hline
      \text{Name} & \text{Spin} & \text{Self-Conj.} &\ \text{FR name}\ & \quad\text{FR mass}\quad & \text{Value [GeV]} & \text{FR width}  & \text{Value [GeV]} & \ \text{PDG ID} \
      \\ \hline \hline
      G & 1 & \text{yes} & \texttt{g} & \texttt{MG} & \texttt{0} & \texttt{0} & & 21\\
      A & 1 & \text{yes} & \texttt{A} & \texttt{MA} & \texttt{0} & \texttt{0} & & 22\\
      Z_L & 1 & \text{yes} & \texttt{Z} & \texttt{MZ} & 91.19 & \texttt{WZ} & 2.495 & 23\\      
      W_L^\pm & 1 & \text{no} & \texttt{W+/W-} & \texttt{MW} &  80.4  & \texttt{WW}  &   2.085   & 24\\
      Z_R & 1 & \text{yes} & \texttt{ZR} & \texttt{MZR} & \texttt{internal}  & \texttt{WZR}  &   334.0 & 33\\
      W_R^\pm & 1 & \text{no} & \texttt{WR+/WR-} & \texttt{MWR} &  6000  & \texttt{WWR}  &   190.5   & 34\\
      \hline
      h & 0 & \text{yes} & \texttt{H} & \texttt{MH} &  125.1 & \texttt{WH} &   0.004115   & 25\\
      \Delta & 0 & \text{yes} & \texttt{DD} & \texttt{MDD} &  500  & \texttt{WDD}  &   0.011   & 45\\
      H & 0 & \text{yes} & \texttt{HH} & \texttt{MHH} & 20000 & \texttt{WHH} &   2667.9  & 35\\ 
       A & 0 & \text{yes} & \texttt{AA} & \texttt{MAA} & 20010 & \texttt{WAA} &   2019.2   & 55\\
      H^+ & 0 & \text{no} & \texttt{H+/H-} & \texttt{MHP} &\texttt{internal}  &   \texttt{WHP} &  2012.4   & 37\\         
      \Delta_R^{++} & 0 & \text{no} & \texttt{DR++/DR-{-}} & \texttt{MDRPP} &  500  & \texttt{WDRPP} &   25.4   & 9000001\\
      \Delta_L^{++} & 0 & \text{no} & \texttt{DL++/DL-{-}} & \texttt{MDLPP} &\texttt{internal}  & \texttt{WDLPP} &   32.4   & 9000002\\
      \Delta_L^+ & 0 & \text{no} & \texttt{DL+/DL-} & \texttt{MDLP} &\texttt{internal} & \texttt{WDLP} &   28.8  & 9000003\\
      \Delta_L^0 & 0 & \text{yes} & \texttt{DL0} & \texttt{MDL} &  500  & \texttt{WDL} &   25.5   & 9000004\\
      \chi_L^0 & 0 & \text{yes} &\texttt{CHIL0} & \texttt{MCHIL0} &\texttt{internal}  & \texttt{WCHIL0} &   25.5  & 9000005\\
      \hline
      d & \frac{1}{2} & \text{no} & \texttt{d/d\~{}} & \texttt{MD} &  0.00504     &\texttt{0} & & 1\\
      u & \frac{1}{2} & \text{no} & \texttt{u/u\~{}} & \texttt{MU} &  0.00255  &\texttt{0} & & 2\\
      s & \frac{1}{2} & \text{no} & \texttt{s/s\~{}} & \texttt{MS} &  0.101    &\texttt{0} & & 3\\
      c & \frac{1}{2} & \text{no} & \texttt{c/c\~{}} & \texttt{MC} &  1.27     &\texttt{0} & & 4\\
      b & \frac{1}{2} & \text{no} & \texttt{b/b\~{}} & \texttt{MB} &   4.7    &\texttt{0} & & 5\\
      t & \frac{1}{2} & \text{no} & \texttt{t/t\~{}} & \texttt{MT} &   172    &\texttt{WT} &   1.508  & 6\\
      \hline
      e & \frac{1}{2} & \text{no} & \texttt{e-/e+} & \texttt{ME} &   0.511\times 10^{-3}     & \texttt{0} & & 11\\
      \mu & \frac{1}{2} & \text{no} & \texttt{mu-/mu+} & \texttt{MMU} &   0.1057     & \texttt{0} & & 13\\
      \tau & \frac{1}{2} & \text{no} & \texttt{ta-/ta+} & \texttt{MTA} &  1.777      & \texttt{0} & & 15\\
      \hline
      \nu_1 & \frac{1}{2} & \text{yes} & \texttt{ve} & \texttt{MNU1} &  1.\times 10^{-12}     & \texttt{0} & & 12\\
      \nu_2 & \frac{1}{2} & \text{yes} & \texttt{vm} & \texttt{MNU2} &   8.9\times 10^{-12}     & \texttt{0} & & 14\\
      \nu_3 & \frac{1}{2} & \text{yes} & \texttt{vt} & \texttt{MNU3} &   50.4 \times 10^{-12}     & \texttt{0} & & 16\\
      N_1 & \frac{1}{2} & \text{yes} & \texttt{N1} & \texttt{MN1} &  100     & \texttt{WN1} &   3.4\times 10^{-10}  & 9900012\\
      N_2 & \frac{1}{2} & \text{yes} & \texttt{N2} & \texttt{MN2} &  2000     & \texttt{WN2} &   5.4\times10^{-4}  & 9900014\\
      N_3 & \frac{1}{2} & \text{yes} & \texttt{N3} & \texttt{MN3} &   10000    & \texttt{WN3} &   85.9  & 9900016\\
      \hline
    \end{array}$
    \caption{Physical particle fields in the LRSM, with their masses, widths, and PDG IDs.}
    \label{tab:Particles}
\end{table*}

\begin{table*}
\small
\renewcommand{\arraystretch}{1.1}
  \centering
$\begin{array}{|c|c|c|c|c|}
  \hline
  \quad\text{FR Name}\quad & \text{Default} & \quad\text{LHA block}\quad &\text{LHA code} & \text{Description}
  \\ \hline\hline
  \texttt{tb}       &   0.1     &   \texttt{LRSMINPUTS}  &    1    & \text{Bidoublet VEVs ratio $t_\beta$} \\
  \texttt{alp}      &   0.1     &   \texttt{LRSMINPUTS}  &    2    & \text{Bidoublet VEV $v_2$ phase $\alpha$}\\
  \texttt{thetamix} &   0.01    &   \texttt{LRSMINPUTS}  &    3    & \text{$h-\Re\Delta_R$ mixing angle $\theta$} \\
  \texttt{etamix}   &   0.00001   &   \texttt{LRSMINPUTS}  &    4    & \text{$H-\Re\Delta_R$ mixing angle $\eta$} \\
  \texttt{phimix}   &   0       &   \texttt{LRSMINPUTS}  &    5    & \text{$h-\Re\phi_{20}$ mixing angle $\phi$} \\
  \texttt{r4}       &   0       &   \texttt{LRSMINPUTS}  &    6    & \text{$\rho_4$ scalar coupling} \\
  \texttt{zetaLR}   &   1       &   \texttt{LRSMINPUTS}  &    7    & \text{Ratio of Left/Right gauge couplings}\\
  \hline
  \texttt{aS}       &   0.1184  &   \texttt{SMINPUTS}    &    1    & \text{Strong coupling} \\
  \texttt{gw}       &   0.65    &   \texttt{SMINPUTS}    &    2    & \text{Weak coupling $g_L$} \\
  \texttt{aEWM1}    &   127.9   &   \texttt{SMINPUTS}    &    3    & \text{Inverse of electromagnetic coupling} \\
  \texttt{Gf}       &   11.66/\text{TeV}^2  & \texttt{SMINPUTS} & 4 & \text{Fermi coupling} \\
  \texttt{MU\_R}    &    91.19\ \GeV        &   \texttt{LOOP}   & 1 & \text{NLO renormalization scale} \\
\hline
  \texttt{CKMlam}   &   0.2245  & \texttt{CKMBLOCK}  &    1      &     \\
  \texttt{CKMA}     &    0.836  & \texttt{CKMBLOCK}  &    2      &     \text{Left CKM mixing entries}    \\
  \texttt{CKMrho}   &   0.122   & \texttt{CKMBLOCK}  &    3      &     \\
  \texttt{CKMeta}   &   0.355   & \texttt{CKMBLOCK}  &    4      &      \\
  \hline
  \texttt{CKMRs12}  &   0.2245  & \texttt{CKMBLOCK}  &    5      &      \\
  \texttt{CKMRs13}  &   0.0036    & \texttt{CKMBLOCK}  &    6      &     \text{Right CKM mixing entries} \\
  \texttt{CKMRs23}  &   0.0421    & \texttt{CKMBLOCK}  &    7      &      \\
  \texttt{CKMRdel}  &   1.2404     & \texttt{CKMBLOCK}  &    8      &      \\
  \hline
  \texttt{PMNSLs12} &   0.544   &   \texttt{PMNSBLOCK}   &    1      &      \\
  \texttt{PMNSLs23} &   0.756   &   \texttt{PMNSBLOCK}   &    2      &   \text{Left PMNS mixing entries}    \\
  \texttt{PMNSLs13} &   0.148   &   \texttt{PMNSBLOCK}   &    3      &      \\
  \texttt{PMNSLdel} &   3.437   &   \texttt{PMNSBLOCK}   &    4      &      \\
  \hline
  \texttt{PMNSRs12} &   0.544   &   \texttt{PMNSBLOCK}   &    11     &      \\
  \texttt{PMNSRs23} &   0.756   &   \texttt{PMNSBLOCK}   &    12     &    \text{Right PMNS mixing entries}  \\
  \texttt{PMNSRs13} &   0.148   &   \texttt{PMNSBLOCK}   &    13     &     \\
  \texttt{PMNSRdel} &   3.437   &   \texttt{PMNSBLOCK}   &    14     &    \\
  \hline
  \texttt{PMNSLphi1} &   0      &   \texttt{PMNSBLOCK}   &    15      &     \text{Left Majorana CP phases}  \\
  \texttt{PMNSLphi2} &   0      &   \texttt{PMNSBLOCK}   &    16      &     \\
  \hline
  \texttt{PMNSRphi1} &   0      &   \texttt{PMNSBLOCK}   &    17      &      \\
  \texttt{PMNSRphi2} &   0      &   \texttt{PMNSBLOCK}   &    18      &    \text{Right Majorana CP phases} \\
  \texttt{PMNSRphi3} & 0   &   \texttt{PMNSBLOCK}   &    19      &     \\
  \hline
  \texttt{MNU0}     &   0.05\, \eV   &   \texttt{PMNSBLOCK}   &    20    &  \text{Lightest neutrino mass} \\
  \texttt{DMsol}    &  7.41\times 10^{-5}\, \eV^2 &   \texttt{PMNSBLOCK}   &    21    &  \Delta m_{sol}^2 \\
  \texttt{DMatm}    &  2.51\times 10^{-3}\, \eV^2 &   \texttt{PMNSBLOCK}   &    22    &  \Delta m_{atm}^2\\
  \texttt{NUH}      &  1  &   \texttt{PMNSBLOCK}   &    23    &   \text{Neutrino Hierarchy ($\geq0$ Normal, $<0$ Inverted)}   \\
  \hline
  \end{array}$
  \caption{All external parameters (other than particle masses).\label{tab:External}}
\end{table*}

\subsection{UFO, parallelization and speedup}

\noindent
The FR model file can be used to produce the UFO package by using~\textsf{FeynRules};
in this work we used the version \textsf{Feynrules-2.3.49}. 
The complexity of the present model largely exceeds that of previous ones and requires the highest possible 
level of parallelization.
This was not fully present in the current version of~\textsf{FeynRules}, where a number of issues prevents 
optimal running in parallel. 
A patched version is then provided as \textsf{Feynrules-2.3.49\_fne}. 
With this patched version the time required to generate the LO UFO decreases from circa 1 week to $\mathcal O(1)$ hour on 10 cores. 
Similar speedup patches had to be applied to the \textsf{MoGRe} package~\cite{Frixione:2019fxg}, which 
is also provided as \textsf{MoGRe\_v1.1\_fne.m}, as well as to the \textsf{Feynarts} interface within FR.
These speedup patches are essential for the generation of the NLO version, even with no neutrino Dirac parameters, see next subsection.

The \textsf{Mathematica} notebooks used to generate the UFOs as well as the patched packages are available along with the model files as supplementary material, and uploaded to \cite{LRSM-tools,mLRSM-1.0-model}.

\subsection{NLO and restrictions} \label{subsec:NLO}
\noindent
UFO packages at NLO in the QCD coupling can be generated from the above model files, by feeding first the 
relevant (QCD) subset of the Lagrangian into~\textsf{MoGRe}~\cite{Frixione:2019fxg}.
The~\textsf{MoGRe} package calculates the needed one-loop QCD couterterms, which are then processed 
with~\textsf{FeynArts}~\cite{Hahn:2000kx} and~\textsf{NLOCT}~\cite{Degrande:2014vpa} to generate the one loop amplitudes, and these are used 
into~\textsf{FeynRules} to generate the UFO at NLO.

We generated UFOs for a number of relevant model restrictions and cases, as listed in Table~\ref{tab:ufos}. 
The first three lines list models in the Feynman gauge, and the last three lines the models in the unitary gauge, 
devoid of ghosts and wbGs.
The UG models should be sufficient for collider event generation at LO, while at NLO  only Feynman gauge should be used with \textsf{Madgraph}, because loop diagrams are not correctly evaluated in the unitary gauge, at the moment~\cite{Alwall:2014hca}.

Also, at NLO the $u$, $d$, $s$, $c$, $b$ quarks are always restricted to be massless, which is required for numerical stability of the renormalization procedure. 
In case the light neutrino masses, as well as the Dirac couplings, are not needed, one may use 
the~\texttt{mlrsm-nu-loop} model, where they were set to zero, retaining only masses for heavy $N_i$. 
The~\texttt{mlrsm-loop} NLO model permits otherwise nonzero neutrino masses, parametrized by mixing and 
oscillation parameters, as well as lightest neutrino mass scale~\texttt{MNU0} and neutrino hierarchy~\texttt{NUH}. 
We recall that in the model neutrino masses drive the Dirac mass matrix and Yukawa couplings, via Eq.~\eqref{eq:MD} 
and Eqs~\eqref{eq:Y}.
Finally, in the NLO UFOs we removed the four-scalars interactions.

\subsection{Use in \textsf{Madgraph}}

\noindent
The models were tested in~\textsf{Madgraph 3.5.3} both at LO and NLO order, and results for benchmark processes
were obtained as reported in the next section. 
As a basic example one can generate the KS process at the NLO level:
\begin{align}
    &\texttt{import model mlrsm-loop}   \nonumber \\
    &\texttt{define l = l+ l-}          \nonumber \\
    &\texttt{define n = n1 n2 n3}       \nonumber \\
    &\texttt{generate p p > l n [QCD]}. \nonumber
\end{align}
This generates $pp \to \ell N$ events, including processes mediated by $W_R$, by LR gauge boson mixing, and Dirac 
couplings, at NLO order in QCD.

\medskip

\begin{figure*}[t]
  \centering
  \includegraphics[width=0.33\textwidth]{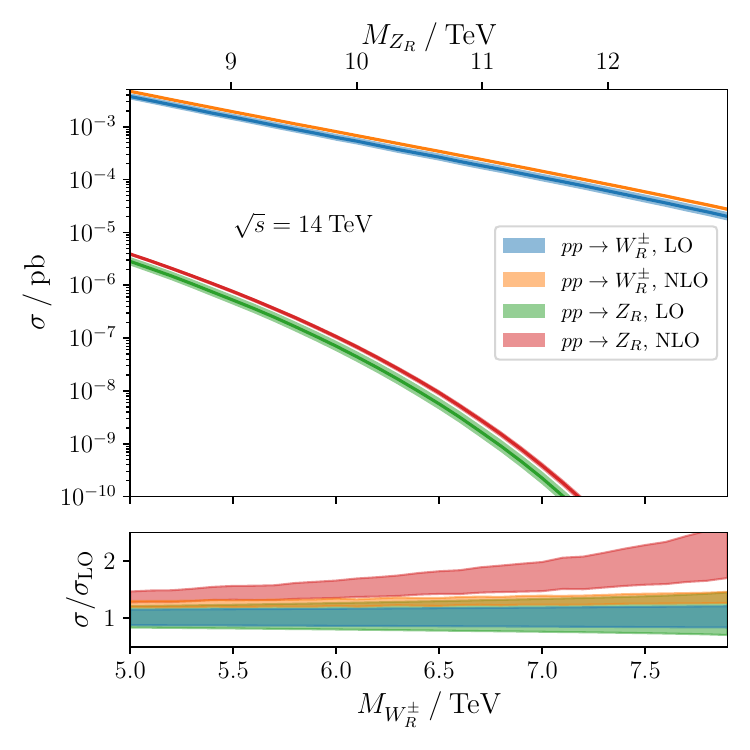}%
  \includegraphics[width=0.33\textwidth]{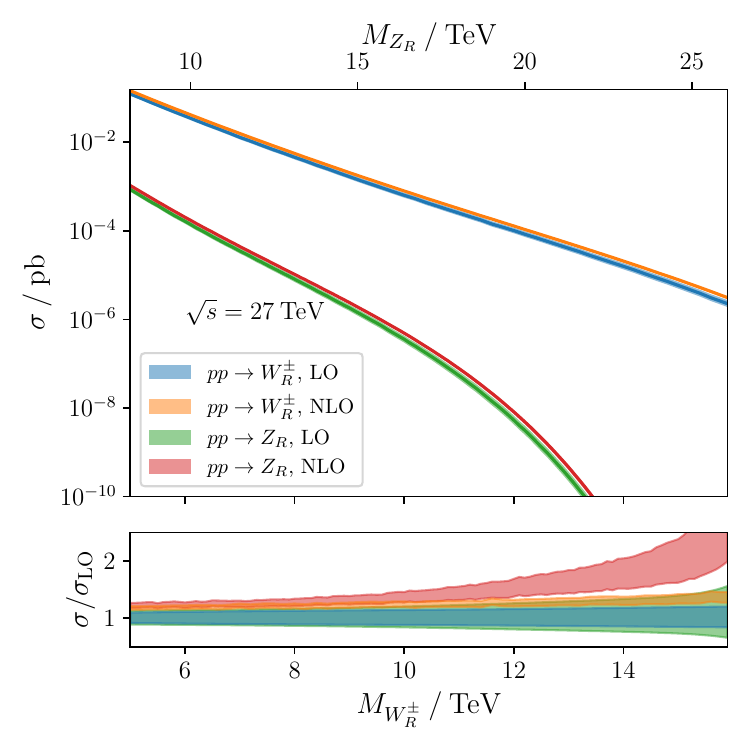}%
  \includegraphics[width=0.33\textwidth]{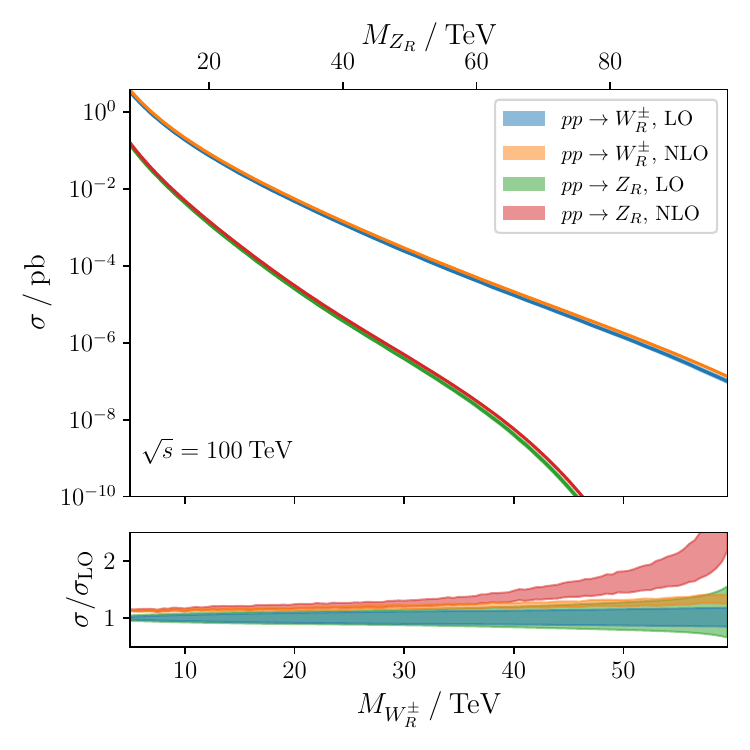}%
    \vspace*{-2ex}
  \caption{%
    Production cross-sections for $W_R$ and $Z_R$ at LO and NLO for $\sqrt s = 14, 27$ and $100$ TeV.
    The lower part shows the $K$-factor, in which the shaded regions denote the scale- and pdf-uncertainties
    added in quadrature.
    Note that $M_{Z_R}$ is shown on the top $x$ axis and explains the kinematic suppression of $pp \to Z_R$ 
    for large $M_{Z_R}$.
    }
    \label{fig:WRProduction}
\end{figure*}

A few technical notes are useful:
\begin{itemize}
  \item In case the chosen physical inputs lead to non-perturbative couplings, \textsf{Madgraph} may issue 
  a \texttt{WARNING: Failed to update dependent parameter.} and results will probably be invalid. 
  We recommend checking the ranges of chose physical inputs with the \texttt{LRSMEVAL} function described above, 
  or with the equivalent Python code. 
  The python code is shipped with a \texttt{jupyter notebook} in which its usage is explained. 
  Apart from providing input parameters in the form of a dictionary, it also includes a 
  ``\texttt{param\_card} reader'' that automatically converts the parameters stored in a \textsf{Madgraph} 
  \texttt{param\_card.dat} to the internal syntax.
  \item For the full model files at NLO, or if neutrinos masses and Dirac couplings are not restricted 
  to zero, update of card parameters in \textsf{Madgraph} hits a time constraint and the following 
  warning is issued: 
  \texttt{WARNING: The model takes too long to load so we bypass the updating of dependent parameter. ...}. 
  In this case the user should insert an "\texttt{update dependent}" command just after card edition, before 
  event generation.
  \item The coupling orders of all interactions are just \texttt{QED} or \texttt{QCD}, with usual 
  hierarchy \texttt{QED=2}, \texttt{QCD=1}. 
\end{itemize}

\begin{figure*}[t]
  \centering
  \includegraphics[width=0.33\textwidth]{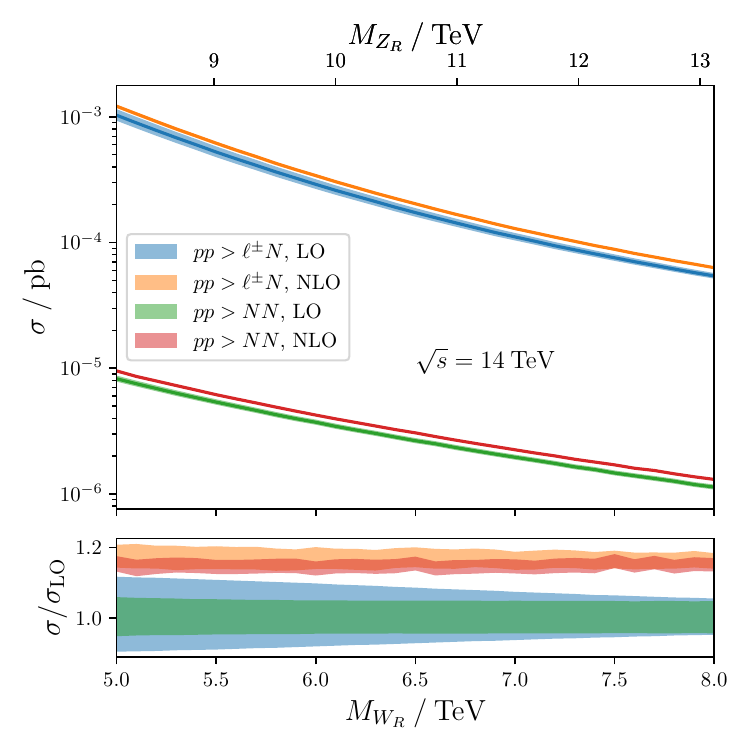}%
  \includegraphics[width=0.33\textwidth]{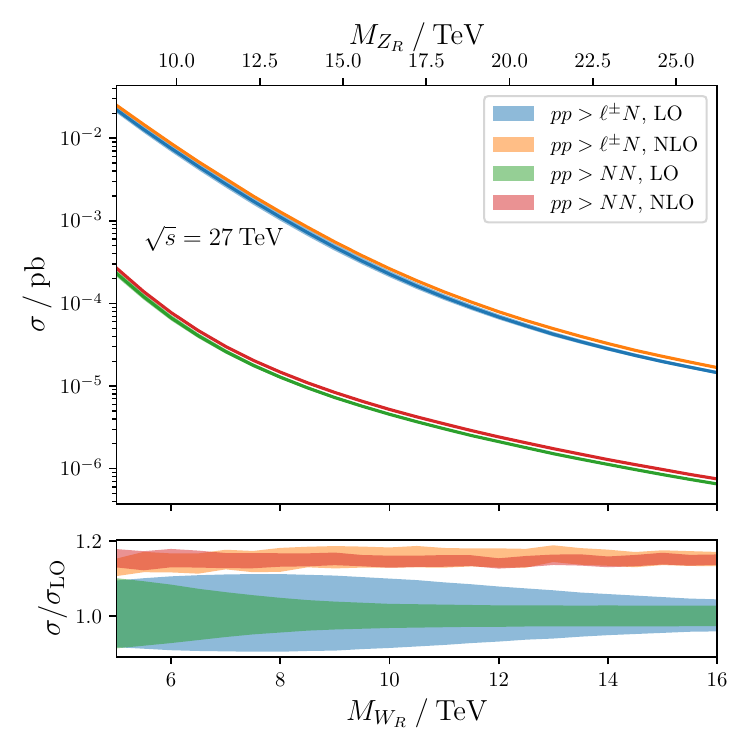}%
  \includegraphics[width=0.33\textwidth]{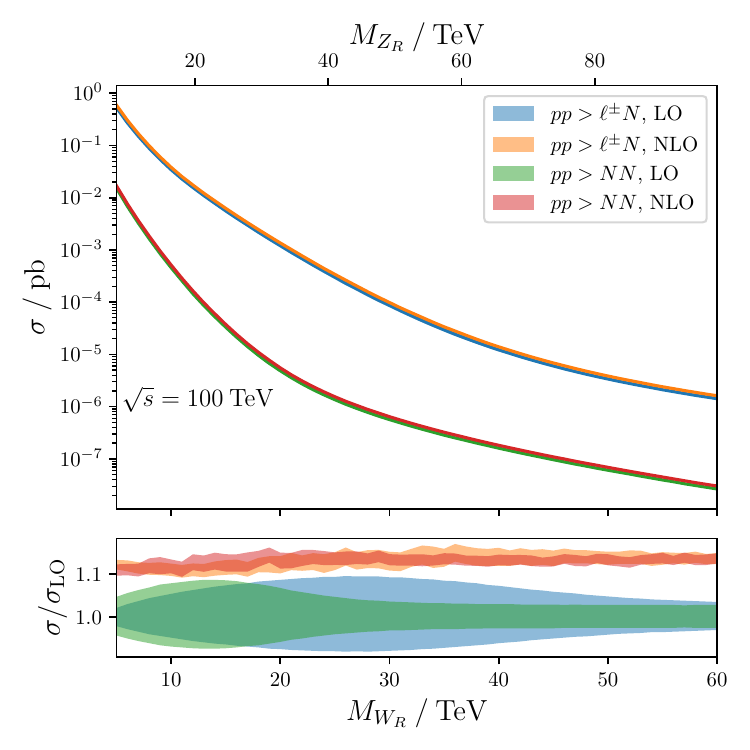}%
    \vspace*{-2ex}
  \caption{%
    Inclusive production cross-sections for $pp\to\ell^\pm N$ ($\ell^\pm = e^\pm, \mu^\pm, \tau^\pm$, 
    $N = N_1, N_2, N_3$) and $pp\to NN$ at LO and NLO for $\sqrt s = 14, 27, 100\,\TeV$. 
    The lower part shows the $K$-factor in which the shaded regions denote the scale- and pdf-uncertainties 
    added in quadrature.
  }
  \label{fig:LNProduction}
\end{figure*}
\begin{figure}[t]
  \centering
  \includegraphics[width=\columnwidth]{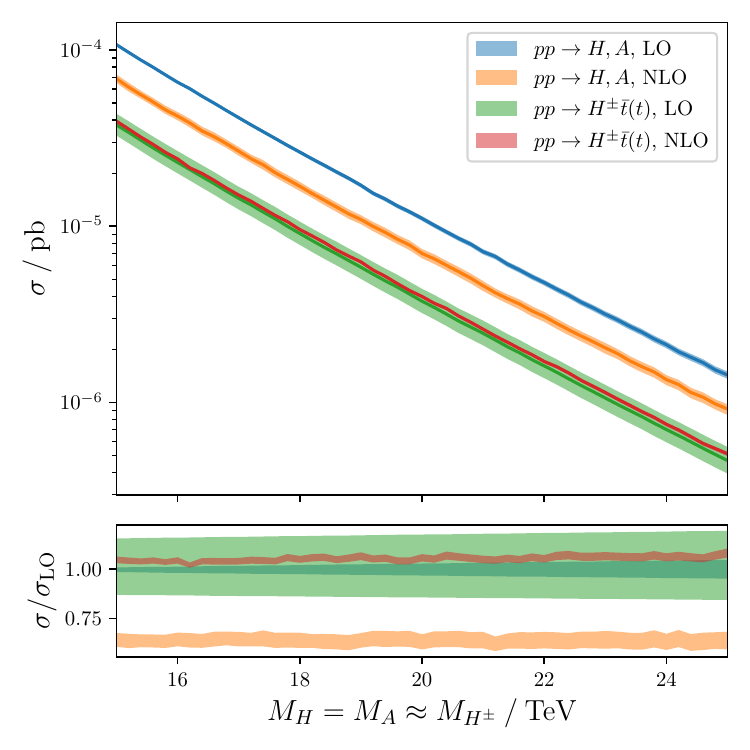}%
    \vspace*{-2ex}
  \caption{%
    Production cross-sections for $H$, $A$ and $H^\pm$ (together with a (anti-)top quark)
    for $\sqrt s = 100\,\TeV$.
    The lower part shows the $K$-factor in which the shaded regions denote the pdf and scale 
    uncertainties added in quadrature.
  }
  \label{fig:AAHHProduction}
\end{figure}

\section{Benchmarks} \label{sec:Benchmarks}

\noindent
In this section we give an overview of a few example processes that can be consistently studied at 
NLO in QCD with the new model file.
For all processes considered we use the following computational setup:
\begin{itemize}
  \item We use the \texttt{mlrsm-nu-loop} version of the model file (5 dynamical flavours, charged 
  lepton and light neutrino masses/yukawas have been set to zero).
  \item All parameters are fixed to their default values as shown in Table~\ref{tab:External} unless 
  otherwise indicated.
  \item We only compute the fixed-order cross sections without hadron shower at LO and NLO.
  \item We use \textsf{MadGraph5\_aMC\_v3.5.3}~\cite{Alwall:2014hca} with LHAPDF 6.5.4~\cite{Buckley:2014ana} 
  with default settings.
  \item The pdfsets are \texttt{NNPDF40\_nlo\_as\_01190}~\cite{NNPDF:2021njg} (\texttt{lhaid=334100}).
  \item For the factorization and renormalization scales we use the default dynamical scale scheme, e.g. for leading order processes dynamical scale scheme
  \#4 ($\mu_F = \mu_R = \sqrt{\hat s}$); pdf and scale uncertainties are added in quadrature.
\end{itemize}

\noindent
In order to estimate the impact of NLO QCD corrections, we define the $K$-factor as
\begin{equation}
  K = \frac{\sigma \pm \delta\sigma}{\sigma_\text{LO}} \, ,
\end{equation}
in which $\delta \sigma$ denotes the uncertainty due to pdf uncertainties (estimated from the pdf replicas) 
and scale variation.
The factorization scale $\mu_F$ and the renormalization scale $\mu_R$ are taken dynamical as 
$\mu_{F,R} = S_{F,R}\sqrt{\hat s}$, where $\hat s$ is the partonic center of mass energy and $S_{F,R}$ 
is a scale factor.
The residual scale dependence is estimated by varying the scale factors $S_{F,R}$ independently on
the interval $[\frac{1}{2}, 1, 2]$ leading to 9 different values of the cross-section.
As the central value one takes $S_F = S_R = 1$, while the scale variation is quoted as the envelope of 
the largest negative/positive deviation. 
Generally, one can expect that NLO corrections lead to a decrease of this residual scale dependence due 
to an order-by-order absorption of the involved scale-dependent terms.

Let us begin by studying the Drell-Yan production cross sections of the new massive gauge bosons.
In Figure~\ref{fig:WRProduction} we show the $2\to 1$ cross-sections of $pp\to W_R^\pm$ and $pp\to Z_R$ 
for $\sqrt{s} = 14,\:27$ and $100\:\mathrm{TeV}$ for a wide range of masses $M_{W_R^\pm}$ and 
$M_{Z_R} \approx 1.67 M_{W_R^\pm}$ (cf. Eq.\eqref{eqn:MZRMWR}).
Note that the mass $M_{Z_R}$ is given in secondary $x$-axis on top.
In the lower parts of the plots we show the LO and NLO $K$-factors in which the shaded regions denote the 
uncertainty due to pdf uncertainties and scale variation.
The NLO corrections lead to an increase of the cross-section of order $\mathcal O(10-20\%)$ (except for $M_{Z_R}$ close to 
the kinematic threshold) and lead to a reduction of the scale dependence from $\mathcal O(10-15\%)$ to $\mathcal O(2-5\%)$.
For larger masses $M_{Z_R}$, close to the kinematic threshold, the $Z_R$ production cross-section strongly 
decreases while the size of the NLO corrections (and therefore the $k$-factors) and also scale dependence strongly increase.
The production cross section strongly increases with center of mass energy by 
three orders of magnitude between $\sqrt{s}=14$~TeV and $\sqrt{s}=100$~TeV.

\begin{figure}[t]
    \centering
    \includegraphics[width=\columnwidth]{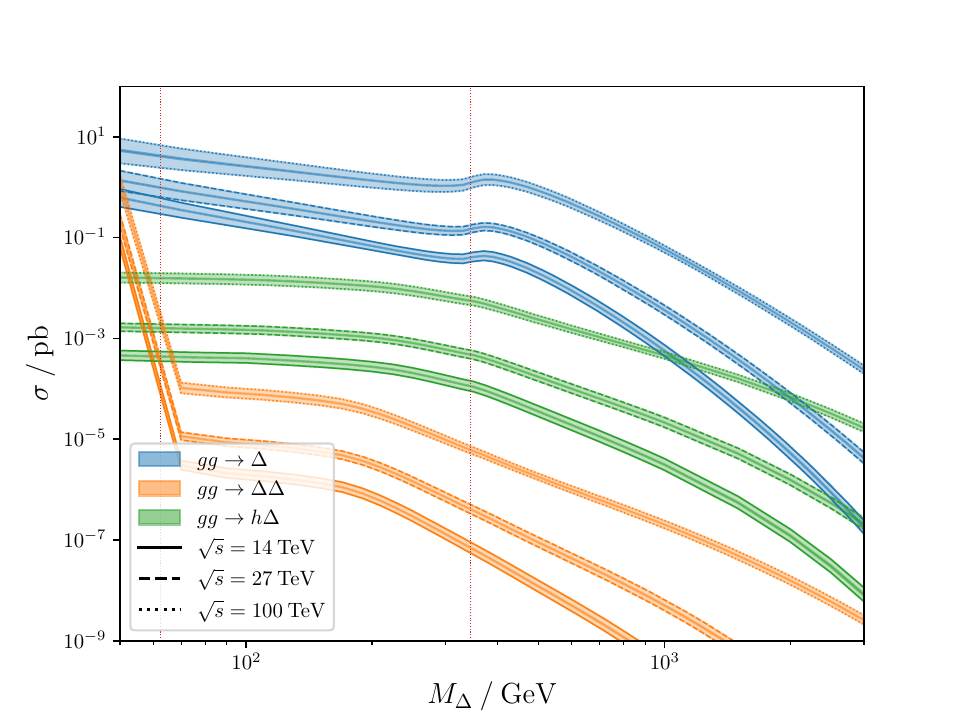}%
    \vspace*{-2ex}
    \caption{Production cross-sections for $g g \to\Delta$ (blue), $g g\to \Delta\Delta$ (orange) 
    and $g g \to h\Delta$ (green) for $\sqrt s = 14$, $27$, $100\,\TeV$ (solid, dashed and dotted 
    boundaries respectively).
    The shaded region denotes the pdf and scale uncertainty added in quadrature, while the vertical 
    red lines denote the Higgs mass threshold ($M_\Delta = \frac{M_h}{2}$) and the top mass threshold 
    ($M_\Delta = 2 M_t$). All cross sections are calculated for $\theta = 0.1$.}
    \label{fig:DeltaProduction}
\end{figure}

In Figure~\ref{fig:LNProduction} we show our results for the inclusive production cross sections 
$pp \to \ell^\pm N$ and $pp \to N N$, in which $\ell^\pm = e^\pm\,,\mu^\pm$ and $\tau^\pm$, 
while $N = N_1\,, N_2$ and $N_3$ (except for $pp\to NN$ at $\sqrt{s}=14$~TeV, since only $N_1$ and $N_2$ 
are kinematically accessible).
Since the $\ell^\pm N$ ($NN$) production is dominantly mediated via the Drell-Yan process 
$p p \to W_R^\pm (Z_R) \to \ell^\pm N (NN)$, we observe a very similar impact of the NLO corrections 
as for the $pp \to W_R^\pm, Z_R$ production cross-sections for most values of $M_{W_R}$ and $M_{Z_R}$. 
For larger $M_{Z_R}$, such that the on-shell production of $Z_R$ as a resonance is suppressed (see also Figure~\ref{fig:WRProduction}), the cross-section is dominated by non-resonant $NN$-production, leading to a different behavior of the cross-sections.
With our choice of the heavy neutrino masses (cf. Table~\ref{tab:External}), both cross sections 
are dominated by having $N_1$ in the final state.

The Drell-Yan production cross-sections for
 heavy (pseudo-) scalars $A$ and $H$ as well as 
the heavy charged scalar $H^\pm$ (with an additional (anti-) top quark in the final state) proceeds mostly via 
initial state $b$-quarks and a gluon, while gluon initiated processes give a 2-3 order of magnitude smaller contribution.
In Figure~\ref{fig:AAHHProduction} we show our results for the  $p p \to H^\pm \bar t (t)$
production cross-sections for $\sqrt{s}=100$~TeV.
The masses $M_H$ and $M_A$ are set equal, while the mass $M_{H^\pm}$ is degenerate with $M_A$ up to order 
$\epsilon^2$ corrections.

Firstly, we notice that for $p p \to A,H$ the NLO corrections lead to a significant {\it decrease} of the 
cross section by $30-40\%$, due to reduced high scale top Yukawa, while the scale dependence slightly increases.
The NLO corrections $p p \to H^\pm \bar t (t)$ lead to a mild increase of the cross section 
and a significant reduction of the scale variation from $\mathcal O(15-20\%)$ to $\mathcal O(1-2\%)$.

Lastly, we consider several production mechanisms of the neutral scalar $\Delta$.
Since $\Delta$ mostly comes from the real part of the neutral component of the right-handed triplet 
$\Delta_R$, it couples to quarks only through mixing with with the SM-like scalar $h$.
Therefore, its largest quark coupling is to the top quark, making different type of gluon fusion 
processes the dominant production mechanisms.
Since gluon fusion is a loop-induced process, we can only consider the leading order cross-sections 
within the scope of \textsf{Madgraph} and this model file.
In particular, we consider single- and pair production via gluon fusion $gg\to \Delta, \Delta\Delta$ 
as well as ``Higgs-$\Delta$-strahlung'' $gg\to h\Delta$ which proceeds both via $gg \to h \to h\Delta$, 
as well as $gg \to \Delta \to h\Delta$ in addition to box-topologies.
In Figure~\ref{fig:DeltaProduction} we show our results for these production mechanisms for 
$m_\Delta \in (50, 3000)$~GeV for $\sqrt{s} = 14 \, , 27$ and $100$~TeV.
In addition, we set here $\theta=0.1$ to attain sizeable production cross sections.
Firstly, we notice that for $M_\Delta \simeq 2 m_t$ (denoted by the right vertical red line) the 
$gg \to \Delta$ cross section increases due to threshold effects in the corresponding ``top-triangle'', 
before it drops off for larger masses.
Next, for $M_\Delta \lesssim M_h/2$ (denoted by the left vertical red line), the $gg \to \Delta \Delta$ 
production cross section becomes large due to an on-shell intermediate Higgs in $gg \to h \to \Delta \Delta$. 
Above threshold, the cross-section strongly decreases and becomes somewhat subdominant.
Lastly, the process $gg \to h \Delta$ has a sizeable cross section for a wide range of masses 
$M_\Delta \in (M_h/2, \sim 500)$~GeV, competing with $gg \to \Delta$ for larger masses.

In addition to the aforementioned ``benchmark processes'', we also checked several SM processes with the 
LRSM model file, such as $p p\to t\bar t$ and $g g \to h \to t \bar t$ production cross-sections, as well as $pp\to VV$ ($V= W^\pm, Z$), finding 
perfect agreement with the built-in \textsf{Madgraph} models.
Finally, in Table~\ref{tab:bench} we summarize our results for the choice of parameters given in 
Table~\ref{tab:External} as a representative benchmark point.
This phenomenological study is by no means complete and rather serves as a proof of concept.
We would like to emphasize that with our model file fully consistent collider studies (cross-section 
calculation and event generation) are now possible with NLO QCD corrections as well as the study of 
loop-induced processes at LO. 
The NLO corrections generally lead to a significant decrease of the scale dependence and therefore to 
a significant reduction of signal modelling uncertainties for collider studies.

\begin{table}
    {\scriptsize
\renewcommand{\arraystretch}{2.05}%
  \centering
  $\begin{array}{|c|c|c|c|c|}  
  \hline
  \multirow{2}{*}{Process} & \sigma_\text{LO}   & \sigma_\text{NLO} & \multirow{2}{*}{$K$} & \sqrt{s} \\[-1.4ex]
     & [\mathrm{fb}]  &  [\mathrm{fb}] &   & [\mathrm{TeV}] \\[1ex]
  \hline
  \hline
   & 0.636_{-13\%}^{+16\%} & 0.806_{-5.4\%}^{+4.4\%} & 1.27 & 14\\
    pp\to W_R^\pm & 38.21_{-9.4\%}^{+11\%} & 46.0_{-3.2\%}^{+2.6\%} & 1.20& 27\\
    & 1572_{-3.0\%}^{+3.2\%} & 1791_{-1.2\%}^{+1.2\%} & 1.14 & 100\\
    \hline
    & 7.0_{-20\%}^{+26\%}\times\! 10^{-5} & 10.6_{-10\%}^{+10\%}\times\! 10^{-5} & 1.52 & 14\\
    pp \to Z_R & 0.16_{-12\%}^{+14\%} & 0.20_{-4.4\%}^{+3.5\%} & 1.25 & 27\\
    & 59.9_{-5.2\%}^{+5.7\%} & 68.6_{-1.6\%}^{+1.5\%} & 1.15 & 100\\
    \hline
    & 0.29_{-8\%}^{+10\%}&0.34_{-3.0\%}^{+2.3\%} &1.17 & 14\\
    p p \to \ell^\pm N &2.79_{-9\%}^{+11\%} &3.22_{-2.7\%}^{+1.8\%} & 1.15& 27\\
    & 271.0_{-2.9\%}^{+3.0\%} & 303.3_{-1.1\%}^{+1.2\%}& 1.12& 100\\
    \hline
    & 3.72_{-4.5\%}^{+5.0\%}\times\! 10^{-3}& 4.24_{-1.8\%}^{+1.7\%}\times\! 10^{-3}& 1.14& 14\\
    pp \to N N & 6.7_{-7\%}^{+8\%}\times\! 10^{-2} & 7.75_{-2.4\%}^{+1.8\%}\times\! 10^{-2}&1.15 & 27\\
    &6.8_{-5.1\%}^{+5.6\%} & 7.59_{-1.3\%}^{+1.1\%}& 1.11& 100\\
    \hline
    \multirow{2}{*}{$p p \to H$} & 8.1_{-16\%}^{+21\%}\times\! 10^{-12} &
    7.2_{-5.2\%}^{+2.2\%}\times\! 10^{-12}& 0.89& 27\\
    & 1.11_{-3.2\%}^{+2.9\%}\times\! 10^{-2}&7.0_{-6.0\%}^{+5.7\%}\times\! 10^{-3} & 0.63& 100\\
    \hline
    \multirow{2}{*}{$p p \to A$} & 8.0_{-16\%}^{+21\%}\times\! 10^{-12} & 7.1_{-5.2\%}^{+2.2\%}\times\! 10^{-12}& 0.89& 27\\
    & 1.10_{-3.2\%}^{+2.9\%}\times\! 10^{-2} & 7.0_{-6.0\%}^{+5.7\%} 10^{-3}& 0.63 & 100\\
    \hline
    \multirow{2}{*}{$p p \to H^\pm \bar t (t)$} & 8.0_{-25\%}^{+36\%}\times\! 10^{-13} & 1.0_{-10\%}^{+7\%}\times\! 10^{-12}& 1.35& 27\\
    & 3.8_{-14\%}^{+18\%}\times\! 10^{-3}& 4.0_{-2.3\%}^{+1.2\%}\times\! 10^{-3}& 1.06& 100\\
    \hline
    & 0.19_{-20\%}^{+26\%}& -- & -- & 14\\
    gg \to \Delta & 0.84_{-17\%}^{+21\%}& -- & -- & 27\\
    & 8.6_{-20\%}^{+25\%}& -- & -- & 100\\
    \hline
    & 3.7_{-22\%}^{+31\%}\times\! 10^{-9} & -- & -- & 14\\
    gg \to \Delta \Delta & 2.7_{-19\%}^{+26\%}\times\! 10^{-8} & -- & -- & 27\\
    & 5.1_{-14\%}^{+17\%}\times\! 10^{-7} & -- & -- & 100\\
    \hline
    & 3.2_{-21\%}^{+29\%} \times 10^{-4}& -- & -- & 14\\
    gg \to \Delta h & 1.7_{-18\%}^{+23\%}\times10^{-3} & -- & -- & 27\\
    & 2.4_{-16\%}^{+20\%}\times 10^{-2} & -- & -- & 100\\
    \hline
  \end{array}$}%
\vspace*{1ex}
  \caption{%
  Example cross sections for the set of input parameters given in Table~\ref{tab:External}. 
  The (super)sub-scripts denote the (upper)lower envelope due to pdf and scale uncertainties. 
  We also quote the central $K$-factor.}
  \label{tab:bench}
\end{table}

%
%
\section{Summary and outlook} \label{sec:Summary}

\noindent The main purpose of this work was to provide an implementation of the LRSM, which is as
complete as possible and can be used for a variety of collider and other phenomenological studies.
The collider frontier in the near future will mostly be focused on the HL-LHC, but in the less 
near future we may expect to see lepton colliders FCC-ee, ILC, CLIC or muon colliders, and finally
high energetic scatterings of protons in FCC-eh or FCC-hh at few tens of TeV (30 or 100 TeV).
With the present model file, the NLO QCD corrections can be reliably calculated for any process within 
the LRSM, properly including the mixings of scalars, gauge bosons and fermions in different gauges, too.
Not all the parts of the model are necessarily needed for simulations, even very precise ones,
therefore we provide a number of restricted versions, which may be faster and easier to use.

Perhaps the most significant analytical developments in this work are the explicit calculation of heavy-light 
neutrino mixing and the solution of the scalar spectrum.
For neutrinos, the Majorana-Dirac connection was found some time ago~\cite{Nemevsek:2012iq}, but with the 
help of the Cayley-Hamilton theorem, we managed to come up with a closed form solution for a general complex 
matrix in terms of its invariants.
This relation may become very useful to clarify the flavor relations in the leptonic sector, e.g.\ how the Dirac induced transitions (in eEDMs, $0\nu2\beta$, collider of $N_1 \to \nu \gamma$) are related
to the flavor structure of $V_R$ and the $m_N$ mass spectrum.
The explicit result allows for a direct insight by taking certain expansions (small angles or masses)
or correction to limits, like type II (sub)dominance.

In the scalar sector, the mass spectrum was derived for the first time without extra assumptions of small mixing 
angles, thus performing an exact diagonalization.
The only phenomenologically required assumption was that the Higgs field $h$ has a limited admixture 
with heavy scalars. 
Remarkably, we were able to derive exact expressions for the model couplings in terms of scalar masses and 
mixings, which can cover both the standard regime of heavy scalars as well as the regime of light $\Delta$.
The latter case is relevant for the Majorana-Higgs program, but the provided solution covers the regimes 
of quasi-degeneracy, too.  
Criteria for perturbativity of the chosen inputs were explicitly identified, and provided as an implemented 
routine. 

The model capabilities were demonstrated with a number of benchmarks concerning the single production of heavy 
resonances, $pp \to W_R, Z_R$, various scalars $pp \to \Delta, H, A, H^\pm$ and production of heavy neutrinos
$pp \to \ell N$, $pp\to NN$, calculated at the NLO level in QCD.

Another  foreseeable improvement might be the inclusion of electroweak corrections.
These could be used to automatize the computation of  electroweak precision observables and
calculate one-loop processes, like radiative production and decays at one loop.
Such an analysis may not be impossible but it would require going beyond the current state of the art, 
given the complexity of the model.
We shall leave it for a future study.
The payoff might be having precise and automatized control over rare processes that might be
possible to explore at future colliders.

\acknowledgments
\noindent We wish to thank Benjamin Fuks for discussions, as well as  for  comments on the manuscript and on the status of the unitary gauge at NLO in \textsf{Madgraph}. We also wish to thank Richard Ruiz for useful discussions. 
MN is supported by the Slovenian Research Agency under the research core funding No.\ P1-0035 
and in part by the research grants N1-0253 and J1-4389, as well as the bilateral project
Proteus PR-12696. 
JK is supported by the Slovenian Research Agency under the research grants N1-0253 and 
in part by J1-4389.

\appendix

\section{Square root(s) of a \texorpdfstring{$3 \times 3$}{3x3} matrix} \label{app:SqrtMatrix}
\noindent
We calculate the closed form analytic expression for the 8 square roots of a general complex
$3 \times 3$ matrix using the Cayley-Hamilton theorem.
For some of the previous work, see~\cite{Hoger:1984sqm, Franca:1989sqm, Ashkartizabi:2016sqm}.

We start by writing the invariants of $A$ and $\sqrt A$
\begin{align}
    T &= \text{tr } A \, ,  &T_{1/2} &= \text{tr } \sqrt A \, ,\\[.7ex]
    T_2 &= \text{tr } A.A \, ,& T_{2, 1/2} &= \text{tr } \sqrt{A}.\sqrt{A} = T \, , \\[.7ex]
    \Delta &= \det A \, , &  \Delta_{1/2} &= \det \sqrt{A} = \sqrt{\Delta} \, ,
\end{align}
where the trace of the square of $\sqrt{A}$ and its determinant are trivially related to 
the invariants of $A$.
Cayley-Hamilton theorem states that matrices themselves are solutions to their characteristic
polynomial equations, which can be written in terms of invariants
\begin{align}
  A^3 - T A^2 + \tilde T A - \Delta \,\mathbb{1} &= 0 \, , 
  & 
  \ \tilde T = \frac{1}{2} \left( T^2 - T_2 \right) .
\end{align}
The same holds for the square root that obeys a polynomial with its own invariants (two of which 
are directly related to the invariants of $A$)
\begin{align}
    \left(\sqrt{A} \right)^3 - T_{1/2} \left(\sqrt{A} \right)^2 + \tilde T_{1/2} \sqrt{A} - 
    \sqrt{\Delta}\, \mathbb{1} &= 0 \, , 
    \\
    \label{eq:Tt12}
    \tilde T_{1/2} = \frac{1}{2} \left( T_{1/2}^2 - T \right)  ,&
    \\[1ex]
    \label{eq:sqAEq0}
    A \sqrt A - T_{1/2} A + \tilde T_{1/2} \sqrt A - \sqrt \Delta \,\mathbb{1} &= 0 \, .
\end{align}

We can get to a linear equation for $\sqrt{A}$ by multiplying~\eqref{eq:sqAEq0} with $\sqrt{A}$
\begin{align} \label{eq:sqAEq1}
  A^2 - T_{1/2} A \sqrt A + \tilde T_{1/2} A - \sqrt \Delta \sqrt A  &= 0 \, ,
  \\[1ex] \label{eq:sqAEq2}
  A^2 - T_{1/2} \left( T_{1/2} A - \tilde T_{1/2} \sqrt A + \sqrt \Delta\, \mathbb{1} \right)+\quad &\nonumber
  \\ 
  {}+ \tilde T_{1/2} A - \sqrt \Delta \sqrt A  &= 0 \, ,
\end{align}
which finally gives
\begin{align} \label{eq:sqASol}
  \sqrt{A} &= \pm\frac{ A^2 + \left( \tilde T_{1/2} - T_{1/2}^2 \right) A - \sqrt{\Delta}\, T_{1/2}\, 
  \mathbb 1}{\sqrt{\Delta} - T_{1/2} \tilde T_{1/2}} \, .
\end{align}
This is already a striking result, where we know the off-diagonal flavor structure of the root directly
from a power expansion of $A$, terminating at the second order (as expected for $3 \times 3$ matrices
from the Cayley-Hamilton theorem).

All the non-linearity is hidden in the equation for the remaining invariant $T_{1/2}$, which we get by
taking the trace of~\eqref{eq:sqAEq2} and using~\eqref{eq:Tt12}, such that
\begin{align}
  T_{1/2}^4 - 2 T \, T_{1/2}^2 - 8 \sqrt{\Delta} \, T_{1/2} + 2 T_2 - T^2 = 0 \, .
\end{align}
This is a depressed quartic with four possible solutions 
\begin{align}
  T_{1/2} &= \frac{\pm \eta_s + s\, \chi}{2 \sqrt{6 \xi}} \, , & s &= \pm \, ,
\end{align}
which in turn give four corresponding $\tilde T_{1/2}$ via (\ref{eq:Tt12}).

The shorthands $\xi$,  $\chi$, $\eta_\pm$, are only functions of the invariants of $A$ and are found 
from these expressions:
\begin{widetext}
\begin{align}
  \xi^3 &= - 32 \left(5 T^3 - 9 T T_2 - 54 \Delta \right) +96 \sqrt{3 \left(T^2 - 2 T_2 \right) 
  \left(T^2 - T_2 \right)^2 + 12 T \left(9 T_2 - 5 T^2 \right) \Delta + 324 \Delta^2} \, ,
  \\
  \chi^2 &= -16 \, \sqrt[3]{2}\, T^2 + 48 \, \sqrt[3]{2}\, T_2 + 8 T \xi + \sqrt[3]{4}\, \xi^2 \, ,
  \\
  \eta_{\pm}^2 &= 16 \, \sqrt[3]{2} \left(T^2 - 3 T_2 \right) + \xi \left(16 T - \sqrt[3]{4}\, \xi \pm 
  96 \sqrt 6\, \chi^{-1}\sqrt{\Delta \xi} \right)  ,
\end{align}
\end{widetext}
where one assumes positive roots only.
Together with the overall sign in~\eqref{eq:sqASol}, we get the total of 8 matrix roots.

With an explicit expression for $\sqrt{A}$, it is also easy to get the inverse root,
by multiplying~\eqref{eq:sqAEq0} with $A^{-1/2}$,
\begin{align}
  A - T_{1/2} \sqrt{A} + \tilde T_{1/2}\, \mathbb{1} - \sqrt \Delta \,A^{-1/2} = 0 \, ,
\end{align}
so that the inverse root is found as
\begin{align}
  \left(\sqrt{A}\right)^{-1} = \frac{1}{\sqrt \Delta} \left( A - T_{1/2} \sqrt{A} + \tilde T_{1/2}\, \mathbb{1} \right)  . 
\end{align}
Again, there are eight possible branches and also the form of this solution is a matrix power expansion
up to $A^2$.

\newpage

\bibliographystyle{apsrev4-2}
\bibliography{references}

\begin{thebibliography}{134}%
\makeatletter
\providecommand \@ifxundefined [1]{%
 \@ifx{#1\undefined}
}%
\providecommand \@ifnum [1]{%
 \ifnum #1\expandafter \@firstoftwo
 \else \expandafter \@secondoftwo
 \fi
}%
\providecommand \@ifx [1]{%
 \ifx #1\expandafter \@firstoftwo
 \else \expandafter \@secondoftwo
 \fi
}%
\providecommand \natexlab [1]{#1}%
\providecommand \enquote  [1]{``#1''}%
\providecommand \bibnamefont  [1]{#1}%
\providecommand \bibfnamefont [1]{#1}%
\providecommand \citenamefont [1]{#1}%
\providecommand \href@noop [0]{\@secondoftwo}%
\providecommand \href [0]{\begingroup \@sanitize@url \@href}%
\providecommand \@href[1]{\@@startlink{#1}\@@href}%
\providecommand \@@href[1]{\endgroup#1\@@endlink}%
\providecommand \@sanitize@url [0]{\catcode `\\12\catcode `\$12\catcode
  `\&12\catcode `\#12\catcode `\^12\catcode `\_12\catcode `\%12\relax}%
\providecommand \@@startlink[1]{}%
\providecommand \@@endlink[0]{}%
\providecommand \url  [0]{\begingroup\@sanitize@url \@url }%
\providecommand \@url [1]{\endgroup\@href {#1}{\urlprefix }}%
\providecommand \urlprefix  [0]{URL }%
\providecommand \Eprint [0]{\href }%
\providecommand \doibase [0]{https://doi.org/}%
\providecommand \selectlanguage [0]{\@gobble}%
\providecommand \bibinfo  [0]{\@secondoftwo}%
\providecommand \bibfield  [0]{\@secondoftwo}%
\providecommand \translation [1]{[#1]}%
\providecommand \BibitemOpen [0]{}%
\providecommand \bibitemStop [0]{}%
\providecommand \bibitemNoStop [0]{.\EOS\space}%
\providecommand \EOS [0]{\spacefactor3000\relax}%
\providecommand \BibitemShut  [1]{\csname bibitem#1\endcsname}%
\let\auto@bib@innerbib\@empty
\bibitem [{\citenamefont {Weinberg}(1967)}]{Weinberg:1967tq}%
  \BibitemOpen
  \bibfield  {author} {\bibinfo {author} {\bibfnamefont {S.}~\bibnamefont
  {Weinberg}},\ }\href {https://doi.org/10.1103/PhysRevLett.19.1264} {\bibfield
   {journal} {\bibinfo  {journal} {Phys. Rev. Lett.}\ }\textbf {\bibinfo
  {volume} {19}},\ \bibinfo {pages} {1264} (\bibinfo {year}
  {1967})}\BibitemShut {NoStop}%
\bibitem [{\citenamefont {Aad}\ \emph {et~al.}(2022)\citenamefont {Aad} \emph
  {et~al.}}]{ATLAS:2022vkf}%
  \BibitemOpen
  \bibfield  {author} {\bibinfo {author} {\bibfnamefont {G.}~\bibnamefont
  {Aad}} \emph {et~al.} (\bibinfo {collaboration} {ATLAS}),\ }\href
  {https://doi.org/10.1038/s41586-022-04893-w} {\bibfield  {journal} {\bibinfo
  {journal} {Nature}\ }\textbf {\bibinfo {volume} {607}},\ \bibinfo {pages}
  {52} (\bibinfo {year} {2022})},\ \bibinfo {note} {[Erratum: Nature 612, E24
  (2022)]},\ \Eprint {https://arxiv.org/abs/2207.00092} {arXiv:2207.00092
  [hep-ex]} \BibitemShut {NoStop}%
\bibitem [{\citenamefont {Tumasyan}\ \emph
  {et~al.}(2022{\natexlab{a}})\citenamefont {Tumasyan} \emph
  {et~al.}}]{CMS:2022dwd}%
  \BibitemOpen
  \bibfield  {author} {\bibinfo {author} {\bibfnamefont {A.}~\bibnamefont
  {Tumasyan}} \emph {et~al.} (\bibinfo {collaboration} {CMS}),\ }\href
  {https://doi.org/10.1038/s41586-022-04892-x} {\bibfield  {journal} {\bibinfo
  {journal} {Nature}\ }\textbf {\bibinfo {volume} {607}},\ \bibinfo {pages}
  {60} (\bibinfo {year} {2022}{\natexlab{a}})},\ \Eprint
  {https://arxiv.org/abs/2207.00043} {arXiv:2207.00043 [hep-ex]} \BibitemShut
  {NoStop}%
\bibitem [{\citenamefont {Pati}\ and\ \citenamefont
  {Salam}(1974)}]{Pati:1974yy}%
  \BibitemOpen
  \bibfield  {author} {\bibinfo {author} {\bibfnamefont {J.~C.}\ \bibnamefont
  {Pati}}\ and\ \bibinfo {author} {\bibfnamefont {A.}~\bibnamefont {Salam}},\
  }\href {https://doi.org/10.1103/PhysRevD.10.275} {\bibfield  {journal}
  {\bibinfo  {journal} {Phys. Rev.}\ }\textbf {\bibinfo {volume} {D10}},\
  \bibinfo {pages} {275} (\bibinfo {year} {1974})},\ \bibinfo {note} {[Erratum:
  Phys. Rev.D11,703(1975)]}\BibitemShut {NoStop}%
\bibitem [{\citenamefont {Mohapatra}\ and\ \citenamefont
  {Pati}(1975)}]{Mohapatra:1974gc}%
  \BibitemOpen
  \bibfield  {author} {\bibinfo {author} {\bibfnamefont {R.~N.}\ \bibnamefont
  {Mohapatra}}\ and\ \bibinfo {author} {\bibfnamefont {J.~C.}\ \bibnamefont
  {Pati}},\ }\href {https://doi.org/10.1103/PhysRevD.11.2558} {\bibfield
  {journal} {\bibinfo  {journal} {Phys. Rev.}\ }\textbf {\bibinfo {volume}
  {D11}},\ \bibinfo {pages} {2558} (\bibinfo {year} {1975})}\BibitemShut
  {NoStop}%
\bibitem [{\citenamefont {Senjanovi\'c}\ and\ \citenamefont
  {Mohapatra}(1975)}]{Senjanovic:1975rk}%
  \BibitemOpen
  \bibfield  {author} {\bibinfo {author} {\bibfnamefont {G.}~\bibnamefont
  {Senjanovi\'c}}\ and\ \bibinfo {author} {\bibfnamefont {R.~N.}\ \bibnamefont
  {Mohapatra}},\ }\href {https://doi.org/10.1103/PhysRevD.12.1502} {\bibfield
  {journal} {\bibinfo  {journal} {Phys. Rev.}\ }\textbf {\bibinfo {volume}
  {D12}},\ \bibinfo {pages} {1502} (\bibinfo {year} {1975})}\BibitemShut
  {NoStop}%
\bibitem [{\citenamefont {Senjanovi\'c}(1979)}]{Senjanovic:1978ev}%
  \BibitemOpen
  \bibfield  {author} {\bibinfo {author} {\bibfnamefont {G.}~\bibnamefont
  {Senjanovi\'c}},\ }\href {https://doi.org/10.1016/0550-3213(79)90604-7}
  {\bibfield  {journal} {\bibinfo  {journal} {Nucl. Phys.}\ }\textbf {\bibinfo
  {volume} {B153}},\ \bibinfo {pages} {334} (\bibinfo {year}
  {1979})}\BibitemShut {NoStop}%
\bibitem [{\citenamefont {Mohapatra}\ and\ \citenamefont
  {Senjanovi\'c}(1980)}]{Mohapatra:1979ia}%
  \BibitemOpen
  \bibfield  {author} {\bibinfo {author} {\bibfnamefont {R.~N.}\ \bibnamefont
  {Mohapatra}}\ and\ \bibinfo {author} {\bibfnamefont {G.}~\bibnamefont
  {Senjanovi\'c}},\ }\href {https://doi.org/10.1103/PhysRevLett.44.912}
  {\bibfield  {journal} {\bibinfo  {journal} {Phys. Rev. Lett.}\ }\textbf
  {\bibinfo {volume} {44}},\ \bibinfo {pages} {912} (\bibinfo {year} {1980})},\
  \bibinfo {note} {[,231(1979)]}\BibitemShut {NoStop}%
\bibitem [{\citenamefont {Senjanovi\'c}\ and\ \citenamefont
  {Senjanovi\'c}(1980)}]{Senjanovic:1979cta}%
  \BibitemOpen
  \bibfield  {author} {\bibinfo {author} {\bibfnamefont {G.}~\bibnamefont
  {Senjanovi\'c}}\ and\ \bibinfo {author} {\bibfnamefont {P.}~\bibnamefont
  {Senjanovi\'c}},\ }\href {https://doi.org/10.1103/PhysRevD.21.3253}
  {\bibfield  {journal} {\bibinfo  {journal} {Phys. Rev.}\ }\textbf {\bibinfo
  {volume} {D21}},\ \bibinfo {pages} {3253} (\bibinfo {year}
  {1980})}\BibitemShut {NoStop}%
\bibitem [{\citenamefont {Minkowski}(1977)}]{Minkowski:1977sc}%
  \BibitemOpen
  \bibfield  {author} {\bibinfo {author} {\bibfnamefont {P.}~\bibnamefont
  {Minkowski}},\ }\href {https://doi.org/10.1016/0370-2693(77)90435-X}
  {\bibfield  {journal} {\bibinfo  {journal} {Phys. Lett.}\ }\textbf {\bibinfo
  {volume} {67B}},\ \bibinfo {pages} {421} (\bibinfo {year}
  {1977})}\BibitemShut {NoStop}%
\bibitem [{\citenamefont {Mohapatra}\ and\ \citenamefont
  {Senjanovi\'c}(1981)}]{Mohapatra:1980yp}%
  \BibitemOpen
  \bibfield  {author} {\bibinfo {author} {\bibfnamefont {R.~N.}\ \bibnamefont
  {Mohapatra}}\ and\ \bibinfo {author} {\bibfnamefont {G.}~\bibnamefont
  {Senjanovi\'c}},\ }\href {https://doi.org/10.1103/PhysRevD.23.165} {\bibfield
   {journal} {\bibinfo  {journal} {Phys. Rev.}\ }\textbf {\bibinfo {volume}
  {D23}},\ \bibinfo {pages} {165} (\bibinfo {year} {1981})}\BibitemShut
  {NoStop}%
\bibitem [{\citenamefont {Gell-Mann}\ \emph {et~al.}(1979)\citenamefont
  {Gell-Mann}, \citenamefont {Ramond},\ and\ \citenamefont
  {Slansky}}]{Gell-Mann:1979vob}%
  \BibitemOpen
  \bibfield  {author} {\bibinfo {author} {\bibfnamefont {M.}~\bibnamefont
  {Gell-Mann}}, \bibinfo {author} {\bibfnamefont {P.}~\bibnamefont {Ramond}},\
  and\ \bibinfo {author} {\bibfnamefont {R.}~\bibnamefont {Slansky}},\
  }\href@noop {} {\bibfield  {journal} {\bibinfo  {journal} {Conf. Proc. C}\
  }\textbf {\bibinfo {volume} {790927}},\ \bibinfo {pages} {315} (\bibinfo
  {year} {1979})},\ \Eprint {https://arxiv.org/abs/1306.4669} {arXiv:1306.4669
  [hep-th]} \BibitemShut {NoStop}%
\bibitem [{\citenamefont {Glashow}(1980)}]{Glashow:1979nm}%
  \BibitemOpen
  \bibfield  {author} {\bibinfo {author} {\bibfnamefont {S.~L.}\ \bibnamefont
  {Glashow}},\ }\href {https://doi.org/10.1007/978-1-4684-7197-7_15} {\bibfield
   {journal} {\bibinfo  {journal} {NATO Sci. Ser. B}\ }\textbf {\bibinfo
  {volume} {61}},\ \bibinfo {pages} {687} (\bibinfo {year} {1980})}\BibitemShut
  {NoStop}%
\bibitem [{\citenamefont {Beall}\ \emph {et~al.}(1982)\citenamefont {Beall},
  \citenamefont {Bander},\ and\ \citenamefont {Soni}}]{Beall:1981ze}%
  \BibitemOpen
  \bibfield  {author} {\bibinfo {author} {\bibfnamefont {G.}~\bibnamefont
  {Beall}}, \bibinfo {author} {\bibfnamefont {M.}~\bibnamefont {Bander}},\ and\
  \bibinfo {author} {\bibfnamefont {A.}~\bibnamefont {Soni}},\ }\bibfield
  {booktitle} {\emph {\bibinfo {booktitle} {{11th International Symposium on
  Lepton and Photon Interactions at High Energies Ithaca, New York, August 4-9,
  1983}}},\ }\href {https://doi.org/10.1103/PhysRevLett.48.848} {\bibfield
  {journal} {\bibinfo  {journal} {Phys. Rev. Lett.}\ }\textbf {\bibinfo
  {volume} {48}},\ \bibinfo {pages} {848} (\bibinfo {year} {1982})}\BibitemShut
  {NoStop}%
\bibitem [{\citenamefont {Mohapatra}\ \emph {et~al.}(1983)\citenamefont
  {Mohapatra}, \citenamefont {Senjanovic},\ and\ \citenamefont
  {Tran}}]{Mohapatra:1983ae}%
  \BibitemOpen
  \bibfield  {author} {\bibinfo {author} {\bibfnamefont {R.~N.}\ \bibnamefont
  {Mohapatra}}, \bibinfo {author} {\bibfnamefont {G.}~\bibnamefont
  {Senjanovic}},\ and\ \bibinfo {author} {\bibfnamefont {M.~D.}\ \bibnamefont
  {Tran}},\ }\href {https://doi.org/10.1103/PhysRevD.28.546} {\bibfield
  {journal} {\bibinfo  {journal} {Phys. Rev. D}\ }\textbf {\bibinfo {volume}
  {28}},\ \bibinfo {pages} {546} (\bibinfo {year} {1983})}\BibitemShut
  {NoStop}%
\bibitem [{\citenamefont {Ecker}\ and\ \citenamefont
  {Grimus}(1985)}]{Ecker:1985vv}%
  \BibitemOpen
  \bibfield  {author} {\bibinfo {author} {\bibfnamefont {G.}~\bibnamefont
  {Ecker}}\ and\ \bibinfo {author} {\bibfnamefont {W.}~\bibnamefont {Grimus}},\
  }\href {https://doi.org/10.1016/0550-3213(85)90616-9} {\bibfield  {journal}
  {\bibinfo  {journal} {Nucl. Phys.}\ }\textbf {\bibinfo {volume} {B258}},\
  \bibinfo {pages} {328} (\bibinfo {year} {1985})}\BibitemShut {NoStop}%
\bibitem [{\citenamefont {Zhang}\ \emph {et~al.}(2008)\citenamefont {Zhang},
  \citenamefont {An}, \citenamefont {Ji},\ and\ \citenamefont
  {Mohapatra}}]{Zhang:2007da}%
  \BibitemOpen
  \bibfield  {author} {\bibinfo {author} {\bibfnamefont {Y.}~\bibnamefont
  {Zhang}}, \bibinfo {author} {\bibfnamefont {H.}~\bibnamefont {An}}, \bibinfo
  {author} {\bibfnamefont {X.}~\bibnamefont {Ji}},\ and\ \bibinfo {author}
  {\bibfnamefont {R.~N.}\ \bibnamefont {Mohapatra}},\ }\href
  {https://doi.org/10.1016/j.nuclphysb.2008.05.019} {\bibfield  {journal}
  {\bibinfo  {journal} {Nucl. Phys.}\ }\textbf {\bibinfo {volume} {B802}},\
  \bibinfo {pages} {247} (\bibinfo {year} {2008})},\ \Eprint
  {https://arxiv.org/abs/0712.4218} {arXiv:0712.4218 [hep-ph]} \BibitemShut
  {NoStop}%
\bibitem [{\citenamefont {Zhang}\ \emph {et~al.}(2007)\citenamefont {Zhang},
  \citenamefont {An}, \citenamefont {Ji},\ and\ \citenamefont
  {Mohapatra}}]{Zhang:2007fn}%
  \BibitemOpen
  \bibfield  {author} {\bibinfo {author} {\bibfnamefont {Y.}~\bibnamefont
  {Zhang}}, \bibinfo {author} {\bibfnamefont {H.}~\bibnamefont {An}}, \bibinfo
  {author} {\bibfnamefont {X.}~\bibnamefont {Ji}},\ and\ \bibinfo {author}
  {\bibfnamefont {R.~N.}\ \bibnamefont {Mohapatra}},\ }\href
  {https://doi.org/10.1103/PhysRevD.76.091301} {\bibfield  {journal} {\bibinfo
  {journal} {Phys. Rev. D}\ }\textbf {\bibinfo {volume} {76}},\ \bibinfo
  {pages} {091301} (\bibinfo {year} {2007})},\ \Eprint
  {https://arxiv.org/abs/0704.1662} {arXiv:0704.1662 [hep-ph]} \BibitemShut
  {NoStop}%
\bibitem [{\citenamefont {Maiezza}\ \emph {et~al.}(2010)\citenamefont
  {Maiezza}, \citenamefont {Nemev\v{s}ek}, \citenamefont {Nesti},\ and\
  \citenamefont {Senjanovi\'c}}]{Maiezza:2010ic}%
  \BibitemOpen
  \bibfield  {author} {\bibinfo {author} {\bibfnamefont {A.}~\bibnamefont
  {Maiezza}}, \bibinfo {author} {\bibfnamefont {M.}~\bibnamefont
  {Nemev\v{s}ek}}, \bibinfo {author} {\bibfnamefont {F.}~\bibnamefont
  {Nesti}},\ and\ \bibinfo {author} {\bibfnamefont {G.}~\bibnamefont
  {Senjanovi\'c}},\ }\href {https://doi.org/10.1103/PhysRevD.82.055022}
  {\bibfield  {journal} {\bibinfo  {journal} {Phys. Rev.}\ }\textbf {\bibinfo
  {volume} {D82}},\ \bibinfo {pages} {055022} (\bibinfo {year} {2010})},\
  \Eprint {https://arxiv.org/abs/1005.5160} {arXiv:1005.5160 [hep-ph]}
  \BibitemShut {NoStop}%
\bibitem [{\citenamefont {Bertolini}\ \emph {et~al.}(2013)\citenamefont
  {Bertolini}, \citenamefont {Maiezza},\ and\ \citenamefont
  {Nesti}}]{Bertolini:2013noa}%
  \BibitemOpen
  \bibfield  {author} {\bibinfo {author} {\bibfnamefont {S.}~\bibnamefont
  {Bertolini}}, \bibinfo {author} {\bibfnamefont {A.}~\bibnamefont {Maiezza}},\
  and\ \bibinfo {author} {\bibfnamefont {F.}~\bibnamefont {Nesti}},\ }\href
  {https://doi.org/10.1103/PhysRevD.88.034014} {\bibfield  {journal} {\bibinfo
  {journal} {Phys. Rev.}\ }\textbf {\bibinfo {volume} {D88}},\ \bibinfo {pages}
  {034014} (\bibinfo {year} {2013})},\ \Eprint
  {https://arxiv.org/abs/1305.5739} {arXiv:1305.5739 [hep-ph]} \BibitemShut
  {NoStop}%
\bibitem [{\citenamefont {Maiezza}\ and\ \citenamefont
  {Nemev\v{s}ek}(2014)}]{Maiezza:2014ala}%
  \BibitemOpen
  \bibfield  {author} {\bibinfo {author} {\bibfnamefont {A.}~\bibnamefont
  {Maiezza}}\ and\ \bibinfo {author} {\bibfnamefont {M.}~\bibnamefont
  {Nemev\v{s}ek}},\ }\href {https://doi.org/10.1103/PhysRevD.90.095002}
  {\bibfield  {journal} {\bibinfo  {journal} {Phys. Rev.}\ }\textbf {\bibinfo
  {volume} {D90}},\ \bibinfo {pages} {095002} (\bibinfo {year} {2014})},\
  \Eprint {https://arxiv.org/abs/1407.3678} {arXiv:1407.3678 [hep-ph]}
  \BibitemShut {NoStop}%
\bibitem [{\citenamefont {Bertolini}\ \emph {et~al.}(2020)\citenamefont
  {Bertolini}, \citenamefont {Maiezza},\ and\ \citenamefont
  {Nesti}}]{Bertolini:2019out}%
  \BibitemOpen
  \bibfield  {author} {\bibinfo {author} {\bibfnamefont {S.}~\bibnamefont
  {Bertolini}}, \bibinfo {author} {\bibfnamefont {A.}~\bibnamefont {Maiezza}},\
  and\ \bibinfo {author} {\bibfnamefont {F.}~\bibnamefont {Nesti}},\ }\href
  {https://doi.org/10.1103/PhysRevD.101.035036} {\bibfield  {journal} {\bibinfo
   {journal} {Phys. Rev. D}\ }\textbf {\bibinfo {volume} {101}},\ \bibinfo
  {pages} {035036} (\bibinfo {year} {2020})},\ \Eprint
  {https://arxiv.org/abs/1911.09472} {arXiv:1911.09472 [hep-ph]} \BibitemShut
  {NoStop}%
\bibitem [{\citenamefont {Bertolini}\ \emph {et~al.}(2021)\citenamefont
  {Bertolini}, \citenamefont {Di~Luzio},\ and\ \citenamefont
  {Nesti}}]{Bertolini:2020hjc}%
  \BibitemOpen
  \bibfield  {author} {\bibinfo {author} {\bibfnamefont {S.}~\bibnamefont
  {Bertolini}}, \bibinfo {author} {\bibfnamefont {L.}~\bibnamefont
  {Di~Luzio}},\ and\ \bibinfo {author} {\bibfnamefont {F.}~\bibnamefont
  {Nesti}},\ }\href {https://doi.org/10.1103/PhysRevLett.126.081801} {\bibfield
   {journal} {\bibinfo  {journal} {Phys. Rev. Lett.}\ }\textbf {\bibinfo
  {volume} {126}},\ \bibinfo {pages} {081801} (\bibinfo {year} {2021})},\
  \Eprint {https://arxiv.org/abs/2006.12508} {arXiv:2006.12508 [hep-ph]}
  \BibitemShut {NoStop}%
\bibitem [{\citenamefont {Bernard}\ \emph {et~al.}(2020)\citenamefont
  {Bernard}, \citenamefont {Descotes-Genon},\ and\ \citenamefont
  {Vale~Silva}}]{Bernard:2020cyi}%
  \BibitemOpen
  \bibfield  {author} {\bibinfo {author} {\bibfnamefont {V.}~\bibnamefont
  {Bernard}}, \bibinfo {author} {\bibfnamefont {S.}~\bibnamefont
  {Descotes-Genon}},\ and\ \bibinfo {author} {\bibfnamefont {L.}~\bibnamefont
  {Vale~Silva}},\ }\href {https://doi.org/10.1007/JHEP09(2020)088} {\bibfield
  {journal} {\bibinfo  {journal} {JHEP}\ }\textbf {\bibinfo {volume}
  {2020}}\bibfield  {number} {\bibinfo  {number} { (09)},\ \bibinfo {pages}
  {088}},\ }\Eprint {https://arxiv.org/abs/2001.00886} {arXiv:2001.00886
  [hep-ph]} \BibitemShut {NoStop}%
\bibitem [{\citenamefont {Dekens}\ \emph {et~al.}(2021)\citenamefont {Dekens},
  \citenamefont {Andreoli}, \citenamefont {de~Vries}, \citenamefont
  {Mereghetti},\ and\ \citenamefont {Oosterhof}}]{Dekens:2021bro}%
  \BibitemOpen
  \bibfield  {author} {\bibinfo {author} {\bibfnamefont {W.}~\bibnamefont
  {Dekens}}, \bibinfo {author} {\bibfnamefont {L.}~\bibnamefont {Andreoli}},
  \bibinfo {author} {\bibfnamefont {J.}~\bibnamefont {de~Vries}}, \bibinfo
  {author} {\bibfnamefont {E.}~\bibnamefont {Mereghetti}},\ and\ \bibinfo
  {author} {\bibfnamefont {F.}~\bibnamefont {Oosterhof}},\ }\href
  {https://doi.org/10.1007/JHEP11(2021)127} {\bibfield  {journal} {\bibinfo
  {journal} {JHEP}\ }\textbf {\bibinfo {volume} {2021}}\bibfield  {number}
  {\bibinfo  {number} { (11)},\ \bibinfo {pages} {127}},\ }\Eprint
  {https://arxiv.org/abs/2107.10852} {arXiv:2107.10852 [hep-ph]} \BibitemShut
  {NoStop}%
\bibitem [{\citenamefont {Atre}\ \emph {et~al.}(2009)\citenamefont {Atre},
  \citenamefont {Han}, \citenamefont {Pascoli},\ and\ \citenamefont
  {Zhang}}]{Atre:2009rg}%
  \BibitemOpen
  \bibfield  {author} {\bibinfo {author} {\bibfnamefont {A.}~\bibnamefont
  {Atre}}, \bibinfo {author} {\bibfnamefont {T.}~\bibnamefont {Han}}, \bibinfo
  {author} {\bibfnamefont {S.}~\bibnamefont {Pascoli}},\ and\ \bibinfo {author}
  {\bibfnamefont {B.}~\bibnamefont {Zhang}},\ }\href
  {https://doi.org/10.1088/1126-6708/2009/05/030} {\bibfield  {journal}
  {\bibinfo  {journal} {JHEP}\ }\textbf {\bibinfo {volume} {2009}}\bibfield
  {number} {\bibinfo  {number} { (05)},\ \bibinfo {pages} {030}},\ }\Eprint
  {https://arxiv.org/abs/0901.3589} {arXiv:0901.3589 [hep-ph]} \BibitemShut
  {NoStop}%
\bibitem [{\citenamefont {Bhupal~Dev}\ \emph {et~al.}(2012)\citenamefont
  {Bhupal~Dev}, \citenamefont {Franceschini},\ and\ \citenamefont
  {Mohapatra}}]{BhupalDev:2012zg}%
  \BibitemOpen
  \bibfield  {author} {\bibinfo {author} {\bibfnamefont {P.~S.}\ \bibnamefont
  {Bhupal~Dev}}, \bibinfo {author} {\bibfnamefont {R.}~\bibnamefont
  {Franceschini}},\ and\ \bibinfo {author} {\bibfnamefont {R.~N.}\ \bibnamefont
  {Mohapatra}},\ }\href {https://doi.org/10.1103/PhysRevD.86.093010} {\bibfield
   {journal} {\bibinfo  {journal} {Phys. Rev. D}\ }\textbf {\bibinfo {volume}
  {86}},\ \bibinfo {pages} {093010} (\bibinfo {year} {2012})},\ \Eprint
  {https://arxiv.org/abs/1207.2756} {arXiv:1207.2756 [hep-ph]} \BibitemShut
  {NoStop}%
\bibitem [{\citenamefont {Cai}\ \emph {et~al.}(2018)\citenamefont {Cai},
  \citenamefont {Han}, \citenamefont {Li},\ and\ \citenamefont
  {Ruiz}}]{Cai:2017mow}%
  \BibitemOpen
  \bibfield  {author} {\bibinfo {author} {\bibfnamefont {Y.}~\bibnamefont
  {Cai}}, \bibinfo {author} {\bibfnamefont {T.}~\bibnamefont {Han}}, \bibinfo
  {author} {\bibfnamefont {T.}~\bibnamefont {Li}},\ and\ \bibinfo {author}
  {\bibfnamefont {R.}~\bibnamefont {Ruiz}},\ }\href
  {https://doi.org/10.3389/fphy.2018.00040} {\bibfield  {journal} {\bibinfo
  {journal} {Front. in Phys.}\ }\textbf {\bibinfo {volume} {6}},\ \bibinfo
  {pages} {40} (\bibinfo {year} {2018})},\ \Eprint
  {https://arxiv.org/abs/1711.02180} {arXiv:1711.02180 [hep-ph]} \BibitemShut
  {NoStop}%
\bibitem [{\citenamefont {Abdullahi}\ \emph {et~al.}(2023)\citenamefont
  {Abdullahi} \emph {et~al.}}]{Abdullahi:2022jlv}%
  \BibitemOpen
  \bibfield  {author} {\bibinfo {author} {\bibfnamefont {A.~M.}\ \bibnamefont
  {Abdullahi}} \emph {et~al.},\ }\href
  {https://doi.org/10.1088/1361-6471/ac98f9} {\bibfield  {journal} {\bibinfo
  {journal} {J. Phys. G}\ }\textbf {\bibinfo {volume} {50}},\ \bibinfo {pages}
  {020501} (\bibinfo {year} {2023})},\ \Eprint
  {https://arxiv.org/abs/2203.08039} {arXiv:2203.08039 [hep-ph]} \BibitemShut
  {NoStop}%
\bibitem [{\citenamefont {Keung}\ and\ \citenamefont
  {Senjanovi\'c}(1983)}]{Keung:1983uu}%
  \BibitemOpen
  \bibfield  {author} {\bibinfo {author} {\bibfnamefont {W.-Y.}\ \bibnamefont
  {Keung}}\ and\ \bibinfo {author} {\bibfnamefont {G.}~\bibnamefont
  {Senjanovi\'c}},\ }\href {https://doi.org/10.1103/PhysRevLett.50.1427}
  {\bibfield  {journal} {\bibinfo  {journal} {Phys. Rev. Lett.}\ }\textbf
  {\bibinfo {volume} {50}},\ \bibinfo {pages} {1427} (\bibinfo {year}
  {1983})}\BibitemShut {NoStop}%
\bibitem [{\citenamefont {Nemev\v{s}ek}\ \emph
  {et~al.}(2011{\natexlab{a}})\citenamefont {Nemev\v{s}ek}, \citenamefont
  {Nesti}, \citenamefont {Senjanovi\'c},\ and\ \citenamefont
  {Zhang}}]{Nemevsek:2011hz}%
  \BibitemOpen
  \bibfield  {author} {\bibinfo {author} {\bibfnamefont {M.}~\bibnamefont
  {Nemev\v{s}ek}}, \bibinfo {author} {\bibfnamefont {F.}~\bibnamefont {Nesti}},
  \bibinfo {author} {\bibfnamefont {G.}~\bibnamefont {Senjanovi\'c}},\ and\
  \bibinfo {author} {\bibfnamefont {Y.}~\bibnamefont {Zhang}},\ }\href
  {https://doi.org/10.1103/PhysRevD.83.115014} {\bibfield  {journal} {\bibinfo
  {journal} {Phys. Rev. D}\ }\textbf {\bibinfo {volume} {83}},\ \bibinfo
  {pages} {115014} (\bibinfo {year} {2011}{\natexlab{a}})},\ \Eprint
  {https://arxiv.org/abs/1103.1627} {arXiv:1103.1627 [hep-ph]} \BibitemShut
  {NoStop}%
\bibitem [{\citenamefont {Nemev\v{s}ek}\ \emph {et~al.}(2018)\citenamefont
  {Nemev\v{s}ek}, \citenamefont {Nesti},\ and\ \citenamefont
  {Popara}}]{Nemevsek:2018bbt}%
  \BibitemOpen
  \bibfield  {author} {\bibinfo {author} {\bibfnamefont {M.}~\bibnamefont
  {Nemev\v{s}ek}}, \bibinfo {author} {\bibfnamefont {F.}~\bibnamefont
  {Nesti}},\ and\ \bibinfo {author} {\bibfnamefont {G.}~\bibnamefont
  {Popara}},\ }\href {https://doi.org/10.1103/PhysRevD.97.115018} {\bibfield
  {journal} {\bibinfo  {journal} {Phys. Rev.}\ }\textbf {\bibinfo {volume}
  {D97}},\ \bibinfo {pages} {115018} (\bibinfo {year} {2018})},\ \Eprint
  {https://arxiv.org/abs/1801.05813} {arXiv:1801.05813 [hep-ph]} \BibitemShut
  {NoStop}%
\bibitem [{\citenamefont {Mitra}\ \emph {et~al.}(2016)\citenamefont {Mitra},
  \citenamefont {Ruiz}, \citenamefont {Scott},\ and\ \citenamefont
  {Spannowsky}}]{Mitra:2016kov}%
  \BibitemOpen
  \bibfield  {author} {\bibinfo {author} {\bibfnamefont {M.}~\bibnamefont
  {Mitra}}, \bibinfo {author} {\bibfnamefont {R.}~\bibnamefont {Ruiz}},
  \bibinfo {author} {\bibfnamefont {D.~J.}\ \bibnamefont {Scott}},\ and\
  \bibinfo {author} {\bibfnamefont {M.}~\bibnamefont {Spannowsky}},\ }\href
  {https://doi.org/10.1103/PhysRevD.94.095016} {\bibfield  {journal} {\bibinfo
  {journal} {Phys. Rev. D}\ }\textbf {\bibinfo {volume} {94}},\ \bibinfo
  {pages} {095016} (\bibinfo {year} {2016})},\ \Eprint
  {https://arxiv.org/abs/1607.03504} {arXiv:1607.03504 [hep-ph]} \BibitemShut
  {NoStop}%
\bibitem [{\citenamefont {Dey}\ \emph {et~al.}(2024)\citenamefont {Dey},
  \citenamefont {Rahaman},\ and\ \citenamefont {Rai}}]{Dey:2022tbp}%
  \BibitemOpen
  \bibfield  {author} {\bibinfo {author} {\bibfnamefont {A.}~\bibnamefont
  {Dey}}, \bibinfo {author} {\bibfnamefont {R.}~\bibnamefont {Rahaman}},\ and\
  \bibinfo {author} {\bibfnamefont {S.~K.}\ \bibnamefont {Rai}},\ }\href
  {https://doi.org/10.1140/epjc/s10052-024-12493-3} {\bibfield  {journal}
  {\bibinfo  {journal} {Eur. Phys. J. C}\ }\textbf {\bibinfo {volume} {84}},\
  \bibinfo {pages} {132} (\bibinfo {year} {2024})},\ \Eprint
  {https://arxiv.org/abs/2207.06857} {arXiv:2207.06857 [hep-ph]} \BibitemShut
  {NoStop}%
\bibitem [{\citenamefont {Alimena}\ \emph {et~al.}(2020)\citenamefont {Alimena}
  \emph {et~al.}}]{Alimena:2019zri}%
  \BibitemOpen
  \bibfield  {author} {\bibinfo {author} {\bibfnamefont {J.}~\bibnamefont
  {Alimena}} \emph {et~al.},\ }\href {https://doi.org/10.1088/1361-6471/ab4574}
  {\bibfield  {journal} {\bibinfo  {journal} {J. Phys. G}\ }\textbf {\bibinfo
  {volume} {47}},\ \bibinfo {pages} {090501} (\bibinfo {year} {2020})},\
  \Eprint {https://arxiv.org/abs/1903.04497} {arXiv:1903.04497 [hep-ex]}
  \BibitemShut {NoStop}%
\bibitem [{\citenamefont {Langacker}\ and\ \citenamefont
  {Sankar}(1989)}]{Langacker:1989xa}%
  \BibitemOpen
  \bibfield  {author} {\bibinfo {author} {\bibfnamefont {P.}~\bibnamefont
  {Langacker}}\ and\ \bibinfo {author} {\bibfnamefont {S.~U.}\ \bibnamefont
  {Sankar}},\ }\href {https://doi.org/10.1103/PhysRevD.40.1569} {\bibfield
  {journal} {\bibinfo  {journal} {Phys. Rev. D}\ }\textbf {\bibinfo {volume}
  {40}},\ \bibinfo {pages} {1569} (\bibinfo {year} {1989})}\BibitemShut
  {NoStop}%
\bibitem [{\citenamefont {Frank}\ \emph {et~al.}(2019)\citenamefont {Frank},
  \citenamefont {\"Ozdal},\ and\ \citenamefont {Poulose}}]{Frank:2018ifw}%
  \BibitemOpen
  \bibfield  {author} {\bibinfo {author} {\bibfnamefont {M.}~\bibnamefont
  {Frank}}, \bibinfo {author} {\bibfnamefont {O.}~\bibnamefont {\"Ozdal}},\
  and\ \bibinfo {author} {\bibfnamefont {P.}~\bibnamefont {Poulose}},\ }\href
  {https://doi.org/10.1103/PhysRevD.99.035001} {\bibfield  {journal} {\bibinfo
  {journal} {Phys. Rev. D}\ }\textbf {\bibinfo {volume} {99}},\ \bibinfo
  {pages} {035001} (\bibinfo {year} {2019})},\ \Eprint
  {https://arxiv.org/abs/1812.05681} {arXiv:1812.05681 [hep-ph]} \BibitemShut
  {NoStop}%
\bibitem [{\citenamefont {Solera}\ \emph {et~al.}(2024)\citenamefont {Solera},
  \citenamefont {Pich},\ and\ \citenamefont {Vale~Silva}}]{Solera:2023kwt}%
  \BibitemOpen
  \bibfield  {author} {\bibinfo {author} {\bibfnamefont {S.~F.}\ \bibnamefont
  {Solera}}, \bibinfo {author} {\bibfnamefont {A.}~\bibnamefont {Pich}},\ and\
  \bibinfo {author} {\bibfnamefont {L.}~\bibnamefont {Vale~Silva}},\ }\href
  {https://doi.org/10.1007/JHEP02(2024)027} {\bibfield  {journal} {\bibinfo
  {journal} {JHEP}\ }\textbf {\bibinfo {volume} {2024}}\bibfield  {number}
  {\bibinfo  {number} { (02)},\ \bibinfo {pages} {027}},\ }\Eprint
  {https://arxiv.org/abs/2309.06094} {arXiv:2309.06094 [hep-ph]} \BibitemShut
  {NoStop}%
\bibitem [{\citenamefont {Rizzo}(2007)}]{Rizzo:2007xs}%
  \BibitemOpen
  \bibfield  {author} {\bibinfo {author} {\bibfnamefont {T.~G.}\ \bibnamefont
  {Rizzo}},\ }\href {https://doi.org/10.1088/1126-6708/2007/05/037} {\bibfield
  {journal} {\bibinfo  {journal} {JHEP}\ }\textbf {\bibinfo {volume}
  {2007}}\bibfield  {number} {\bibinfo  {number} { (05)},\ \bibinfo {pages}
  {037}},\ }\Eprint {https://arxiv.org/abs/0704.0235} {arXiv:0704.0235
  [hep-ph]} \BibitemShut {NoStop}%
\bibitem [{\citenamefont {Han}\ \emph {et~al.}(2013)\citenamefont {Han},
  \citenamefont {Lewis}, \citenamefont {Ruiz},\ and\ \citenamefont
  {Si}}]{Han:2012vk}%
  \BibitemOpen
  \bibfield  {author} {\bibinfo {author} {\bibfnamefont {T.}~\bibnamefont
  {Han}}, \bibinfo {author} {\bibfnamefont {I.}~\bibnamefont {Lewis}}, \bibinfo
  {author} {\bibfnamefont {R.}~\bibnamefont {Ruiz}},\ and\ \bibinfo {author}
  {\bibfnamefont {Z.-g.}\ \bibnamefont {Si}},\ }\href
  {https://doi.org/10.1103/PhysRevD.87.035011} {\bibfield  {journal} {\bibinfo
  {journal} {Phys. Rev. D}\ }\textbf {\bibinfo {volume} {87}},\ \bibinfo
  {pages} {035011} (\bibinfo {year} {2013})},\ \bibinfo {note} {[Erratum:
  Phys.Rev.D 87, 039906 (2013)]},\ \Eprint {https://arxiv.org/abs/1211.6447}
  {arXiv:1211.6447 [hep-ph]} \BibitemShut {NoStop}%
\bibitem [{\citenamefont {Dev}\ \emph {et~al.}(2016{\natexlab{a}})\citenamefont
  {Dev}, \citenamefont {Kim},\ and\ \citenamefont {Mohapatra}}]{Dev:2015kca}%
  \BibitemOpen
  \bibfield  {author} {\bibinfo {author} {\bibfnamefont {P.~S.~B.}\
  \bibnamefont {Dev}}, \bibinfo {author} {\bibfnamefont {D.}~\bibnamefont
  {Kim}},\ and\ \bibinfo {author} {\bibfnamefont {R.~N.}\ \bibnamefont
  {Mohapatra}},\ }\href {https://doi.org/10.1007/JHEP01(2016)118} {\bibfield
  {journal} {\bibinfo  {journal} {JHEP}\ }\textbf {\bibinfo {volume}
  {2016}}\bibfield  {number} {\bibinfo  {number} { (01)},\ \bibinfo {pages}
  {118}},\ }\Eprint {https://arxiv.org/abs/1510.04328} {arXiv:1510.04328
  [hep-ph]} \BibitemShut {NoStop}%
\bibitem [{\citenamefont {Gluza}\ \emph {et~al.}(2016)\citenamefont {Gluza},
  \citenamefont {Jelinski},\ and\ \citenamefont {Szafron}}]{Gluza:2016qqv}%
  \BibitemOpen
  \bibfield  {author} {\bibinfo {author} {\bibfnamefont {J.}~\bibnamefont
  {Gluza}}, \bibinfo {author} {\bibfnamefont {T.}~\bibnamefont {Jelinski}},\
  and\ \bibinfo {author} {\bibfnamefont {R.}~\bibnamefont {Szafron}},\ }\href
  {https://doi.org/10.1103/PhysRevD.93.113017} {\bibfield  {journal} {\bibinfo
  {journal} {Phys. Rev. D}\ }\textbf {\bibinfo {volume} {93}},\ \bibinfo
  {pages} {113017} (\bibinfo {year} {2016})},\ \Eprint
  {https://arxiv.org/abs/1604.01388} {arXiv:1604.01388 [hep-ph]} \BibitemShut
  {NoStop}%
\bibitem [{\citenamefont {Frank}\ \emph {et~al.}(2023)\citenamefont {Frank},
  \citenamefont {Fuks}, \citenamefont {Jueid}, \citenamefont {Moretti},\ and\
  \citenamefont {Ozdal}}]{Frank:2023epx}%
  \BibitemOpen
  \bibfield  {author} {\bibinfo {author} {\bibfnamefont {M.}~\bibnamefont
  {Frank}}, \bibinfo {author} {\bibfnamefont {B.}~\bibnamefont {Fuks}},
  \bibinfo {author} {\bibfnamefont {A.}~\bibnamefont {Jueid}}, \bibinfo
  {author} {\bibfnamefont {S.}~\bibnamefont {Moretti}},\ and\ \bibinfo {author}
  {\bibfnamefont {O.}~\bibnamefont {Ozdal}},\ }\href@noop {} {\bibfield
  {journal} {\bibinfo  {journal} {arXiv}\ } (\bibinfo {year} {2023})},\ \Eprint
  {https://arxiv.org/abs/2312.08521} {arXiv:2312.08521 [hep-ph]} \BibitemShut
  {NoStop}%
\bibitem [{\citenamefont {Mandal}\ \emph {et~al.}(2017)\citenamefont {Mandal},
  \citenamefont {Mitra},\ and\ \citenamefont {Sinha}}]{Mandal:2017tab}%
  \BibitemOpen
  \bibfield  {author} {\bibinfo {author} {\bibfnamefont {S.}~\bibnamefont
  {Mandal}}, \bibinfo {author} {\bibfnamefont {M.}~\bibnamefont {Mitra}},\ and\
  \bibinfo {author} {\bibfnamefont {N.}~\bibnamefont {Sinha}},\ }\href
  {https://doi.org/10.1103/PhysRevD.96.035023} {\bibfield  {journal} {\bibinfo
  {journal} {Phys. Rev. D}\ }\textbf {\bibinfo {volume} {96}},\ \bibinfo
  {pages} {035023} (\bibinfo {year} {2017})},\ \Eprint
  {https://arxiv.org/abs/1705.01932} {arXiv:1705.01932 [hep-ph]} \BibitemShut
  {NoStop}%
\bibitem [{\citenamefont {Godbole}\ \emph {et~al.}(2021)\citenamefont
  {Godbole}, \citenamefont {Maharathy}, \citenamefont {Mandal}, \citenamefont
  {Mitra},\ and\ \citenamefont {Sinha}}]{Godbole:2020jqw}%
  \BibitemOpen
  \bibfield  {author} {\bibinfo {author} {\bibfnamefont {R.~M.}\ \bibnamefont
  {Godbole}}, \bibinfo {author} {\bibfnamefont {S.~P.}\ \bibnamefont
  {Maharathy}}, \bibinfo {author} {\bibfnamefont {S.}~\bibnamefont {Mandal}},
  \bibinfo {author} {\bibfnamefont {M.}~\bibnamefont {Mitra}},\ and\ \bibinfo
  {author} {\bibfnamefont {N.}~\bibnamefont {Sinha}},\ }\href
  {https://doi.org/10.1103/PhysRevD.104.095009} {\bibfield  {journal} {\bibinfo
   {journal} {Phys. Rev. D}\ }\textbf {\bibinfo {volume} {104}},\ \bibinfo
  {pages} {095009} (\bibinfo {year} {2021})},\ \Eprint
  {https://arxiv.org/abs/2008.05467} {arXiv:2008.05467 [hep-ph]} \BibitemShut
  {NoStop}%
\bibitem [{\citenamefont {Alves}\ \emph {et~al.}(2024)\citenamefont {Alves},
  \citenamefont {Fong}, \citenamefont {Leal},\ and\ \citenamefont
  {Zukanovich~Funchal}}]{Alves:2023znq}%
  \BibitemOpen
  \bibfield  {author} {\bibinfo {author} {\bibfnamefont {G.~F.~S.}\
  \bibnamefont {Alves}}, \bibinfo {author} {\bibfnamefont {C.~S.}\ \bibnamefont
  {Fong}}, \bibinfo {author} {\bibfnamefont {L.~P.~S.}\ \bibnamefont {Leal}},\
  and\ \bibinfo {author} {\bibfnamefont {R.}~\bibnamefont
  {Zukanovich~Funchal}},\ }\href
  {https://doi.org/10.1103/PhysRevLett.133.161802} {\bibfield  {journal}
  {\bibinfo  {journal} {Phys. Rev. Lett.}\ }\textbf {\bibinfo {volume} {133}},\
  \bibinfo {pages} {161802} (\bibinfo {year} {2024})},\ \Eprint
  {https://arxiv.org/abs/2307.04862} {arXiv:2307.04862 [hep-ph]} \BibitemShut
  {NoStop}%
\bibitem [{\citenamefont {Ruiz}(2017)}]{Ruiz:2017nip}%
  \BibitemOpen
  \bibfield  {author} {\bibinfo {author} {\bibfnamefont {R.}~\bibnamefont
  {Ruiz}},\ }\href {https://doi.org/10.1140/epjc/s10052-017-4950-2} {\bibfield
  {journal} {\bibinfo  {journal} {Eur. Phys. J.}\ }\textbf {\bibinfo {volume}
  {C77}},\ \bibinfo {pages} {375} (\bibinfo {year} {2017})},\ \Eprint
  {https://arxiv.org/abs/1703.04669} {arXiv:1703.04669 [hep-ph]} \BibitemShut
  {NoStop}%
\bibitem [{\citenamefont {Nemev\v{s}ek}\ and\ \citenamefont
  {Nesti}(2023)}]{Nemevsek:2023hwx}%
  \BibitemOpen
  \bibfield  {author} {\bibinfo {author} {\bibfnamefont {M.}~\bibnamefont
  {Nemev\v{s}ek}}\ and\ \bibinfo {author} {\bibfnamefont {F.}~\bibnamefont
  {Nesti}},\ }\href {https://doi.org/10.1103/PhysRevD.108.015030} {\bibfield
  {journal} {\bibinfo  {journal} {Phys. Rev. D}\ }\textbf {\bibinfo {volume}
  {108}},\ \bibinfo {pages} {015030} (\bibinfo {year} {2023})},\ \Eprint
  {https://arxiv.org/abs/2306.12104} {arXiv:2306.12104 [hep-ph]} \BibitemShut
  {NoStop}%
\bibitem [{\citenamefont {Rizzo}(2014)}]{Rizzo:2014xma}%
  \BibitemOpen
  \bibfield  {author} {\bibinfo {author} {\bibfnamefont {T.~G.}\ \bibnamefont
  {Rizzo}},\ }\href {https://doi.org/10.1103/PhysRevD.89.095022} {\bibfield
  {journal} {\bibinfo  {journal} {Phys. Rev. D}\ }\textbf {\bibinfo {volume}
  {89}},\ \bibinfo {pages} {095022} (\bibinfo {year} {2014})},\ \Eprint
  {https://arxiv.org/abs/1403.5465} {arXiv:1403.5465 [hep-ph]} \BibitemShut
  {NoStop}%
\bibitem [{\citenamefont {Ng}\ \emph {et~al.}(2015)\citenamefont {Ng},
  \citenamefont {de~la Puente},\ and\ \citenamefont {Pan}}]{Ng:2015hba}%
  \BibitemOpen
  \bibfield  {author} {\bibinfo {author} {\bibfnamefont {J.~N.}\ \bibnamefont
  {Ng}}, \bibinfo {author} {\bibfnamefont {A.}~\bibnamefont {de~la Puente}},\
  and\ \bibinfo {author} {\bibfnamefont {B.~W.-P.}\ \bibnamefont {Pan}},\
  }\href {https://doi.org/10.1007/JHEP12(2015)172} {\bibfield  {journal}
  {\bibinfo  {journal} {JHEP}\ }\textbf {\bibinfo {volume} {2015}}\bibfield
  {number} {\bibinfo  {number} { (12)},\ \bibinfo {pages} {172}},\ }\Eprint
  {https://arxiv.org/abs/1505.01934} {arXiv:1505.01934 [hep-ph]} \BibitemShut
  {NoStop}%
\bibitem [{\citenamefont {Golling}\ \emph {et~al.}(2016)\citenamefont {Golling}
  \emph {et~al.}}]{Golling:2016gvc}%
  \BibitemOpen
  \bibfield  {author} {\bibinfo {author} {\bibfnamefont {T.}~\bibnamefont
  {Golling}} \emph {et~al.},\ }\bibfield  {journal} {\bibinfo  {journal}
  {arXiv}\ }\href {https://doi.org/10.23731/CYRM-2017-003.441}
  {10.23731/CYRM-2017-003.441} (\bibinfo {year} {2016}),\ \Eprint
  {https://arxiv.org/abs/1606.00947} {arXiv:1606.00947 [hep-ph]} \BibitemShut
  {NoStop}%
\bibitem [{\citenamefont
  {Urqu\'\i{}a-Calder\'on}(2024)}]{Urquia-Calderon:2023dkf}%
  \BibitemOpen
  \bibfield  {author} {\bibinfo {author} {\bibfnamefont {K.~A.}\ \bibnamefont
  {Urqu\'\i{}a-Calder\'on}},\ }\href
  {https://doi.org/10.1103/PhysRevD.109.055002} {\bibfield  {journal} {\bibinfo
   {journal} {Phys. Rev. D}\ }\textbf {\bibinfo {volume} {109}},\ \bibinfo
  {pages} {055002} (\bibinfo {year} {2024})},\ \Eprint
  {https://arxiv.org/abs/2310.17406} {arXiv:2310.17406 [hep-ph]} \BibitemShut
  {NoStop}%
\bibitem [{\citenamefont {Tumasyan}\ \emph
  {et~al.}(2022{\natexlab{b}})\citenamefont {Tumasyan} \emph
  {et~al.}}]{CMS:2021dzb}%
  \BibitemOpen
  \bibfield  {author} {\bibinfo {author} {\bibfnamefont {A.}~\bibnamefont
  {Tumasyan}} \emph {et~al.} (\bibinfo {collaboration} {CMS}),\ }\href
  {https://doi.org/10.1007/JHEP04(2022)047} {\bibfield  {journal} {\bibinfo
  {journal} {JHEP}\ }\textbf {\bibinfo {volume} {2022}}\bibfield  {number}
  {\bibinfo  {number} { (04)},\ \bibinfo {pages} {047}},\ }\Eprint
  {https://arxiv.org/abs/2112.03949} {arXiv:2112.03949 [hep-ex]} \BibitemShut
  {NoStop}%
\bibitem [{\citenamefont {Aad}\ \emph {et~al.}(2023{\natexlab{a}})\citenamefont
  {Aad} \emph {et~al.}}]{ATLAS:2023cjo}%
  \BibitemOpen
  \bibfield  {author} {\bibinfo {author} {\bibfnamefont {G.}~\bibnamefont
  {Aad}} \emph {et~al.} (\bibinfo {collaboration} {ATLAS}),\ }\href@noop {}
  {\bibfield  {journal} {\bibinfo  {journal} {arXiv}\ } (\bibinfo {year}
  {2023}{\natexlab{a}})},\ \Eprint {https://arxiv.org/abs/2304.09553}
  {arXiv:2304.09553 [hep-ex]} \BibitemShut {NoStop}%
\bibitem [{\citenamefont {Aad}\ \emph {et~al.}(2019)\citenamefont {Aad} \emph
  {et~al.}}]{ATLAS:2019lsy}%
  \BibitemOpen
  \bibfield  {author} {\bibinfo {author} {\bibfnamefont {G.}~\bibnamefont
  {Aad}} \emph {et~al.} (\bibinfo {collaboration} {ATLAS}),\ }\href
  {https://doi.org/10.1103/PhysRevD.100.052013} {\bibfield  {journal} {\bibinfo
   {journal} {Phys. Rev. D}\ }\textbf {\bibinfo {volume} {100}},\ \bibinfo
  {pages} {052013} (\bibinfo {year} {2019})},\ \Eprint
  {https://arxiv.org/abs/1906.05609} {arXiv:1906.05609 [hep-ex]} \BibitemShut
  {NoStop}%
\bibitem [{\citenamefont {Tumasyan}\ \emph
  {et~al.}(2022{\natexlab{c}})\citenamefont {Tumasyan} \emph
  {et~al.}}]{CMS:2022krd}%
  \BibitemOpen
  \bibfield  {author} {\bibinfo {author} {\bibfnamefont {A.}~\bibnamefont
  {Tumasyan}} \emph {et~al.} (\bibinfo {collaboration} {CMS}),\ }\href
  {https://doi.org/10.1007/JHEP07(2022)067} {\bibfield  {journal} {\bibinfo
  {journal} {JHEP}\ }\textbf {\bibinfo {volume} {2022}}\bibfield  {number}
  {\bibinfo  {number} { (07)},\ \bibinfo {pages} {067}},\ }\Eprint
  {https://arxiv.org/abs/2202.06075} {arXiv:2202.06075 [hep-ex]} \BibitemShut
  {NoStop}%
\bibitem [{\citenamefont {Aaboud}\ \emph {et~al.}(2018)\citenamefont {Aaboud}
  \emph {et~al.}}]{ATLAS:2018ihk}%
  \BibitemOpen
  \bibfield  {author} {\bibinfo {author} {\bibfnamefont {M.}~\bibnamefont
  {Aaboud}} \emph {et~al.} (\bibinfo {collaboration} {ATLAS}),\ }\href
  {https://doi.org/10.1103/PhysRevLett.120.161802} {\bibfield  {journal}
  {\bibinfo  {journal} {Phys. Rev. Lett.}\ }\textbf {\bibinfo {volume} {120}},\
  \bibinfo {pages} {161802} (\bibinfo {year} {2018})},\ \Eprint
  {https://arxiv.org/abs/1801.06992} {arXiv:1801.06992 [hep-ex]} \BibitemShut
  {NoStop}%
\bibitem [{\citenamefont {Collaboration}(2021)}]{ATLAS:2021bjk}%
  \BibitemOpen
  \bibfield  {author} {\bibinfo {author} {\bibfnamefont {A.}~\bibnamefont
  {Collaboration}} (\bibinfo {collaboration} {ATLAS}),\ }\href@noop {}
  {\bibfield  {journal} {\bibinfo  {journal} {ATLAS-CONF}\ } (\bibinfo {year}
  {2021})}\BibitemShut {NoStop}%
\bibitem [{\citenamefont {Tumasyan}\ \emph {et~al.}(2023)\citenamefont
  {Tumasyan} \emph {et~al.}}]{CMS:2022ncp}%
  \BibitemOpen
  \bibfield  {author} {\bibinfo {author} {\bibfnamefont {A.}~\bibnamefont
  {Tumasyan}} \emph {et~al.} (\bibinfo {collaboration} {CMS}),\ }\href
  {https://doi.org/10.1007/JHEP09(2023)051} {\bibfield  {journal} {\bibinfo
  {journal} {JHEP}\ }\textbf {\bibinfo {volume} {2023}}\bibfield  {number}
  {\bibinfo  {number} { (09)},\ \bibinfo {pages} {051}},\ }\Eprint
  {https://arxiv.org/abs/2212.12604} {arXiv:2212.12604 [hep-ex]} \BibitemShut
  {NoStop}%
\bibitem [{\citenamefont {Aad}\ \emph {et~al.}(2020)\citenamefont {Aad} \emph
  {et~al.}}]{ATLAS:2019fgd}%
  \BibitemOpen
  \bibfield  {author} {\bibinfo {author} {\bibfnamefont {G.}~\bibnamefont
  {Aad}} \emph {et~al.} (\bibinfo {collaboration} {ATLAS}),\ }\href
  {https://doi.org/10.1007/JHEP03(2020)145} {\bibfield  {journal} {\bibinfo
  {journal} {JHEP}\ }\textbf {\bibinfo {volume} {2020}}\bibfield  {number}
  {\bibinfo  {number} { (03)},\ \bibinfo {pages} {145}},\ }\Eprint
  {https://arxiv.org/abs/1910.08447} {arXiv:1910.08447 [hep-ex]} \BibitemShut
  {NoStop}%
\bibitem [{\citenamefont {Sirunyan}\ \emph {et~al.}(2020)\citenamefont
  {Sirunyan} \emph {et~al.}}]{CMS:2019gwf}%
  \BibitemOpen
  \bibfield  {author} {\bibinfo {author} {\bibfnamefont {A.~M.}\ \bibnamefont
  {Sirunyan}} \emph {et~al.} (\bibinfo {collaboration} {CMS}),\ }\href
  {https://doi.org/10.1007/JHEP05(2020)033} {\bibfield  {journal} {\bibinfo
  {journal} {JHEP}\ }\textbf {\bibinfo {volume} {2020}}\bibfield  {number}
  {\bibinfo  {number} { (05)},\ \bibinfo {pages} {033}},\ }\Eprint
  {https://arxiv.org/abs/1911.03947} {arXiv:1911.03947 [hep-ex]} \BibitemShut
  {NoStop}%
\bibitem [{\citenamefont {Hayrapetyan}\ \emph {et~al.}(2023)\citenamefont
  {Hayrapetyan} \emph {et~al.}}]{CMS:2023gte}%
  \BibitemOpen
  \bibfield  {author} {\bibinfo {author} {\bibfnamefont {A.}~\bibnamefont
  {Hayrapetyan}} \emph {et~al.} (\bibinfo {collaboration} {CMS}),\ }\href@noop
  {} {\bibfield  {journal} {\bibinfo  {journal} {arXiv}\ } (\bibinfo {year}
  {2023})},\ \Eprint {https://arxiv.org/abs/2310.19893} {arXiv:2310.19893
  [hep-ex]} \BibitemShut {NoStop}%
\bibitem [{\citenamefont {Aad}\ \emph {et~al.}(2023{\natexlab{b}})\citenamefont
  {Aad} \emph {et~al.}}]{ATLAS:2023ibb}%
  \BibitemOpen
  \bibfield  {author} {\bibinfo {author} {\bibfnamefont {G.}~\bibnamefont
  {Aad}} \emph {et~al.} (\bibinfo {collaboration} {ATLAS}),\ }\href
  {https://doi.org/10.1007/JHEP12(2023)073} {\bibfield  {journal} {\bibinfo
  {journal} {JHEP}\ }\textbf {\bibinfo {volume} {2023}}\bibfield  {number}
  {\bibinfo  {number} { (12)},\ \bibinfo {pages} {073}},\ }\Eprint
  {https://arxiv.org/abs/2308.08521} {arXiv:2308.08521 [hep-ex]} \BibitemShut
  {NoStop}%
\bibitem [{\citenamefont {Aad}\ \emph {et~al.}(2023{\natexlab{c}})\citenamefont
  {Aad} \emph {et~al.}}]{ATLAS:2022enb}%
  \BibitemOpen
  \bibfield  {author} {\bibinfo {author} {\bibfnamefont {G.}~\bibnamefont
  {Aad}} \emph {et~al.} (\bibinfo {collaboration} {ATLAS}),\ }\href
  {https://doi.org/10.1007/JHEP06(2023)016} {\bibfield  {journal} {\bibinfo
  {journal} {JHEP}\ }\textbf {\bibinfo {volume} {2023}}\bibfield  {number}
  {\bibinfo  {number} { (06)},\ \bibinfo {pages} {016}},\ }\Eprint
  {https://arxiv.org/abs/2207.00230} {arXiv:2207.00230 [hep-ex]} \BibitemShut
  {NoStop}%
\bibitem [{\citenamefont {Collaboration}(2022)}]{CMS:2022pjv}%
  \BibitemOpen
  \bibfield  {author} {\bibinfo {author} {\bibfnamefont {C.}~\bibnamefont
  {Collaboration}} (\bibinfo {collaboration} {CMS}),\ }\href@noop {} {\bibfield
   {journal} {\bibinfo  {journal} {arXiv}\ } (\bibinfo {year} {2022})},\
  \Eprint {https://arxiv.org/abs/2210.00043} {arXiv:2210.00043 [hep-ex]}
  \BibitemShut {NoStop}%
\bibitem [{\citenamefont {Nemev\v{s}ek}\ \emph {et~al.}(2013)\citenamefont
  {Nemev\v{s}ek}, \citenamefont {Senjanovi\'c},\ and\ \citenamefont
  {Tello}}]{Nemevsek:2012iq}%
  \BibitemOpen
  \bibfield  {author} {\bibinfo {author} {\bibfnamefont {M.}~\bibnamefont
  {Nemev\v{s}ek}}, \bibinfo {author} {\bibfnamefont {G.}~\bibnamefont
  {Senjanovi\'c}},\ and\ \bibinfo {author} {\bibfnamefont {V.}~\bibnamefont
  {Tello}},\ }\href {https://doi.org/10.1103/PhysRevLett.110.151802} {\bibfield
   {journal} {\bibinfo  {journal} {Phys. Rev. Lett.}\ }\textbf {\bibinfo
  {volume} {110}},\ \bibinfo {pages} {151802} (\bibinfo {year} {2013})},\
  \Eprint {https://arxiv.org/abs/1211.2837} {arXiv:1211.2837 [hep-ph]}
  \BibitemShut {NoStop}%
\bibitem [{\citenamefont {Chen}\ \emph {et~al.}(2013)\citenamefont {Chen},
  \citenamefont {Dev},\ and\ \citenamefont {Mohapatra}}]{Chen:2013foz}%
  \BibitemOpen
  \bibfield  {author} {\bibinfo {author} {\bibfnamefont {C.-Y.}\ \bibnamefont
  {Chen}}, \bibinfo {author} {\bibfnamefont {P.~S.~B.}\ \bibnamefont {Dev}},\
  and\ \bibinfo {author} {\bibfnamefont {R.~N.}\ \bibnamefont {Mohapatra}},\
  }\href {https://doi.org/10.1103/PhysRevD.88.033014} {\bibfield  {journal}
  {\bibinfo  {journal} {Phys. Rev. D}\ }\textbf {\bibinfo {volume} {88}},\
  \bibinfo {pages} {033014} (\bibinfo {year} {2013})},\ \Eprint
  {https://arxiv.org/abs/1306.2342} {arXiv:1306.2342 [hep-ph]} \BibitemShut
  {NoStop}%
\bibitem [{\citenamefont {Arbela\'ez}\ \emph {et~al.}(2018)\citenamefont
  {Arbela\'ez}, \citenamefont {Dib}, \citenamefont {Schmidt},\ and\
  \citenamefont {Vasquez}}]{Arbelaez:2017zqq}%
  \BibitemOpen
  \bibfield  {author} {\bibinfo {author} {\bibfnamefont {C.}~\bibnamefont
  {Arbela\'ez}}, \bibinfo {author} {\bibfnamefont {C.}~\bibnamefont {Dib}},
  \bibinfo {author} {\bibfnamefont {I.}~\bibnamefont {Schmidt}},\ and\ \bibinfo
  {author} {\bibfnamefont {J.~C.}\ \bibnamefont {Vasquez}},\ }\href
  {https://doi.org/10.1103/PhysRevD.97.055011} {\bibfield  {journal} {\bibinfo
  {journal} {Phys. Rev. D}\ }\textbf {\bibinfo {volume} {97}},\ \bibinfo
  {pages} {055011} (\bibinfo {year} {2018})},\ \Eprint
  {https://arxiv.org/abs/1712.08704} {arXiv:1712.08704 [hep-ph]} \BibitemShut
  {NoStop}%
\bibitem [{\citenamefont {Li}\ \emph {et~al.}(2022)\citenamefont {Li},
  \citenamefont {Ramsey-Musolf},\ and\ \citenamefont {Vasquez}}]{Li:2022cuq}%
  \BibitemOpen
  \bibfield  {author} {\bibinfo {author} {\bibfnamefont {G.}~\bibnamefont
  {Li}}, \bibinfo {author} {\bibfnamefont {M.~J.}\ \bibnamefont
  {Ramsey-Musolf}},\ and\ \bibinfo {author} {\bibfnamefont {J.~C.}\
  \bibnamefont {Vasquez}},\ }\href
  {https://doi.org/10.1103/PhysRevD.105.115021} {\bibfield  {journal} {\bibinfo
   {journal} {Phys. Rev. D}\ }\textbf {\bibinfo {volume} {105}},\ \bibinfo
  {pages} {115021} (\bibinfo {year} {2022})},\ \Eprint
  {https://arxiv.org/abs/2202.01789} {arXiv:2202.01789 [hep-ph]} \BibitemShut
  {NoStop}%
\bibitem [{\citenamefont {Tello}\ \emph {et~al.}(2011)\citenamefont {Tello},
  \citenamefont {Nemev\v{s}ek}, \citenamefont {Nesti}, \citenamefont
  {Senjanovi\'c},\ and\ \citenamefont {Vissani}}]{Tello:2010am}%
  \BibitemOpen
  \bibfield  {author} {\bibinfo {author} {\bibfnamefont {V.}~\bibnamefont
  {Tello}}, \bibinfo {author} {\bibfnamefont {M.}~\bibnamefont {Nemev\v{s}ek}},
  \bibinfo {author} {\bibfnamefont {F.}~\bibnamefont {Nesti}}, \bibinfo
  {author} {\bibfnamefont {G.}~\bibnamefont {Senjanovi\'c}},\ and\ \bibinfo
  {author} {\bibfnamefont {F.}~\bibnamefont {Vissani}},\ }\href
  {https://doi.org/10.1103/PhysRevLett.106.151801} {\bibfield  {journal}
  {\bibinfo  {journal} {Phys. Rev. Lett.}\ }\textbf {\bibinfo {volume} {106}},\
  \bibinfo {pages} {151801} (\bibinfo {year} {2011})},\ \Eprint
  {https://arxiv.org/abs/1011.3522} {arXiv:1011.3522 [hep-ph]} \BibitemShut
  {NoStop}%
\bibitem [{\citenamefont {Nemev\v{s}ek}\ \emph
  {et~al.}(2011{\natexlab{b}})\citenamefont {Nemev\v{s}ek}, \citenamefont
  {Nesti}, \citenamefont {Senjanovi\'c},\ and\ \citenamefont
  {Tello}}]{Nemevsek:2011aa}%
  \BibitemOpen
  \bibfield  {author} {\bibinfo {author} {\bibfnamefont {M.}~\bibnamefont
  {Nemev\v{s}ek}}, \bibinfo {author} {\bibfnamefont {F.}~\bibnamefont {Nesti}},
  \bibinfo {author} {\bibfnamefont {G.}~\bibnamefont {Senjanovi\'c}},\ and\
  \bibinfo {author} {\bibfnamefont {V.}~\bibnamefont {Tello}},\ }\href@noop {}
  {\bibfield  {journal} {\bibinfo  {journal} {arXiv}\ } (\bibinfo {year}
  {2011}{\natexlab{b}})},\ \Eprint {https://arxiv.org/abs/1112.3061}
  {arXiv:1112.3061 [hep-ph]} \BibitemShut {NoStop}%
\bibitem [{\citenamefont {Barry}\ and\ \citenamefont
  {Rodejohann}(2013)}]{Barry:2013xxa}%
  \BibitemOpen
  \bibfield  {author} {\bibinfo {author} {\bibfnamefont {J.}~\bibnamefont
  {Barry}}\ and\ \bibinfo {author} {\bibfnamefont {W.}~\bibnamefont
  {Rodejohann}},\ }\href {https://doi.org/10.1007/JHEP09(2013)153} {\bibfield
  {journal} {\bibinfo  {journal} {Journal of High Energy Physics}\ }\textbf
  {\bibinfo {volume} {2013}},\ \bibinfo {pages} {153} (\bibinfo {year}
  {2013})}\BibitemShut {NoStop}%
\bibitem [{\citenamefont {de~Vries}\ \emph {et~al.}(2022)\citenamefont
  {de~Vries}, \citenamefont {Li}, \citenamefont {Ramsey-Musolf},\ and\
  \citenamefont {Vasquez}}]{deVries:2022nyh}%
  \BibitemOpen
  \bibfield  {author} {\bibinfo {author} {\bibfnamefont {J.}~\bibnamefont
  {de~Vries}}, \bibinfo {author} {\bibfnamefont {G.}~\bibnamefont {Li}},
  \bibinfo {author} {\bibfnamefont {M.~J.}\ \bibnamefont {Ramsey-Musolf}},\
  and\ \bibinfo {author} {\bibfnamefont {J.~C.}\ \bibnamefont {Vasquez}},\
  }\href {https://doi.org/10.1007/JHEP11(2022)056} {\bibfield  {journal}
  {\bibinfo  {journal} {JHEP}\ }\textbf {\bibinfo {volume} {2022}}\bibfield
  {number} {\bibinfo  {number} { (11)},\ \bibinfo {pages} {056}},\ }\Eprint
  {https://arxiv.org/abs/2209.03031} {arXiv:2209.03031 [hep-ph]} \BibitemShut
  {NoStop}%
\bibitem [{\citenamefont {Nemev\v{s}ek}\ \emph {et~al.}(2012)\citenamefont
  {Nemev\v{s}ek}, \citenamefont {Senjanovi\'c},\ and\ \citenamefont
  {Zhang}}]{Nemevsek:2012cd}%
  \BibitemOpen
  \bibfield  {author} {\bibinfo {author} {\bibfnamefont {M.}~\bibnamefont
  {Nemev\v{s}ek}}, \bibinfo {author} {\bibfnamefont {G.}~\bibnamefont
  {Senjanovi\'c}},\ and\ \bibinfo {author} {\bibfnamefont {Y.}~\bibnamefont
  {Zhang}},\ }\href {https://doi.org/10.1088/1475-7516/2012/07/006} {\bibfield
  {journal} {\bibinfo  {journal} {JCAP}\ }\textbf {\bibinfo {volume}
  {2012}}\bibfield  {number} {\bibinfo  {number} { (07)},\ \bibinfo {pages}
  {006}},\ }\Eprint {https://arxiv.org/abs/1205.0844} {arXiv:1205.0844
  [hep-ph]} \BibitemShut {NoStop}%
\bibitem [{\citenamefont {Nemev\v{s}ek}\ and\ \citenamefont
  {Zhang}(2023{\natexlab{a}})}]{Nemevsek:2022anh}%
  \BibitemOpen
  \bibfield  {author} {\bibinfo {author} {\bibfnamefont {M.}~\bibnamefont
  {Nemev\v{s}ek}}\ and\ \bibinfo {author} {\bibfnamefont {Y.}~\bibnamefont
  {Zhang}},\ }\href {https://doi.org/10.1103/PhysRevLett.130.121002} {\bibfield
   {journal} {\bibinfo  {journal} {Phys. Rev. Lett.}\ }\textbf {\bibinfo
  {volume} {130}},\ \bibinfo {pages} {121002} (\bibinfo {year}
  {2023}{\natexlab{a}})},\ \Eprint {https://arxiv.org/abs/2206.11293}
  {arXiv:2206.11293 [hep-ph]} \BibitemShut {NoStop}%
\bibitem [{\citenamefont {Nemev\v{s}ek}\ and\ \citenamefont
  {Zhang}(2023{\natexlab{b}})}]{Nemevsek:2023yjl}%
  \BibitemOpen
  \bibfield  {author} {\bibinfo {author} {\bibfnamefont {M.}~\bibnamefont
  {Nemev\v{s}ek}}\ and\ \bibinfo {author} {\bibfnamefont {Y.}~\bibnamefont
  {Zhang}},\ }\href@noop {} {\bibfield  {journal} {\bibinfo  {journal} {arXiv}\
  } (\bibinfo {year} {2023}{\natexlab{b}})},\ \Eprint
  {https://arxiv.org/abs/2312.00129} {arXiv:2312.00129 [hep-ph]} \BibitemShut
  {NoStop}%
\bibitem [{\citenamefont {Senjanovi\'c}\ and\ \citenamefont
  {Tello}(2019)}]{Senjanovic:2018xtu}%
  \BibitemOpen
  \bibfield  {author} {\bibinfo {author} {\bibfnamefont {G.}~\bibnamefont
  {Senjanovi\'c}}\ and\ \bibinfo {author} {\bibfnamefont {V.}~\bibnamefont
  {Tello}},\ }\href {https://doi.org/10.1103/PhysRevD.100.115031} {\bibfield
  {journal} {\bibinfo  {journal} {Phys. Rev.}\ }\textbf {\bibinfo {volume}
  {D100}},\ \bibinfo {pages} {115031} (\bibinfo {year} {2019})},\ \Eprint
  {https://arxiv.org/abs/1812.03790} {arXiv:1812.03790 [hep-ph]} \BibitemShut
  {NoStop}%
\bibitem [{\citenamefont {Senjanovi\'c}\ and\ \citenamefont
  {Tello}(2020)}]{Senjanovic:2020int}%
  \BibitemOpen
  \bibfield  {author} {\bibinfo {author} {\bibfnamefont {G.}~\bibnamefont
  {Senjanovi\'c}}\ and\ \bibinfo {author} {\bibfnamefont {V.}~\bibnamefont
  {Tello}},\ }\href@noop {} {\bibfield  {journal} {\bibinfo  {journal} {arXiv}\
  } (\bibinfo {year} {2020})},\ \Eprint {https://arxiv.org/abs/2004.04036}
  {arXiv:2004.04036 [hep-ph]} \BibitemShut {NoStop}%
\bibitem [{\citenamefont {Kiers}\ \emph {et~al.}(2023)\citenamefont {Kiers},
  \citenamefont {Kiers}, \citenamefont {Szynkman},\ and\ \citenamefont
  {Tarutina}}]{Kiers:2022cyc}%
  \BibitemOpen
  \bibfield  {author} {\bibinfo {author} {\bibfnamefont {J.}~\bibnamefont
  {Kiers}}, \bibinfo {author} {\bibfnamefont {K.}~\bibnamefont {Kiers}},
  \bibinfo {author} {\bibfnamefont {A.}~\bibnamefont {Szynkman}},\ and\
  \bibinfo {author} {\bibfnamefont {T.}~\bibnamefont {Tarutina}},\ }\href
  {https://doi.org/10.1103/PhysRevD.107.075001} {\bibfield  {journal} {\bibinfo
   {journal} {Phys. Rev. D}\ }\textbf {\bibinfo {volume} {107}},\ \bibinfo
  {pages} {075001} (\bibinfo {year} {2023})},\ \Eprint
  {https://arxiv.org/abs/2212.14837} {arXiv:2212.14837 [hep-ph]} \BibitemShut
  {NoStop}%
\bibitem [{\citenamefont {Olness}\ and\ \citenamefont
  {Ebel}(1985)}]{Olness:1985bg}%
  \BibitemOpen
  \bibfield  {author} {\bibinfo {author} {\bibfnamefont {F.~I.}\ \bibnamefont
  {Olness}}\ and\ \bibinfo {author} {\bibfnamefont {M.~E.}\ \bibnamefont
  {Ebel}},\ }\href {https://doi.org/10.1103/PhysRevD.32.1769} {\bibfield
  {journal} {\bibinfo  {journal} {Phys. Rev. D}\ }\textbf {\bibinfo {volume}
  {32}},\ \bibinfo {pages} {1769} (\bibinfo {year} {1985})}\BibitemShut
  {NoStop}%
\bibitem [{\citenamefont {Gunion}\ \emph {et~al.}(1989)\citenamefont {Gunion},
  \citenamefont {Grifols}, \citenamefont {Mendez}, \citenamefont {Kayser},\
  and\ \citenamefont {Olness}}]{Gunion:1989in}%
  \BibitemOpen
  \bibfield  {author} {\bibinfo {author} {\bibfnamefont {J.~F.}\ \bibnamefont
  {Gunion}}, \bibinfo {author} {\bibfnamefont {J.}~\bibnamefont {Grifols}},
  \bibinfo {author} {\bibfnamefont {A.}~\bibnamefont {Mendez}}, \bibinfo
  {author} {\bibfnamefont {B.}~\bibnamefont {Kayser}},\ and\ \bibinfo {author}
  {\bibfnamefont {F.~I.}\ \bibnamefont {Olness}},\ }\href
  {https://doi.org/10.1103/PhysRevD.40.1546} {\bibfield  {journal} {\bibinfo
  {journal} {Phys. Rev. D}\ }\textbf {\bibinfo {volume} {40}},\ \bibinfo
  {pages} {1546} (\bibinfo {year} {1989})}\BibitemShut {NoStop}%
\bibitem [{\citenamefont {Deshpande}\ \emph {et~al.}(1991)\citenamefont
  {Deshpande}, \citenamefont {Gunion}, \citenamefont {Kayser},\ and\
  \citenamefont {Olness}}]{Deshpande:1990ip}%
  \BibitemOpen
  \bibfield  {author} {\bibinfo {author} {\bibfnamefont {N.~G.}\ \bibnamefont
  {Deshpande}}, \bibinfo {author} {\bibfnamefont {J.~F.}\ \bibnamefont
  {Gunion}}, \bibinfo {author} {\bibfnamefont {B.}~\bibnamefont {Kayser}},\
  and\ \bibinfo {author} {\bibfnamefont {F.~I.}\ \bibnamefont {Olness}},\
  }\href {https://doi.org/10.1103/PhysRevD.44.837} {\bibfield  {journal}
  {\bibinfo  {journal} {Phys. Rev. D}\ }\textbf {\bibinfo {volume} {44}},\
  \bibinfo {pages} {837} (\bibinfo {year} {1991})}\BibitemShut {NoStop}%
\bibitem [{\citenamefont {Kiers}\ \emph {et~al.}(2005)\citenamefont {Kiers},
  \citenamefont {Assis},\ and\ \citenamefont {Petrov}}]{Kiers:2005gh}%
  \BibitemOpen
  \bibfield  {author} {\bibinfo {author} {\bibfnamefont {K.}~\bibnamefont
  {Kiers}}, \bibinfo {author} {\bibfnamefont {M.}~\bibnamefont {Assis}},\ and\
  \bibinfo {author} {\bibfnamefont {A.~A.}\ \bibnamefont {Petrov}},\ }\href
  {https://doi.org/10.1103/PhysRevD.71.115015} {\bibfield  {journal} {\bibinfo
  {journal} {Phys. Rev. D}\ }\textbf {\bibinfo {volume} {71}},\ \bibinfo
  {pages} {115015} (\bibinfo {year} {2005})},\ \Eprint
  {https://arxiv.org/abs/hep-ph/0503115} {arXiv:hep-ph/0503115} \BibitemShut
  {NoStop}%
\bibitem [{\citenamefont {Khasanov}\ and\ \citenamefont
  {Perez}(2002)}]{Khasanov:2001tu}%
  \BibitemOpen
  \bibfield  {author} {\bibinfo {author} {\bibfnamefont {O.}~\bibnamefont
  {Khasanov}}\ and\ \bibinfo {author} {\bibfnamefont {G.}~\bibnamefont
  {Perez}},\ }\href {https://doi.org/10.1103/PhysRevD.65.053007} {\bibfield
  {journal} {\bibinfo  {journal} {Phys. Rev. D}\ }\textbf {\bibinfo {volume}
  {65}},\ \bibinfo {pages} {053007} (\bibinfo {year} {2002})},\ \Eprint
  {https://arxiv.org/abs/hep-ph/0108176} {arXiv:hep-ph/0108176} \BibitemShut
  {NoStop}%
\bibitem [{\citenamefont {Dekens}\ and\ \citenamefont
  {Boer}(2014)}]{Dekens:2014ina}%
  \BibitemOpen
  \bibfield  {author} {\bibinfo {author} {\bibfnamefont {W.}~\bibnamefont
  {Dekens}}\ and\ \bibinfo {author} {\bibfnamefont {D.}~\bibnamefont {Boer}},\
  }\href {https://doi.org/10.1016/j.nuclphysb.2014.10.025} {\bibfield
  {journal} {\bibinfo  {journal} {Nucl. Phys. B}\ }\textbf {\bibinfo {volume}
  {889}},\ \bibinfo {pages} {727} (\bibinfo {year} {2014})},\ \Eprint
  {https://arxiv.org/abs/1409.4052} {arXiv:1409.4052 [hep-ph]} \BibitemShut
  {NoStop}%
\bibitem [{\citenamefont {Bambhaniya}\ \emph {et~al.}(2015)\citenamefont
  {Bambhaniya}, \citenamefont {Chakrabortty}, \citenamefont {Gluza},
  \citenamefont {Jelinski},\ and\ \citenamefont
  {Szafron}}]{Bambhaniya:2015wna}%
  \BibitemOpen
  \bibfield  {author} {\bibinfo {author} {\bibfnamefont {G.}~\bibnamefont
  {Bambhaniya}}, \bibinfo {author} {\bibfnamefont {J.}~\bibnamefont
  {Chakrabortty}}, \bibinfo {author} {\bibfnamefont {J.}~\bibnamefont {Gluza}},
  \bibinfo {author} {\bibfnamefont {T.}~\bibnamefont {Jelinski}},\ and\
  \bibinfo {author} {\bibfnamefont {R.}~\bibnamefont {Szafron}},\ }\href
  {https://doi.org/10.1103/PhysRevD.92.015016} {\bibfield  {journal} {\bibinfo
  {journal} {Phys. Rev. D}\ }\textbf {\bibinfo {volume} {92}},\ \bibinfo
  {pages} {015016} (\bibinfo {year} {2015})},\ \Eprint
  {https://arxiv.org/abs/1504.03999} {arXiv:1504.03999 [hep-ph]} \BibitemShut
  {NoStop}%
\bibitem [{\citenamefont {Maiezza}\ \emph {et~al.}(2017)\citenamefont
  {Maiezza}, \citenamefont {Senjanovi\'c},\ and\ \citenamefont
  {Vasquez}}]{Maiezza:2016ybz}%
  \BibitemOpen
  \bibfield  {author} {\bibinfo {author} {\bibfnamefont {A.}~\bibnamefont
  {Maiezza}}, \bibinfo {author} {\bibfnamefont {G.}~\bibnamefont
  {Senjanovi\'c}},\ and\ \bibinfo {author} {\bibfnamefont {J.~C.}\ \bibnamefont
  {Vasquez}},\ }\href {https://doi.org/10.1103/PhysRevD.95.095004} {\bibfield
  {journal} {\bibinfo  {journal} {Phys. Rev.}\ }\textbf {\bibinfo {volume}
  {D95}},\ \bibinfo {pages} {095004} (\bibinfo {year} {2017})},\ \Eprint
  {https://arxiv.org/abs/1612.09146} {arXiv:1612.09146 [hep-ph]} \BibitemShut
  {NoStop}%
\bibitem [{\citenamefont {Dev}\ \emph {et~al.}(2016{\natexlab{b}})\citenamefont
  {Dev}, \citenamefont {Mohapatra},\ and\ \citenamefont {Zhang}}]{Dev:2016dja}%
  \BibitemOpen
  \bibfield  {author} {\bibinfo {author} {\bibfnamefont {P.~S.~B.}\
  \bibnamefont {Dev}}, \bibinfo {author} {\bibfnamefont {R.~N.}\ \bibnamefont
  {Mohapatra}},\ and\ \bibinfo {author} {\bibfnamefont {Y.}~\bibnamefont
  {Zhang}},\ }\href {https://doi.org/10.1007/JHEP05(2016)174} {\bibfield
  {journal} {\bibinfo  {journal} {JHEP}\ }\textbf {\bibinfo {volume}
  {2016}}\bibfield  {number} {\bibinfo  {number} { (05)},\ \bibinfo {pages}
  {174}},\ }\Eprint {https://arxiv.org/abs/1602.05947} {arXiv:1602.05947
  [hep-ph]} \BibitemShut {NoStop}%
\bibitem [{\citenamefont {Maiezza}\ \emph {et~al.}(2016)\citenamefont
  {Maiezza}, \citenamefont {Nemev\v{s}ek},\ and\ \citenamefont
  {Nesti}}]{Maiezza:2016bzp}%
  \BibitemOpen
  \bibfield  {author} {\bibinfo {author} {\bibfnamefont {A.}~\bibnamefont
  {Maiezza}}, \bibinfo {author} {\bibfnamefont {M.}~\bibnamefont
  {Nemev\v{s}ek}},\ and\ \bibinfo {author} {\bibfnamefont {F.}~\bibnamefont
  {Nesti}},\ }\href {https://doi.org/10.1103/PhysRevD.94.035008} {\bibfield
  {journal} {\bibinfo  {journal} {Phys. Rev.}\ }\textbf {\bibinfo {volume}
  {D94}},\ \bibinfo {pages} {035008} (\bibinfo {year} {2016})},\ \Eprint
  {https://arxiv.org/abs/1603.00360} {arXiv:1603.00360 [hep-ph]} \BibitemShut
  {NoStop}%
\bibitem [{\citenamefont {Chauhan}\ \emph {et~al.}(2019)\citenamefont
  {Chauhan}, \citenamefont {Dev}, \citenamefont {Mohapatra},\ and\
  \citenamefont {Zhang}}]{Chauhan:2018uuy}%
  \BibitemOpen
  \bibfield  {author} {\bibinfo {author} {\bibfnamefont {G.}~\bibnamefont
  {Chauhan}}, \bibinfo {author} {\bibfnamefont {P.~S.~B.}\ \bibnamefont {Dev}},
  \bibinfo {author} {\bibfnamefont {R.~N.}\ \bibnamefont {Mohapatra}},\ and\
  \bibinfo {author} {\bibfnamefont {Y.}~\bibnamefont {Zhang}},\ }\href
  {https://doi.org/10.1007/JHEP01(2019)208} {\bibfield  {journal} {\bibinfo
  {journal} {JHEP}\ }\textbf {\bibinfo {volume} {2019}}\bibfield  {number}
  {\bibinfo  {number} { (01)},\ \bibinfo {pages} {208}},\ }\Eprint
  {https://arxiv.org/abs/1811.08789} {arXiv:1811.08789 [hep-ph]} \BibitemShut
  {NoStop}%
\bibitem [{\citenamefont {Mohapatra}\ \emph {et~al.}(2019)\citenamefont
  {Mohapatra}, \citenamefont {Yan},\ and\ \citenamefont
  {Zhang}}]{Mohapatra:2019qid}%
  \BibitemOpen
  \bibfield  {author} {\bibinfo {author} {\bibfnamefont {R.~N.}\ \bibnamefont
  {Mohapatra}}, \bibinfo {author} {\bibfnamefont {G.}~\bibnamefont {Yan}},\
  and\ \bibinfo {author} {\bibfnamefont {Y.}~\bibnamefont {Zhang}},\ }\href
  {https://doi.org/10.1016/j.nuclphysb.2019.114764} {\bibfield  {journal}
  {\bibinfo  {journal} {Nucl. Phys. B}\ }\textbf {\bibinfo {volume} {948}},\
  \bibinfo {pages} {114764} (\bibinfo {year} {2019})},\ \Eprint
  {https://arxiv.org/abs/1902.08601} {arXiv:1902.08601 [hep-ph]} \BibitemShut
  {NoStop}%
\bibitem [{\citenamefont {Nemev\v{s}ek}\ \emph {et~al.}(2017)\citenamefont
  {Nemev\v{s}ek}, \citenamefont {Nesti},\ and\ \citenamefont
  {Vasquez}}]{Nemevsek:2016enw}%
  \BibitemOpen
  \bibfield  {author} {\bibinfo {author} {\bibfnamefont {M.}~\bibnamefont
  {Nemev\v{s}ek}}, \bibinfo {author} {\bibfnamefont {F.}~\bibnamefont
  {Nesti}},\ and\ \bibinfo {author} {\bibfnamefont {J.~C.}\ \bibnamefont
  {Vasquez}},\ }\href {https://doi.org/10.1007/JHEP04(2017)114} {\bibfield
  {journal} {\bibinfo  {journal} {JHEP}\ }\textbf {\bibinfo {volume}
  {2017}}\bibfield  {number} {\bibinfo  {number} { (04)},\ \bibinfo {pages}
  {114}},\ }\Eprint {https://arxiv.org/abs/1612.06840} {arXiv:1612.06840
  [hep-ph]} \BibitemShut {NoStop}%
\bibitem [{\citenamefont {Bhupal~Dev}\ \emph {et~al.}(2019)\citenamefont
  {Bhupal~Dev}, \citenamefont {Mohapatra}, \citenamefont {Rodejohann},\ and\
  \citenamefont {Xu}}]{BhupalDev:2018xya}%
  \BibitemOpen
  \bibfield  {author} {\bibinfo {author} {\bibfnamefont {P.~S.}\ \bibnamefont
  {Bhupal~Dev}}, \bibinfo {author} {\bibfnamefont {R.~N.}\ \bibnamefont
  {Mohapatra}}, \bibinfo {author} {\bibfnamefont {W.}~\bibnamefont
  {Rodejohann}},\ and\ \bibinfo {author} {\bibfnamefont {X.-J.}\ \bibnamefont
  {Xu}},\ }\href {https://doi.org/10.1007/JHEP02(2019)154} {\bibfield
  {journal} {\bibinfo  {journal} {JHEP}\ }\textbf {\bibinfo {volume}
  {2019}}\bibfield  {number} {\bibinfo  {number} { (02)},\ \bibinfo {pages}
  {154}},\ }\Eprint {https://arxiv.org/abs/1811.06869} {arXiv:1811.06869
  [hep-ph]} \BibitemShut {NoStop}%
\bibitem [{\citenamefont {Chauhan}(2019)}]{Chauhan:2019fji}%
  \BibitemOpen
  \bibfield  {author} {\bibinfo {author} {\bibfnamefont {G.}~\bibnamefont
  {Chauhan}},\ }\href {https://doi.org/10.1007/JHEP12(2019)137} {\bibfield
  {journal} {\bibinfo  {journal} {JHEP}\ }\textbf {\bibinfo {volume}
  {2019}}\bibfield  {number} {\bibinfo  {number} { (12)},\ \bibinfo {pages}
  {137}},\ }\Eprint {https://arxiv.org/abs/1907.07153} {arXiv:1907.07153
  [hep-ph]} \BibitemShut {NoStop}%
\bibitem [{\citenamefont {Brdar}\ \emph {et~al.}(2019)\citenamefont {Brdar},
  \citenamefont {Graf}, \citenamefont {Helmboldt},\ and\ \citenamefont
  {Xu}}]{Brdar:2019fur}%
  \BibitemOpen
  \bibfield  {author} {\bibinfo {author} {\bibfnamefont {V.}~\bibnamefont
  {Brdar}}, \bibinfo {author} {\bibfnamefont {L.}~\bibnamefont {Graf}},
  \bibinfo {author} {\bibfnamefont {A.~J.}\ \bibnamefont {Helmboldt}},\ and\
  \bibinfo {author} {\bibfnamefont {X.-J.}\ \bibnamefont {Xu}},\ }\href
  {https://doi.org/10.1088/1475-7516/2019/12/027} {\bibfield  {journal}
  {\bibinfo  {journal} {JCAP}\ }\textbf {\bibinfo {volume} {2019}}\bibfield
  {number} {\bibinfo  {number} { (12)},\ \bibinfo {pages} {027}},\ }\Eprint
  {https://arxiv.org/abs/1909.02018} {arXiv:1909.02018 [hep-ph]} \BibitemShut
  {NoStop}%
\bibitem [{\citenamefont {Li}\ \emph {et~al.}(2021)\citenamefont {Li},
  \citenamefont {Yan}, \citenamefont {Zhang},\ and\ \citenamefont
  {Zhao}}]{Li:2020eun}%
  \BibitemOpen
  \bibfield  {author} {\bibinfo {author} {\bibfnamefont {M.}~\bibnamefont
  {Li}}, \bibinfo {author} {\bibfnamefont {Q.-S.}\ \bibnamefont {Yan}},
  \bibinfo {author} {\bibfnamefont {Y.}~\bibnamefont {Zhang}},\ and\ \bibinfo
  {author} {\bibfnamefont {Z.}~\bibnamefont {Zhao}},\ }\href
  {https://doi.org/10.1007/JHEP03(2021)267} {\bibfield  {journal} {\bibinfo
  {journal} {JHEP}\ }\textbf {\bibinfo {volume} {2021}}\bibfield  {number}
  {\bibinfo  {number} { (03)},\ \bibinfo {pages} {267}},\ }\Eprint
  {https://arxiv.org/abs/2012.13686} {arXiv:2012.13686 [hep-ph]} \BibitemShut
  {NoStop}%
\bibitem [{\citenamefont {Maiezza}\ \emph {et~al.}(2015)\citenamefont
  {Maiezza}, \citenamefont {Nemev\v{s}ek},\ and\ \citenamefont
  {Nesti}}]{Maiezza:2015lza}%
  \BibitemOpen
  \bibfield  {author} {\bibinfo {author} {\bibfnamefont {A.}~\bibnamefont
  {Maiezza}}, \bibinfo {author} {\bibfnamefont {M.}~\bibnamefont
  {Nemev\v{s}ek}},\ and\ \bibinfo {author} {\bibfnamefont {F.}~\bibnamefont
  {Nesti}},\ }\href {https://doi.org/10.1103/PhysRevLett.115.081802} {\bibfield
   {journal} {\bibinfo  {journal} {Phys. Rev. Lett.}\ }\textbf {\bibinfo
  {volume} {115}},\ \bibinfo {pages} {081802} (\bibinfo {year} {2015})},\
  \Eprint {https://arxiv.org/abs/1503.06834} {arXiv:1503.06834 [hep-ph]}
  \BibitemShut {NoStop}%
\bibitem [{\citenamefont {Melfo}\ \emph {et~al.}(2012)\citenamefont {Melfo},
  \citenamefont {Nemevsek}, \citenamefont {Nesti}, \citenamefont {Senjanovic},\
  and\ \citenamefont {Zhang}}]{Melfo:2011nx}%
  \BibitemOpen
  \bibfield  {author} {\bibinfo {author} {\bibfnamefont {A.}~\bibnamefont
  {Melfo}}, \bibinfo {author} {\bibfnamefont {M.}~\bibnamefont {Nemevsek}},
  \bibinfo {author} {\bibfnamefont {F.}~\bibnamefont {Nesti}}, \bibinfo
  {author} {\bibfnamefont {G.}~\bibnamefont {Senjanovic}},\ and\ \bibinfo
  {author} {\bibfnamefont {Y.}~\bibnamefont {Zhang}},\ }\href
  {https://doi.org/10.1103/PhysRevD.85.055018} {\bibfield  {journal} {\bibinfo
  {journal} {Phys. Rev. D}\ }\textbf {\bibinfo {volume} {85}},\ \bibinfo
  {pages} {055018} (\bibinfo {year} {2012})},\ \Eprint
  {https://arxiv.org/abs/1108.4416} {arXiv:1108.4416 [hep-ph]} \BibitemShut
  {NoStop}%
\bibitem [{\citenamefont {Maalampi}\ and\ \citenamefont
  {Romanenko}(2002)}]{Maalampi:2002vx}%
  \BibitemOpen
  \bibfield  {author} {\bibinfo {author} {\bibfnamefont {J.}~\bibnamefont
  {Maalampi}}\ and\ \bibinfo {author} {\bibfnamefont {N.}~\bibnamefont
  {Romanenko}},\ }\href {https://doi.org/10.1016/S0370-2693(02)01549-6}
  {\bibfield  {journal} {\bibinfo  {journal} {Phys. Lett. B}\ }\textbf
  {\bibinfo {volume} {532}},\ \bibinfo {pages} {202} (\bibinfo {year}
  {2002})},\ \Eprint {https://arxiv.org/abs/hep-ph/0201196}
  {arXiv:hep-ph/0201196} \BibitemShut {NoStop}%
\bibitem [{\citenamefont {Roitgrund}\ and\ \citenamefont
  {Eilam}(2021)}]{Roitgrund:2020cge}%
  \BibitemOpen
  \bibfield  {author} {\bibinfo {author} {\bibfnamefont {A.}~\bibnamefont
  {Roitgrund}}\ and\ \bibinfo {author} {\bibfnamefont {G.}~\bibnamefont
  {Eilam}},\ }\href {https://doi.org/10.1007/JHEP01(2021)031} {\bibfield
  {journal} {\bibinfo  {journal} {JHEP}\ }\textbf {\bibinfo {volume}
  {2021}}\bibfield  {number} {\bibinfo  {number} { (01)},\ \bibinfo {pages}
  {031}},\ }\bibinfo {note} {[Erratum: JHEP 03, 029 (2021)]},\ \Eprint
  {https://arxiv.org/abs/1704.07772} {arXiv:1704.07772 [hep-ph]} \BibitemShut
  {NoStop}%
\bibitem [{\citenamefont {Alloul}\ \emph {et~al.}(2013)\citenamefont {Alloul},
  \citenamefont {Frank}, \citenamefont {Fuks},\ and\ \citenamefont {Rausch~de
  Traubenberg}}]{Alloul:2013raa}%
  \BibitemOpen
  \bibfield  {author} {\bibinfo {author} {\bibfnamefont {A.}~\bibnamefont
  {Alloul}}, \bibinfo {author} {\bibfnamefont {M.}~\bibnamefont {Frank}},
  \bibinfo {author} {\bibfnamefont {B.}~\bibnamefont {Fuks}},\ and\ \bibinfo
  {author} {\bibfnamefont {M.}~\bibnamefont {Rausch~de Traubenberg}},\ }\href
  {https://doi.org/10.1103/PhysRevD.88.075004} {\bibfield  {journal} {\bibinfo
  {journal} {Phys. Rev. D}\ }\textbf {\bibinfo {volume} {88}},\ \bibinfo
  {pages} {075004} (\bibinfo {year} {2013})},\ \Eprint
  {https://arxiv.org/abs/1307.1711} {arXiv:1307.1711 [hep-ph]} \BibitemShut
  {NoStop}%
\bibitem [{\citenamefont {Akhmedov}\ \emph {et~al.}(2024)\citenamefont
  {Akhmedov}, \citenamefont {Dev}, \citenamefont {Jana},\ and\ \citenamefont
  {Mohapatra}}]{Akhmedov:2024rvp}%
  \BibitemOpen
  \bibfield  {author} {\bibinfo {author} {\bibfnamefont {E.}~\bibnamefont
  {Akhmedov}}, \bibinfo {author} {\bibfnamefont {P.~S.~B.}\ \bibnamefont
  {Dev}}, \bibinfo {author} {\bibfnamefont {S.}~\bibnamefont {Jana}},\ and\
  \bibinfo {author} {\bibfnamefont {R.~N.}\ \bibnamefont {Mohapatra}},\
  }\href@noop {} {\bibfield  {journal} {\bibinfo  {journal} {arXiv}\ }
  (\bibinfo {year} {2024})},\ \Eprint {https://arxiv.org/abs/2401.15145}
  {arXiv:2401.15145 [hep-ph]} \BibitemShut {NoStop}%
\bibitem [{\citenamefont {Alekhin}\ \emph {et~al.}(2016)\citenamefont {Alekhin}
  \emph {et~al.}}]{Alekhin:2015byh}%
  \BibitemOpen
  \bibfield  {author} {\bibinfo {author} {\bibfnamefont {S.}~\bibnamefont
  {Alekhin}} \emph {et~al.},\ }\href
  {https://doi.org/10.1088/0034-4885/79/12/124201} {\bibfield  {journal}
  {\bibinfo  {journal} {Rept. Prog. Phys.}\ }\textbf {\bibinfo {volume} {79}},\
  \bibinfo {pages} {124201} (\bibinfo {year} {2016})},\ \Eprint
  {https://arxiv.org/abs/1504.04855} {arXiv:1504.04855 [hep-ph]} \BibitemShut
  {NoStop}%
\bibitem [{\citenamefont {Kling}\ and\ \citenamefont
  {Trojanowski}(2018)}]{Kling:2018wct}%
  \BibitemOpen
  \bibfield  {author} {\bibinfo {author} {\bibfnamefont {F.}~\bibnamefont
  {Kling}}\ and\ \bibinfo {author} {\bibfnamefont {S.}~\bibnamefont
  {Trojanowski}},\ }\href {https://doi.org/10.1103/PhysRevD.97.095016}
  {\bibfield  {journal} {\bibinfo  {journal} {Phys. Rev. D}\ }\textbf {\bibinfo
  {volume} {97}},\ \bibinfo {pages} {095016} (\bibinfo {year} {2018})},\
  \Eprint {https://arxiv.org/abs/1801.08947} {arXiv:1801.08947 [hep-ph]}
  \BibitemShut {NoStop}%
\bibitem [{\citenamefont {Christensen}\ and\ \citenamefont
  {Duhr}(2009)}]{Christensen:2008py}%
  \BibitemOpen
  \bibfield  {author} {\bibinfo {author} {\bibfnamefont {N.~D.}\ \bibnamefont
  {Christensen}}\ and\ \bibinfo {author} {\bibfnamefont {C.}~\bibnamefont
  {Duhr}},\ }\href {https://doi.org/10.1016/j.cpc.2009.02.018} {\bibfield
  {journal} {\bibinfo  {journal} {Comput. Phys. Commun.}\ }\textbf {\bibinfo
  {volume} {180}},\ \bibinfo {pages} {1614} (\bibinfo {year} {2009})},\ \Eprint
  {https://arxiv.org/abs/0806.4194} {arXiv:0806.4194 [hep-ph]} \BibitemShut
  {NoStop}%
\bibitem [{\citenamefont {Alloul}\ \emph {et~al.}(2014)\citenamefont {Alloul},
  \citenamefont {Christensen}, \citenamefont {Degrande}, \citenamefont {Duhr},\
  and\ \citenamefont {Fuks}}]{Alloul:2013bka}%
  \BibitemOpen
  \bibfield  {author} {\bibinfo {author} {\bibfnamefont {A.}~\bibnamefont
  {Alloul}}, \bibinfo {author} {\bibfnamefont {N.~D.}\ \bibnamefont
  {Christensen}}, \bibinfo {author} {\bibfnamefont {C.}~\bibnamefont
  {Degrande}}, \bibinfo {author} {\bibfnamefont {C.}~\bibnamefont {Duhr}},\
  and\ \bibinfo {author} {\bibfnamefont {B.}~\bibnamefont {Fuks}},\ }\href
  {https://doi.org/10.1016/j.cpc.2014.04.012} {\bibfield  {journal} {\bibinfo
  {journal} {Comput. Phys. Commun.}\ }\textbf {\bibinfo {volume} {185}},\
  \bibinfo {pages} {2250} (\bibinfo {year} {2014})},\ \Eprint
  {https://arxiv.org/abs/1310.1921} {arXiv:1310.1921 [hep-ph]} \BibitemShut
  {NoStop}%
\bibitem [{\citenamefont {Degrande}\ \emph {et~al.}(2012)\citenamefont
  {Degrande}, \citenamefont {Duhr}, \citenamefont {Fuks}, \citenamefont
  {Grellscheid}, \citenamefont {Mattelaer},\ and\ \citenamefont
  {Reiter}}]{Degrande:2011ua}%
  \BibitemOpen
  \bibfield  {author} {\bibinfo {author} {\bibfnamefont {C.}~\bibnamefont
  {Degrande}}, \bibinfo {author} {\bibfnamefont {C.}~\bibnamefont {Duhr}},
  \bibinfo {author} {\bibfnamefont {B.}~\bibnamefont {Fuks}}, \bibinfo {author}
  {\bibfnamefont {D.}~\bibnamefont {Grellscheid}}, \bibinfo {author}
  {\bibfnamefont {O.}~\bibnamefont {Mattelaer}},\ and\ \bibinfo {author}
  {\bibfnamefont {T.}~\bibnamefont {Reiter}},\ }\href
  {https://doi.org/10.1016/j.cpc.2012.01.022} {\bibfield  {journal} {\bibinfo
  {journal} {Comput. Phys. Commun.}\ }\textbf {\bibinfo {volume} {183}},\
  \bibinfo {pages} {1201} (\bibinfo {year} {2012})},\ \Eprint
  {https://arxiv.org/abs/1108.2040} {arXiv:1108.2040 [hep-ph]} \BibitemShut
  {NoStop}%
\bibitem [{\citenamefont {Darm\'e}\ \emph {et~al.}(2023)\citenamefont {Darm\'e}
  \emph {et~al.}}]{Darme:2023jdn}%
  \BibitemOpen
  \bibfield  {author} {\bibinfo {author} {\bibfnamefont {L.}~\bibnamefont
  {Darm\'e}} \emph {et~al.},\ }\href
  {https://doi.org/10.1140/epjc/s10052-023-11780-9} {\bibfield  {journal}
  {\bibinfo  {journal} {Eur. Phys. J. C}\ }\textbf {\bibinfo {volume} {83}},\
  \bibinfo {pages} {631} (\bibinfo {year} {2023})},\ \Eprint
  {https://arxiv.org/abs/2304.09883} {arXiv:2304.09883 [hep-ph]} \BibitemShut
  {NoStop}%
\bibitem [{\citenamefont {Duka}\ \emph {et~al.}(2000)\citenamefont {Duka},
  \citenamefont {Gluza},\ and\ \citenamefont {Zralek}}]{Duka:1999uc}%
  \BibitemOpen
  \bibfield  {author} {\bibinfo {author} {\bibfnamefont {P.}~\bibnamefont
  {Duka}}, \bibinfo {author} {\bibfnamefont {J.}~\bibnamefont {Gluza}},\ and\
  \bibinfo {author} {\bibfnamefont {M.}~\bibnamefont {Zralek}},\ }\href
  {https://doi.org/10.1006/aphy.1999.5988} {\bibfield  {journal} {\bibinfo
  {journal} {Annals Phys.}\ }\textbf {\bibinfo {volume} {280}},\ \bibinfo
  {pages} {336} (\bibinfo {year} {2000})},\ \Eprint
  {https://arxiv.org/abs/hep-ph/9910279} {arXiv:hep-ph/9910279} \BibitemShut
  {NoStop}%
\bibitem [{\citenamefont {Roitgrund}\ \emph {et~al.}(2016)\citenamefont
  {Roitgrund}, \citenamefont {Eilam},\ and\ \citenamefont
  {Bar-Shalom}}]{Roitgrund:2014zka}%
  \BibitemOpen
  \bibfield  {author} {\bibinfo {author} {\bibfnamefont {A.}~\bibnamefont
  {Roitgrund}}, \bibinfo {author} {\bibfnamefont {G.}~\bibnamefont {Eilam}},\
  and\ \bibinfo {author} {\bibfnamefont {S.}~\bibnamefont {Bar-Shalom}},\
  }\href {https://doi.org/10.1016/j.cpc.2015.12.009} {\bibfield  {journal}
  {\bibinfo  {journal} {Comput. Phys. Commun.}\ }\textbf {\bibinfo {volume}
  {203}},\ \bibinfo {pages} {18} (\bibinfo {year} {2016})},\ \Eprint
  {https://arxiv.org/abs/1401.3345} {arXiv:1401.3345 [hep-ph]} \BibitemShut
  {NoStop}%
\bibitem [{\citenamefont {Degrande}\ \emph {et~al.}(2016)\citenamefont
  {Degrande}, \citenamefont {Mattelaer}, \citenamefont {Ruiz},\ and\
  \citenamefont {Turner}}]{Degrande:2016aje}%
  \BibitemOpen
  \bibfield  {author} {\bibinfo {author} {\bibfnamefont {C.}~\bibnamefont
  {Degrande}}, \bibinfo {author} {\bibfnamefont {O.}~\bibnamefont {Mattelaer}},
  \bibinfo {author} {\bibfnamefont {R.}~\bibnamefont {Ruiz}},\ and\ \bibinfo
  {author} {\bibfnamefont {J.}~\bibnamefont {Turner}},\ }\href
  {https://doi.org/10.1103/PhysRevD.94.053002} {\bibfield  {journal} {\bibinfo
  {journal} {Phys. Rev. D}\ }\textbf {\bibinfo {volume} {94}},\ \bibinfo
  {pages} {053002} (\bibinfo {year} {2016})},\ \Eprint
  {https://arxiv.org/abs/1602.06957} {arXiv:1602.06957 [hep-ph]} \BibitemShut
  {NoStop}%
\bibitem [{\citenamefont {Fuks}\ \emph {et~al.}(2020)\citenamefont {Fuks},
  \citenamefont {Nemev\v{s}ek},\ and\ \citenamefont {Ruiz}}]{Fuks:2019clu}%
  \BibitemOpen
  \bibfield  {author} {\bibinfo {author} {\bibfnamefont {B.}~\bibnamefont
  {Fuks}}, \bibinfo {author} {\bibfnamefont {M.}~\bibnamefont {Nemev\v{s}ek}},\
  and\ \bibinfo {author} {\bibfnamefont {R.}~\bibnamefont {Ruiz}},\ }\href
  {https://doi.org/10.1103/PhysRevD.101.075022} {\bibfield  {journal} {\bibinfo
   {journal} {Phys. Rev. D}\ }\textbf {\bibinfo {volume} {101}},\ \bibinfo
  {pages} {075022} (\bibinfo {year} {2020})},\ \Eprint
  {https://arxiv.org/abs/1912.08975} {arXiv:1912.08975 [hep-ph]} \BibitemShut
  {NoStop}%
\bibitem [{\citenamefont {A~H}\ \emph {et~al.}(2023)\citenamefont {A~H},
  \citenamefont {Fuks}, \citenamefont {Shao},\ and\ \citenamefont
  {Simon}}]{AH:2023hft}%
  \BibitemOpen
  \bibfield  {author} {\bibinfo {author} {\bibfnamefont {A.}~\bibnamefont
  {A~H}}, \bibinfo {author} {\bibfnamefont {B.}~\bibnamefont {Fuks}}, \bibinfo
  {author} {\bibfnamefont {H.-S.}\ \bibnamefont {Shao}},\ and\ \bibinfo
  {author} {\bibfnamefont {Y.}~\bibnamefont {Simon}},\ }\href
  {https://doi.org/10.1103/PhysRevD.107.075011} {\bibfield  {journal} {\bibinfo
   {journal} {Phys. Rev. D}\ }\textbf {\bibinfo {volume} {107}},\ \bibinfo
  {pages} {075011} (\bibinfo {year} {2023})},\ \Eprint
  {https://arxiv.org/abs/2301.03640} {arXiv:2301.03640 [hep-ph]} \BibitemShut
  {NoStop}%
\bibitem [{\citenamefont {Mattelaer}\ \emph {et~al.}(2016)\citenamefont
  {Mattelaer}, \citenamefont {Mitra},\ and\ \citenamefont
  {Ruiz}}]{Mattelaer:2016ynf}%
  \BibitemOpen
  \bibfield  {author} {\bibinfo {author} {\bibfnamefont {O.}~\bibnamefont
  {Mattelaer}}, \bibinfo {author} {\bibfnamefont {M.}~\bibnamefont {Mitra}},\
  and\ \bibinfo {author} {\bibfnamefont {R.}~\bibnamefont {Ruiz}},\ }\href@noop
  {} {\bibfield  {journal} {\bibinfo  {journal} {arXiv}\ } (\bibinfo {year}
  {2016})},\ \Eprint {https://arxiv.org/abs/1610.08985} {arXiv:1610.08985
  [hep-ph]} \BibitemShut {NoStop}%
\bibitem [{\citenamefont {Fuks}\ and\ \citenamefont
  {Ruiz}(2017)}]{Fuks:2017vtl}%
  \BibitemOpen
  \bibfield  {author} {\bibinfo {author} {\bibfnamefont {B.}~\bibnamefont
  {Fuks}}\ and\ \bibinfo {author} {\bibfnamefont {R.}~\bibnamefont {Ruiz}},\
  }\href {https://doi.org/10.1007/JHEP05(2017)032} {\bibfield  {journal}
  {\bibinfo  {journal} {JHEP}\ }\textbf {\bibinfo {volume} {2017}}\bibfield
  {number} {\bibinfo  {number} { (05)},\ \bibinfo {pages} {032}},\ }\Eprint
  {https://arxiv.org/abs/1701.05263} {arXiv:1701.05263 [hep-ph]} \BibitemShut
  {NoStop}%
\bibitem [{\citenamefont {Fuks}\ \emph {et~al.}(2021)\citenamefont {Fuks},
  \citenamefont {Neundorf}, \citenamefont {Peters}, \citenamefont {Ruiz},\ and\
  \citenamefont {Saimpert}}]{Fuks:2020zbm}%
  \BibitemOpen
  \bibfield  {author} {\bibinfo {author} {\bibfnamefont {B.}~\bibnamefont
  {Fuks}}, \bibinfo {author} {\bibfnamefont {J.}~\bibnamefont {Neundorf}},
  \bibinfo {author} {\bibfnamefont {K.}~\bibnamefont {Peters}}, \bibinfo
  {author} {\bibfnamefont {R.}~\bibnamefont {Ruiz}},\ and\ \bibinfo {author}
  {\bibfnamefont {M.}~\bibnamefont {Saimpert}},\ }\href
  {https://doi.org/10.1103/PhysRevD.103.115014} {\bibfield  {journal} {\bibinfo
   {journal} {Phys. Rev. D}\ }\textbf {\bibinfo {volume} {103}},\ \bibinfo
  {pages} {115014} (\bibinfo {year} {2021})},\ \Eprint
  {https://arxiv.org/abs/2012.09882} {arXiv:2012.09882 [hep-ph]} \BibitemShut
  {NoStop}%
\bibitem [{\citenamefont {Romao}\ and\ \citenamefont
  {Silva}(2012)}]{Romao:2012pq}%
  \BibitemOpen
  \bibfield  {author} {\bibinfo {author} {\bibfnamefont {J.~C.}\ \bibnamefont
  {Romao}}\ and\ \bibinfo {author} {\bibfnamefont {J.~P.}\ \bibnamefont
  {Silva}},\ }\href {https://doi.org/10.1142/S0217751X12300256} {\bibfield
  {journal} {\bibinfo  {journal} {Int. J. Mod. Phys. A}\ }\textbf {\bibinfo
  {volume} {27}},\ \bibinfo {pages} {1230025} (\bibinfo {year} {2012})},\
  \Eprint {https://arxiv.org/abs/1209.6213} {arXiv:1209.6213 [hep-ph]}
  \BibitemShut {NoStop}%
\bibitem [{\citenamefont {Dawson}\ \emph {et~al.}(2021)\citenamefont {Dawson},
  \citenamefont {Giardino},\ and\ \citenamefont {Homiller}}]{Dawson:2021jcl}%
  \BibitemOpen
  \bibfield  {author} {\bibinfo {author} {\bibfnamefont {S.}~\bibnamefont
  {Dawson}}, \bibinfo {author} {\bibfnamefont {P.~P.}\ \bibnamefont
  {Giardino}},\ and\ \bibinfo {author} {\bibfnamefont {S.}~\bibnamefont
  {Homiller}},\ }\href {https://doi.org/10.1103/PhysRevD.103.075016} {\bibfield
   {journal} {\bibinfo  {journal} {Phys. Rev. D}\ }\textbf {\bibinfo {volume}
  {103}},\ \bibinfo {pages} {075016} (\bibinfo {year} {2021})},\ \Eprint
  {https://arxiv.org/abs/2102.02823} {arXiv:2102.02823 [hep-ph]} \BibitemShut
  {NoStop}%
\bibitem [{\citenamefont {Nemev\v{s}ek}\ \emph
  {et~al.}(2024{\natexlab{a}})\citenamefont {Nemev\v{s}ek}, \citenamefont
  {Nesti},\ and\ \citenamefont {Kriewald}}]{LRSM-tools}%
  \BibitemOpen
  \bibfield  {author} {\bibinfo {author} {\bibfnamefont {M.}~\bibnamefont
  {Nemev\v{s}ek}}, \bibinfo {author} {\bibfnamefont {F.}~\bibnamefont
  {Nesti}},\ and\ \bibinfo {author} {\bibfnamefont {J.}~\bibnamefont
  {Kriewald}},\ }\href
  {https://sites.google.com/site/leftrighthep/1-lrsm-feynrules} {\bibinfo
  {title} {{LRSM tools page}}} (\bibinfo {year}
  {2024}{\natexlab{a}})\BibitemShut {NoStop}%
\bibitem [{\citenamefont {Nemev\v{s}ek}\ \emph
  {et~al.}(2024{\natexlab{b}})\citenamefont {Nemev\v{s}ek}, \citenamefont
  {Nesti},\ and\ \citenamefont {Kriewald}}]{mLRSM-1.0-model}%
  \BibitemOpen
  \bibfield  {author} {\bibinfo {author} {\bibfnamefont {M.}~\bibnamefont
  {Nemev\v{s}ek}}, \bibinfo {author} {\bibfnamefont {F.}~\bibnamefont
  {Nesti}},\ and\ \bibinfo {author} {\bibfnamefont {J.}~\bibnamefont
  {Kriewald}},\ }\href {https://feynrules.irmp.ucl.ac.be/wiki/LRSM_NLO}
  {\bibinfo {title} {{LRSM FeynRules model file, version 1.0 NLO}}} (\bibinfo
  {year} {2024}{\natexlab{b}})\BibitemShut {NoStop}%
\bibitem [{\citenamefont {Grimus}\ and\ \citenamefont
  {Lavoura}(2000)}]{Grimus:2000vj}%
  \BibitemOpen
  \bibfield  {author} {\bibinfo {author} {\bibfnamefont {W.}~\bibnamefont
  {Grimus}}\ and\ \bibinfo {author} {\bibfnamefont {L.}~\bibnamefont
  {Lavoura}},\ }\href {https://doi.org/10.1088/1126-6708/2000/11/042}
  {\bibfield  {journal} {\bibinfo  {journal} {JHEP}\ }\textbf {\bibinfo
  {volume} {2000}}\bibfield  {number} {\bibinfo  {number} { (11)},\ \bibinfo
  {pages} {042}},\ }\Eprint {https://arxiv.org/abs/hep-ph/0008179}
  {arXiv:hep-ph/0008179} \BibitemShut {NoStop}%
\bibitem [{\citenamefont {Akhmedov}\ and\ \citenamefont
  {Rodejohann}(2008)}]{Akhmedov:2008tb}%
  \BibitemOpen
  \bibfield  {author} {\bibinfo {author} {\bibfnamefont {E.~K.}\ \bibnamefont
  {Akhmedov}}\ and\ \bibinfo {author} {\bibfnamefont {W.}~\bibnamefont
  {Rodejohann}},\ }\href {https://doi.org/10.1088/1126-6708/2008/06/106}
  {\bibfield  {journal} {\bibinfo  {journal} {JHEP}\ }\textbf {\bibinfo
  {volume} {2008}}\bibfield  {number} {\bibinfo  {number} { (06)},\ \bibinfo
  {pages} {106}},\ }\Eprint {https://arxiv.org/abs/0803.2417} {arXiv:0803.2417
  [hep-ph]} \BibitemShut {NoStop}%
\bibitem [{\citenamefont {Peskin}\ and\ \citenamefont
  {Schroeder}(1995)}]{Peskin:1995ev}%
  \BibitemOpen
  \bibfield  {author} {\bibinfo {author} {\bibfnamefont {M.~E.}\ \bibnamefont
  {Peskin}}\ and\ \bibinfo {author} {\bibfnamefont {D.~V.}\ \bibnamefont
  {Schroeder}},\ }\href@noop {} {\emph {\bibinfo {title} {{An Introduction to
  quantum field theory}}}}\ (\bibinfo  {publisher} {Addison-Wesley},\ \bibinfo
  {address} {Reading, USA},\ \bibinfo {year} {1995})\BibitemShut {NoStop}%
\bibitem [{\citenamefont {Cheng}\ and\ \citenamefont
  {Li}(1984)}]{Cheng:1984vwu}%
  \BibitemOpen
  \bibfield  {author} {\bibinfo {author} {\bibfnamefont {T.-P.}\ \bibnamefont
  {Cheng}}\ and\ \bibinfo {author} {\bibfnamefont {L.-F.}\ \bibnamefont {Li}},\
  }\href@noop {} {\emph {\bibinfo {title} {{Gauge Theory of Elementary Particle
  Physics}}}}\ (\bibinfo  {publisher} {Oxford University Press},\ \bibinfo
  {address} {Oxford, UK},\ \bibinfo {year} {1984})\BibitemShut {NoStop}%
\bibitem [{\citenamefont {Belyaev}\ \emph {et~al.}(2013)\citenamefont
  {Belyaev}, \citenamefont {Christensen},\ and\ \citenamefont
  {Pukhov}}]{Belyaev:2012qa}%
  \BibitemOpen
  \bibfield  {author} {\bibinfo {author} {\bibfnamefont {A.}~\bibnamefont
  {Belyaev}}, \bibinfo {author} {\bibfnamefont {N.~D.}\ \bibnamefont
  {Christensen}},\ and\ \bibinfo {author} {\bibfnamefont {A.}~\bibnamefont
  {Pukhov}},\ }\href {https://doi.org/10.1016/j.cpc.2013.01.014} {\bibfield
  {journal} {\bibinfo  {journal} {Comput. Phys. Commun.}\ }\textbf {\bibinfo
  {volume} {184}},\ \bibinfo {pages} {1729} (\bibinfo {year} {2013})},\ \Eprint
  {https://arxiv.org/abs/1207.6082} {arXiv:1207.6082 [hep-ph]} \BibitemShut
  {NoStop}%
\bibitem [{\citenamefont {Frixione}\ \emph {et~al.}(2019)\citenamefont
  {Frixione}, \citenamefont {Fuks}, \citenamefont {Hirschi}, \citenamefont
  {Mawatari}, \citenamefont {Shao}, \citenamefont {Sunder},\ and\ \citenamefont
  {Zaro}}]{Frixione:2019fxg}%
  \BibitemOpen
  \bibfield  {author} {\bibinfo {author} {\bibfnamefont {S.}~\bibnamefont
  {Frixione}}, \bibinfo {author} {\bibfnamefont {B.}~\bibnamefont {Fuks}},
  \bibinfo {author} {\bibfnamefont {V.}~\bibnamefont {Hirschi}}, \bibinfo
  {author} {\bibfnamefont {K.}~\bibnamefont {Mawatari}}, \bibinfo {author}
  {\bibfnamefont {H.-S.}\ \bibnamefont {Shao}}, \bibinfo {author}
  {\bibfnamefont {P.~A.}\ \bibnamefont {Sunder}},\ and\ \bibinfo {author}
  {\bibfnamefont {M.}~\bibnamefont {Zaro}},\ }\href
  {https://doi.org/10.1007/JHEP12(2019)008} {\bibfield  {journal} {\bibinfo
  {journal} {JHEP}\ }\textbf {\bibinfo {volume} {2019}}\bibfield  {number}
  {\bibinfo  {number} { (12)},\ \bibinfo {pages} {008}},\ }\Eprint
  {https://arxiv.org/abs/1907.04898} {arXiv:1907.04898 [hep-ph]} \BibitemShut
  {NoStop}%
\bibitem [{\citenamefont {Hahn}(2001)}]{Hahn:2000kx}%
  \BibitemOpen
  \bibfield  {author} {\bibinfo {author} {\bibfnamefont {T.}~\bibnamefont
  {Hahn}},\ }\href {https://doi.org/10.1016/S0010-4655(01)00290-9} {\bibfield
  {journal} {\bibinfo  {journal} {Comput. Phys. Commun.}\ }\textbf {\bibinfo
  {volume} {140}},\ \bibinfo {pages} {418} (\bibinfo {year} {2001})},\ \Eprint
  {https://arxiv.org/abs/hep-ph/0012260} {arXiv:hep-ph/0012260} \BibitemShut
  {NoStop}%
\bibitem [{\citenamefont {Degrande}(2015)}]{Degrande:2014vpa}%
  \BibitemOpen
  \bibfield  {author} {\bibinfo {author} {\bibfnamefont {C.}~\bibnamefont
  {Degrande}},\ }\href {https://doi.org/10.1016/j.cpc.2015.08.015} {\bibfield
  {journal} {\bibinfo  {journal} {Comput. Phys. Commun.}\ }\textbf {\bibinfo
  {volume} {197}},\ \bibinfo {pages} {239} (\bibinfo {year} {2015})},\ \Eprint
  {https://arxiv.org/abs/1406.3030} {arXiv:1406.3030 [hep-ph]} \BibitemShut
  {NoStop}%
\bibitem [{\citenamefont {Alwall}\ \emph {et~al.}(2014)\citenamefont {Alwall},
  \citenamefont {Frederix}, \citenamefont {Frixione}, \citenamefont {Hirschi},
  \citenamefont {Maltoni}, \citenamefont {Mattelaer}, \citenamefont {Shao},
  \citenamefont {Stelzer}, \citenamefont {Torrielli},\ and\ \citenamefont
  {Zaro}}]{Alwall:2014hca}%
  \BibitemOpen
  \bibfield  {author} {\bibinfo {author} {\bibfnamefont {J.}~\bibnamefont
  {Alwall}}, \bibinfo {author} {\bibfnamefont {R.}~\bibnamefont {Frederix}},
  \bibinfo {author} {\bibfnamefont {S.}~\bibnamefont {Frixione}}, \bibinfo
  {author} {\bibfnamefont {V.}~\bibnamefont {Hirschi}}, \bibinfo {author}
  {\bibfnamefont {F.}~\bibnamefont {Maltoni}}, \bibinfo {author} {\bibfnamefont
  {O.}~\bibnamefont {Mattelaer}}, \bibinfo {author} {\bibfnamefont {H.~S.}\
  \bibnamefont {Shao}}, \bibinfo {author} {\bibfnamefont {T.}~\bibnamefont
  {Stelzer}}, \bibinfo {author} {\bibfnamefont {P.}~\bibnamefont {Torrielli}},\
  and\ \bibinfo {author} {\bibfnamefont {M.}~\bibnamefont {Zaro}},\ }\href
  {https://doi.org/10.1007/JHEP07(2014)079} {\bibfield  {journal} {\bibinfo
  {journal} {JHEP}\ }\textbf {\bibinfo {volume} {2014}}\bibfield  {number}
  {\bibinfo  {number} { (07)},\ \bibinfo {pages} {079}},\ }\Eprint
  {https://arxiv.org/abs/1405.0301} {arXiv:1405.0301 [hep-ph]} \BibitemShut
  {NoStop}%
\bibitem [{\citenamefont {Buckley}\ \emph {et~al.}(2015)\citenamefont
  {Buckley}, \citenamefont {Ferrando}, \citenamefont {Lloyd}, \citenamefont
  {Nordstr\"om}, \citenamefont {Page}, \citenamefont {R\"ufenacht},
  \citenamefont {Sch\"onherr},\ and\ \citenamefont {Watt}}]{Buckley:2014ana}%
  \BibitemOpen
  \bibfield  {author} {\bibinfo {author} {\bibfnamefont {A.}~\bibnamefont
  {Buckley}}, \bibinfo {author} {\bibfnamefont {J.}~\bibnamefont {Ferrando}},
  \bibinfo {author} {\bibfnamefont {S.}~\bibnamefont {Lloyd}}, \bibinfo
  {author} {\bibfnamefont {K.}~\bibnamefont {Nordstr\"om}}, \bibinfo {author}
  {\bibfnamefont {B.}~\bibnamefont {Page}}, \bibinfo {author} {\bibfnamefont
  {M.}~\bibnamefont {R\"ufenacht}}, \bibinfo {author} {\bibfnamefont
  {M.}~\bibnamefont {Sch\"onherr}},\ and\ \bibinfo {author} {\bibfnamefont
  {G.}~\bibnamefont {Watt}},\ }\bibfield  {journal} {\bibinfo  {journal} {Eur.
  Phys. J. C}\ }\textbf {\bibinfo {volume} {75}},\ \href
  {https://doi.org/10.1140/epjc/s10052-015-3318-8}
  {10.1140/epjc/s10052-015-3318-8} (\bibinfo {year} {2015}),\ \Eprint
  {https://arxiv.org/abs/1412.7420} {arXiv:1412.7420 [hep-ph]} \BibitemShut
  {NoStop}%
\bibitem [{\citenamefont {Ball}\ \emph {et~al.}(2022)\citenamefont {Ball} \emph
  {et~al.}}]{NNPDF:2021njg}%
  \BibitemOpen
  \bibfield  {author} {\bibinfo {author} {\bibfnamefont {R.~D.}\ \bibnamefont
  {Ball}} \emph {et~al.} (\bibinfo {collaboration} {NNPDF}),\ }\href
  {https://doi.org/10.1140/epjc/s10052-022-10328-7} {\bibfield  {journal}
  {\bibinfo  {journal} {Eur. Phys. J. C}\ }\textbf {\bibinfo {volume} {82}},\
  \bibinfo {pages} {428} (\bibinfo {year} {2022})},\ \Eprint
  {https://arxiv.org/abs/2109.02653} {arXiv:2109.02653 [hep-ph]} \BibitemShut
  {NoStop}%
\bibitem [{\citenamefont {Hoger}\ and\ \citenamefont
  {Carlson}(1984)}]{Hoger:1984sqm}%
  \BibitemOpen
  \bibfield  {author} {\bibinfo {author} {\bibfnamefont {A.}~\bibnamefont
  {Hoger}}\ and\ \bibinfo {author} {\bibfnamefont {D.~E.}\ \bibnamefont
  {Carlson}},\ }\href {https://doi.org/10.1016/0898-1221(89)90240-X} {\bibfield
   {journal} {\bibinfo  {journal} {Quarterly of Applied Mathematics}\ }\textbf
  {\bibinfo {volume} {42}},\ \bibinfo {pages} {113} (\bibinfo {year}
  {1984})}\BibitemShut {NoStop}%
\bibitem [{\citenamefont {Franca}(1989)}]{Franca:1989sqm}%
  \BibitemOpen
  \bibfield  {author} {\bibinfo {author} {\bibfnamefont {L.~P.}\ \bibnamefont
  {Franca}},\ }\href@noop {} {\bibfield  {journal} {\bibinfo  {journal}
  {Computers \& Mathematics with Applications}\ }\textbf {\bibinfo {volume}
  {18}},\ \bibinfo {pages} {459} (\bibinfo {year} {1989})}\BibitemShut
  {NoStop}%
\bibitem [{\citenamefont {M.~Ashkartizabi}\ and\ \citenamefont
  {Mohammadpour}(2017)}]{Ashkartizabi:2016sqm}%
  \BibitemOpen
  \bibfield  {author} {\bibinfo {author} {\bibfnamefont {M.~A.}\ \bibnamefont
  {M.~Ashkartizabi}}\ and\ \bibinfo {author} {\bibfnamefont {A.}~\bibnamefont
  {Mohammadpour}},\ }\href {https://doi.org/10.1007/s10092-016-0202-3}
  {\bibfield  {journal} {\bibinfo  {journal} {Calcolo}\ }\textbf {\bibinfo
  {volume} {54}},\ \bibinfo {pages} {643} (\bibinfo {year} {2017})}\BibitemShut
  {NoStop}%
\end{thebibliography}%

\end{document}